\documentclass[a4paper,11pt]{article}
\pdfoutput=1 

\usepackage{jheppub} 

\usepackage{graphicx}
\usepackage{dcolumn}
\usepackage{bm}
\usepackage{color}

\usepackage{feynmf}
\usepackage{slashed}
\usepackage{enumerate}
\usepackage{caption}

\usepackage[utf8]{inputenc}

\usepackage[T1]{fontenc} 

\title{\boldmath Dark Matter Bound State Formation in Fermionic $Z_2$ DM model with 
Light Dark Photon and Dark Higgs Boson}

\author[b,c]{Pyungwon Ko}
\author[b,d]{Toshinori Matsui}
\author[a,c,1]{Yi-Lei Tang\note{Corresponding author.}}

\affiliation[a]{School of Physics, Sun Yat-Sen University, Guangzhou 510275, China}
\affiliation[b]{School of Physics, KIAS, 85 Hoegiro, Seoul 02455, Republic of Korea}
\affiliation[c]{Quantum Universe Center, KIAS, 85 Hoegiro, Seoul 02455, Republic of Korea}
\affiliation[d]{Department of Physics, KAIST, 291 Daehakro, Daejeon 34141, Republic of Korea}

\emailAdd{pko@kias.re.kr}
\emailAdd{matsui@kaist.ac.kr}
\emailAdd{tangylei@mail.sysu.edu.cn}

\abstract{
If fermionic dark matter (DM) is stabilized by dark $U(1)$ gauge symmetry that is spontaneously 
broken into its subgroup $Z_2$,  the particle contents of the model becomes very rich: 
DM and  excited DM, both of them are Majorana fermions, as well as two dark force mediators, 
dark photon and dark Higgs boson are naturally present due to the underlying dark gauge symmetry.  
In this paper,
we study the DM bound state formation processes within this scenario, assuming  both dark photon 
and dark Higgs are light mediators and including the effects of excited DM.  
The Goldstone boson contributions to the potential matrix in the Schr\"{o}dinger equations are found to be important.  The emissions of a longitudinal vector boson 
(or somehow equivalently a Goldstone boson) during the DM bound state formations are crucial to induce
a significant reannihilation process, reducing the dark matter relic abundance.
Most of the stringent constraints for this kind of dark matter considered in the literature are simply evaded.}

\begin{document} 
\maketitle
\flushbottom

\section{Introduction}

A number of cosmological observations through gravitational interaction indicate that about 
25\% of the energy budget of the current Universe consists of nonbaryonic dark matter (DM). 
So far almost nothing is known about the physical nature of DM: the number of DM species in the universe, their masses and spins, and their interactions among themselves and with the Standard Model (SM) particles. These can be revealed only by nongravitational observation of physical effects related with DM particles.  And various types of DM searches have been performed all around the globe.

From the view point of  particle physics described by quantum field theory,  one of the most important 
and fundamental properties of DM is that it should be absolutely stable or long-lived enough 
in order to make DM of the Universe. 
Let us remind ourselves that electron stability in the SM is related with unbroken $U(1)_{\rm em}$ 
and massless photon.  The longevity of proton is also attributed to the baryon number being an accidental 
global symmetry of the SM and being broken only by dim-6 operators. Likewise, one can assume  
that the absolute stability of DM is due to some local dark gauge symmetry, and long-lived DM 
is due to some accidental global symmetry of the underlying dark gauge symmetry 
\footnote{There are other possibilities:  lightest supersymmetric particles (including massive 
gravitino case) become good cold dark matter (CDM) if  $R$-parity is assumed to be conserved.
Or lightness of DM particles can make them long-lived enough, e.g. light axions or sterile neutrinos.
We do not consider these possibilities since there are no light force mediators that can make DM 
bound states, which is the main theme of this paper.}.  Then this class of DM models come with 
extra particles such as dark photon (or dark gauge bosons), dark Higgs and sometimes excited 
DM because of the underlying local dark gauge symmetries.  Depending on the mass scales of these
new particles and their interaction strengths, one can imagine new interesting phenomenology would
be anticipated in the dark sectors.  In particular if some of them are light, they can play the role of 
light mediators which are often introduced to the DM phenomenology in order to solve some puzzles
in the vanilla CDM paradigm. In short one can introduce light mediators in order to keep gauge 
invariance, which is well tested principle in the SM.

In the literature, based on the weakly interacting massive particle (WIMP) models \footnote{See Ref.~\cite{DanHooperClassic} for example}, scalar/vector light force mediators interchanged 
among the dark matter particles are sometimes introduced as a solution to some problems that 
the vanilla CDM models encounter. Such models are called the self-interacting dark matter 
(SIDM). The attractive/repulsive forces between the two dark matter particles can enhance/reduce 
the annihilation cross section times the relative speed ($\sigma v$) which is an important input for  
determination of both thermal relic density and indirect detection signatures of DM. 
This is called the Sommerfeld effect \footnote{For the original work by A. Sommerfeld, see \cite{SommerfeldOriginal}. And for some early applications in the dark matter, see \cite{Hisano0, 
Hisano1, Hisano2, Hisano3, EarlySommer1, EarlySommer2, Minimal1, Minimal2, EarlySommer3, 
EarlySommer4, EarlySommer5, EarlySommer6, EarlySommer7, NimaSommerfeld}}. If the annihilation rate of the dark matter particles in our galaxy is boosted by this effect, SIDM might become a solution to the positron excess observed by PAMELA and AMS-02 \cite{PAMELA1, AMS1, AMS2}. SIDM also provides the potential to resolve the ``missing-satellite problem'' \cite{MissingSatellite1, MissingSatellite2}, the ``core-cusp problem'' \cite{CoreCusp1, CoreCusp2}, and the ``too-big-to-fail problem'' \cite{TooBigToFail1, TooBigToFail2, TooBigToFail3}. These problems are beyond the scope of this paper, and due to the controversies on these issues \cite{ThreeProblemsSolution1, ThreeProblemsSolution2, ThreeProblemsSolution3}, we do not consider these effects, but only point out that the SIDM models are stringently constrained by the CMB distortion observations\cite{CMBBound}. The cluster observations and simulations, e.g., the bullet cluster also constrains the self-interaction parameters of the dark matter particles \cite{Bullet, Cluster1, Cluster2, Cluster3, Cluster4, Cluster5, Cluster6}.

The Sommerfeld effects are the resonant effect of the so-called ``zero-energy'' bound state of a dark matter particle pair. If the interaction is sufficiently strong and the mediator is sufficiently light, the dark matter can also form a real bound state while emitting a mediator particle to keep the energy conservation. Many of the models are built and calculated \footnote{See Ref.~\cite{Petraki1} for the references on the early dark matter bound state models therein}.  Ref.~\cite{Petraki1, Petraki2} had shown the general derivations of the dark matter bound state formations on various situations with the tools of Bethe-Salpeter wave functions. For the applications on the WIMP model, Ref.~\cite{EllisFengLuo} calculated the modified Boltzmann equation including the contributions from the bound state formations. Its Eqn.~(34) clearly shows the competition of the decay and dissociation of the bound state particles. Ref.~\cite{DMBS1, DMBS2, DMBS3, DMBS4, DMBS5, DMBS6, DMBS7, DMBS8, DMBS9, DMBS10, DMBS11, DMBS12, DMBS13, DMBS14, DMBS15, DMBS16, DMBS17, DMBS18, DMBS19, DMBS20, DMBS21, DMBS22, DMBS23, DMBS24, DMBS25, DMBS26, DMBS27, DMBS28, DMBS29, DMBS30, DMBS31, DMBS32, DMBS33, DMBSRequested} are the recent papers which had built or calculated the dark matter models in which bound state can be formed.

In this paper, we shall consider a case where DM is absolutely stable due to the unbroken $Z_2$ 
symmetry assumed to be the remnant of an underlying local $U(1)$ dark 
gauge symmetry \footnote{The case of the scalar DM model with local $Z_2$ and $Z_3$ symmetries 
were considered in Refs.~\cite{Baek:2014kna} and \cite{Ko:2014nha,Ko:2014loa}, respectively. 
Similar models have been discussed in Ref.~\cite{GaugeDiscreteEarly1, GaugeDiscreteEarly2, PreModelAdded1, PreModel1, PreModel2, Premodel3, PreModel4} in different contexts. }.
In this case, both the dark photon and dark Higgs boson are mediators of the dark force. 
It should be interesting to study the dark matter bound state formation in addition to the Sommerfeld 
enhancement in such a scenario. We will find that the emission of the longitudinal dark photon 
plays a crucial role compared with the Refs.~\cite{Petraki1, Petraki2}.  
Together with the situation of the dark Higgs boson emission, these processes are controlled by 
the wave function ``overlap'' $\mathcal{I}$, and its zeroth order expansion is no longer zero 
in our case, unlike  in Refs.~\cite{Petraki1, Petraki2}.  
This will affect significantly the DM relic density after the first-step annihilation of the DM particles, 
and such a second epoch process is called the ``reannihilation'' \cite{reannihilation1, reannihilation2, reannihilation3, reannihilation4, reannihilation5} in the literature. In our model, the reannihilation 
mainly occurs in the co-annihilation channel of the dark matter and its nearly-degenerated partner.  
Therefore, the suppression of the relative number density of the heavier component automatically switches off the re-annihilation before $x_f \lesssim 10^6$. Such an early re-annihilation does not 
disturb the Big-Bang Nucleosynthesis (BBN), as well as the cosmological epochs afterwards.
 
Another possibility for the DM stability is to assume a global dark symmetry.
For example in Ref.~\cite{Weinberg:2013kea},  S. Weinberg introduces a global dark $U(1)$ 
symmetry that is spontaneously broken into its $Z_2$ subgroup. In that framework, the Goldstone 
boson becomes the fractional cosmic neutrinos (or dark radiation) and is constrained by CMB and 
other cosmological observations.  However DM-stabilizing global symmetry may be broken by 
non-renormalizable operators, especially due to the gravity effects, which may induce fast decay of the DM 
particle with the O(10) GeV or heavier masses. This issue could be simply evaded by implementing
a global dark gauge symmetry to its local version.
Compared with the global dark $U(1)$ symmetry, the local $U(1)$ dark gauge symmetry could 
guarantee the DM stability even in the presence of the non-renormalizable higher dimensional operators 
(see discussions in Refs.~\cite{Baek:2011aa,Ko:2014nha}).  And due to the existence/absence of 
the dark photon, the resulting phenomenology varies significantly in these two situations. 
For example, the viable mass ranges of the DM particles would be completely different in both cases 
\footnote{Work in preparation, with Seungwon Baek, Toshinori Matsui and Wan Il Park.}.

In some literature, one introduces  the soft-breaking term to explicitly break the local or global 
dark $U(1)$ symmetry without a detailed mechanism~\cite{SoftBreaking1, SoftBreaking2}, 
or considers non-renormalizable interactions~\cite{SoftBreaking3}.   
These models suffer from the potential  risk to break the unitarity
\footnote{See Appendix \ref{UnitarityDetails} for a detailed calculation.}. 
One can cure this problem by introducing the dark Higgs mechanism to spontaneously break the 
dark $U(1)$ symmetry and keeping only the renormalizable couplings between the dark Higgs and 
the fermionic dark matter. 

The paper is organized as follows. In Sec.~2, we show our relied Lagrangian of the $Z_2$ Fermionic model. Some basic features of this model are also discussed. In Sec. 3, we classify the bound states by their quantum numbers. The potentials in the schr\"{o}dinger equations are also derived. These potentials are generated by the mediation of the dark photon and dark Higgs boson exchanged between the (excited)  DM particles, and in particular, we derive the potential terms induced by the longitudinal dark photon, or equivalently the Goldstone boson for the first time to our best knowledge.  
In Sec.~4, we illustrate the methods to calculate the bound state formation cross sections and 
the bound state decay induced by the component annihilation in our model.  
The modified Boltzmann equations are also demonstrated. 
In Sec.~5, we present the numerical results based on the formulas in the previous sections.  
Experimental constraints and comparisons with some earlier literature which ignored the longitudinal 
dark photon are also presented. Finally we summarize  in Sec.~6 with future prospects.  
A number of technical issues are described in detail in Appendices.

\section{Model Setup}
We start from a dark $U(1)$ model, with a Dirac fermion dark matter (DM) $\chi$ appointed with a nonzero dark $U(1)$ charge $Q_\chi$ and dark photon.   We also introduce a complex dark Higgs field $\Phi$, which takes a nonzero vacuum expectation value, generating nonzero mass for the dark photon. We shall consider a special case where $\Phi$ breaks the dark $U(1)$ symmetry into a dark $Z_2$ symmetry with a judicious choice of its dark charge $Q_\Phi$. 

Then the gauge invariant and renormalizable Lagrangian for this system is given by 
\begin{eqnarray}
\mathcal{L} & = & -\frac{1}{4} F^{\prime \mu \nu} F^{\prime}_{\mu \nu} - \frac{\epsilon}{4} 
F_{\mu\nu}^{'} B^{\mu\nu} + \overline{\chi} D\!\!\!\!/ \chi - m_{\chi} \overline{\chi} \chi 
+ D_{\mu} \Phi^\dagger D^{\mu} \Phi 
\label{BasicLag}
\\
& - & \mu^2 \Phi^\dagger \Phi - \lambda_{\Phi} |\Phi|^4 + ( \frac{\sqrt{2}}{2} y \Phi \overline{\chi^C} \chi + \text{h.c.}) 
- \lambda_{\Phi H} \Phi^\dagger \Phi H^\dagger H
\nonumber 
\end{eqnarray}
where $F^{\prime}_{\mu \nu} = \partial_{\mu} A^{\prime}_{\nu} - \partial_{\nu} A^{\prime}_{\mu}$, and $A^{\prime}_{\mu}$ is the dark $U(1)$ gauge field. $D_{\mu} = \partial_{\mu} + i Q g A^{\prime}_{\mu}$ is the covariant derivative, where $g$ is the dark coupling constant, and $Q$ is the dark charge that $\Phi$ and $\chi$ takes. Note that we made a judicious choice,   
\[
Q_{\Phi}  = -2 Q_{\chi} ,
\]
in order to allow the $( y \Phi \overline{\chi^C} \chi + \text{h.c.})$ term, which will give rise to the mass splitting of the dark matter components. Here $\chi^C$ means the charge conjugate of the Dirac field $\chi$, and $B_{\mu \nu} = \partial_{\mu} B_{\nu} - \partial_{\nu} B_{\mu}$ is the field strength tensor associated with the SM $U(1)_Y$ hypercharge gauge field. The kinetic mixing term ($\propto \epsilon$) and the Higgs portal interaction ($\propto \lambda_{\Phi H}$) communicate the dark and the standard model sectors. These parameters are constrained by various experimental results, especially for the $\epsilon$ from the dark photon searches. The $\lambda_{\Phi H}$ is constrained by collider searches for the exotic and invisible Higgs decay widths and the direct detections of fermionic Higgs-portal dark matter. More details on such constraints will be addressed in Sec.~\ref{Constraints}. However, an appropriate value of $\lambda_{\Phi H}$ within the constraints is enough to contact 
the dark and the standard model sectors, keeping them to be in thermal equilibrium in the early 
universe. We will explain the details in later discussions.

Let us first decompose $\chi$ into two Weyl spinors, 
\[
\chi = \left[ \begin{array}{c} \chi_L \\ i \sigma^2 \chi_R^* \end{array}\right] .
\]
For $\mu^2 < 0$, the dark Higgs $\Phi$ will take a nonzero vacuum expectation value $v_\Phi$, and 
we expand it as the following:
\[
\Phi = \frac{v_{\Phi}+R+i I}{\sqrt{2}} . 
\]
Written in the basis of $\chi_L$ and $\chi_R$, the mass matrix of the fermions becomes 
\begin{eqnarray}
\mathcal{L} \supset \frac{1}{2} [ \begin{array}{cc} \chi_L^T & \chi_R^T \end{array} ]
\left[ \begin{array}{cc}
\delta m & m_{\chi}  \\
m_{\chi} & \delta m 
\end{array} \right] \left[
\begin{array}{c}
\chi_L \\
\chi_R
\end{array}  \right] + \text{h.c.},	\label{MassBeforeDiag}
\end{eqnarray}
where $\delta m \equiv y v_{\Phi}$. After diagonalizing (\ref{MassBeforeDiag}), we acquire
\begin{eqnarray}
\chi_1 &=& \frac{1}{\sqrt{2}} (\chi_L - \chi_R ), \nonumber \\
\chi_2 &=& \frac{i}{\sqrt{2}} (\chi_L + \chi_R ),	\label{Spinor2Chi12}
\end{eqnarray}
and the corresponding mass matrix
\begin{eqnarray}
\frac{1}{2} [ \begin{array}{cc} \chi_1^T & \chi_2^T \end{array} ]
\left[ \begin{array}{cc}
m_{\chi}-\delta m & 0 \\
0 & m_{\chi} +  \delta m
\end{array} \right] \left[
\begin{array}{c}
\chi_1 \\
\chi_2
\end{array}  \right] + \text{h.c.}. \label{MassAfterDiag}
\end{eqnarray}
We can clearly see that one Dirac fermion $\chi$ splits into two nearly-degenerate Majorana fermions $\chi_1$ and $\chi_2$ if $\delta m \ll m_{\chi}$. Both of them are odd in a remained dark $Z_2$ symmetry, while all the other particles are even. The lighter one is the candidate of dark matter. Its stability is preserved by this dark $Z_2$ symmetry. In this paper, without loss of generality, we adopt $\delta m >0$ thus $\chi_1$ is the dark matter particle.

After $\Phi (x)$ takes the nonzero vacuum expectation value
\begin{eqnarray}
v_{\Phi} = \sqrt{\frac{\mu^2}{\lambda_{\Phi}}}, \label{vevAcquired}
\end{eqnarray}
$R$ and dark photon acquires a positive mass
\begin{eqnarray}
m_{R} &=& \sqrt{2} \mu, \nonumber \\
m_{\gamma^{\prime}} &=& |Q_{\Phi} g v_{\Phi}|. \label{BosonMasses}
\end{eqnarray}
The Goldstone boson (GB) ``$I$'' is ``eaten'' by the dark photon to become its longitudinal mode. However, it is convenient to apply for the ``Goldstone equivalent theorem'' to calculate and understand some processes. Therefore, the GB $I$ will appear in some following discussions. 

The $\sqrt{2} y \Phi \overline{\chi^C} \chi$ term will induce the following Yukawa interaction terms: 
\begin{eqnarray}
\mathcal{L} \supset \frac{1}{2} [ \begin{array}{cc} \chi_1^T & \chi_2^T \end{array} ]
\left[ \begin{array}{cc}
-y R &  y I\\
y I & y R
\end{array} \right] \left[
\begin{array}{c}
\chi_1 \\
\chi_2
\end{array}  \right] + \text{h.c.}. \label{Yukawa}
\end{eqnarray}
The dark photon interactions with $\chi_1$ and $\chi_2$ become
\begin{eqnarray}
\mathcal{L} \supset  [ \begin{array}{cc} \chi_1^{\dag} & \chi_2^{\dag} \end{array} ]
\left[ \begin{array}{cc}
0 &  Q_{\chi} g A^{\prime} \cdot \sigma \\
-Q_{\chi} g A^{\prime} \cdot \sigma & 0
\end{array} \right] \left[
\begin{array}{c}
\chi_1 \\
\chi_2
\end{array}  \right] + \text{h.c.}. \label{Gauge}
\end{eqnarray}
where $\sigma^\mu = ( 1, \vec{\sigma})$. 
In the following we shall focus on the case of light mediators, 
\[
m_{\gamma^{\prime}}, m_R \ll m_{\chi_1} \lesssim m_{\chi_2}.
\] 

If the dark photon get massive through St\"{u}ckelberg mechanism and the mass difference between $\chi_1$ and $\chi_2$ is generated by soft $U(1)$ breaking term, the dark Higgs degrees of freedom would be absent. And the usual pair annihilation of (excited) DM  occur through the channels, 
\begin{equation}
\chi_i + \chi_i  \rightarrow  \gamma^{\prime} + \gamma^{\prime} ~~~(i=1,2) . 
\end{equation}
On the other hand, if we include the dark Higgs degree of freedom,  we have two more additional channels:
the annihilation into a pair of dark Higgs bosons, 
\begin{equation}
\chi_i + \chi_i  \rightarrow  R + R~~~(i=1,2) , 
\end{equation}
and the co-annihilation channel into dark photon and dark Higgs, 
\begin{equation}
\chi_1 + \chi_2  \rightarrow  \gamma^{\prime} + R . 
\label{Coannihilation} 
\end{equation}
Both of them could be be important during the freeze-out processes.
The $\gamma^{\prime}$-mediated  $s$-channel annihilation, 
\[
\chi_1 + \chi_2  \rightarrow  \gamma^{\prime *} \rightarrow {\rm SM ~particles}
\]  
would be suppressed by the small kinetic mixing between the $\gamma^{\prime}$ and SM vector bosons, so it shall be ignored in our paper. In the current universe, $\chi_2$ had all decayed away, so (\ref{Coannihilation}) is absent when we consider the indirect detection constraints.

\section{Dark Matter Bound State Classification}

In the parameter space $m_{R, \gamma^{\prime}} \ll m_{\chi}$, two $\chi_{1,2}$ particles can form 
bound states by emitting $R$ and/or $\gamma^{'}$. Before calculating the $\chi_{1,2}$ bound state 
formation rates and their relevance to DM phenomenology,  it is beneficial to discuss 
the dark matter bound states when dark $U(1)$ symmetry is strictly conserved. 

\subsection{The case of unbroken $U(1)$}
In the case of unbroken dark $U(1)$, the dark fermion $\chi$, together with its anti-particle $\overline{\chi}$, 
make a pair of Dirac (anti-)fermions. In this case, the $\gamma^{\prime}$ exchange mediates a repulsive interaction 
in the same charged $| \chi \chi \rangle$ or $| \overline{\chi} \overline{\chi} \rangle$ states, 
while it becomes attractive in the different charged $| \overline{\chi} \chi \rangle$ state, just like the situation between electron/electron or electron/positron pairs in ordinary QED. 
Interestingly, the interchange of the boson $\Phi^{(*)}$ will cause the oscillation between 
$| \overline{\chi} \chi \rangle \leftrightarrow | \chi \overline{\chi} \rangle$ states, unlike the usual case that scalars 
will always induce an attractive force. 

The total wave function of such kinds of fermion pairs can be written as
\begin{eqnarray}
\int d^3\vec{x} ( \psi_{\chi \chi} (\vec{x}) | \chi \chi \rangle + \psi_{\overline{\chi} \overline{\chi}} (\vec{x}) |\overline{\chi} \overline{\chi} \rangle ) \otimes  |\vec{x}\rangle \otimes |\text{Spin}\rangle, \nonumber \\
\int d^3\vec{x} ( \psi_{\overline{\chi} \chi} (\vec{x}) | \overline{\chi} \chi \rangle + \psi_{\chi \overline{\chi}} (\vec{x}) |\chi \overline{\chi} \rangle ) \otimes  |\vec{x}\rangle \otimes |\text{Spin}\rangle, \label{TotalWaveOriginal}
\end{eqnarray}
where we define
\begin{eqnarray}
\psi_{\chi \chi \leftrightarrow \overline{\chi} \overline{\chi}}(\vec{x}) &=& \left[
\begin{array}{c}
\psi_{\chi \chi}(\vec{x}) \\
\psi_{\overline{\chi} \overline{\chi}}(\vec{x})
\end{array} \right], \nonumber \\
\psi_{\overline{\chi} \chi \leftrightarrow \chi \overline{\chi}}(\vec{x}) &=& \left[
\begin{array}{c}
\psi_{\overline{\chi} \chi}(\vec{x}) \\
\psi_{\chi \overline{\chi}}(\vec{x})
\end{array} \right]
\end{eqnarray}
as the wave function in the coordinate representation, and $\vec{x}$ is the relative distance between the two particles. The Schr\"{o}dinger equations are given by
\begin{eqnarray}
-\frac{\vec{\nabla}^2}{m_{\chi}} \psi_{\chi \chi \leftrightarrow \overline{\chi} \overline{\chi}}(\vec{x}) + \widetilde{V}_{\text{s}} \psi_{\chi \chi \leftrightarrow \overline{\chi} \overline{\chi}}(\vec{x}) &=& E \psi_{\chi \chi \leftrightarrow \overline{\chi} \overline{\chi}}(\vec{x}), \nonumber \\
-\frac{\vec{\nabla}^2}{m_{\chi}} \psi_{\overline{\chi} \chi \leftrightarrow \chi \overline{\chi}}(\vec{x}) + \widetilde{V}_{\text{d}} \psi_{\overline{\chi} \chi \leftrightarrow \chi \overline{\chi}}(\vec{x}) &=& 
E \psi_{\overline{\chi} \chi \leftrightarrow \chi \overline{\chi}}(\vec{x}),	\label{SchroedingerU1Conserving}
\end{eqnarray}
where
\begin{eqnarray}
\widetilde{V}_{\text{s}} = \left[
\begin{array}{cc}
-V_{\gamma^{\prime}} & \\
& -V_{\gamma^{\prime}}
\end{array} \right],~
\widetilde{V}_{\text{d}} = \left[
\begin{array}{cc}
V_{\gamma^{\prime}} & V_{\Phi} \\
V_{\Phi} & V_{\gamma^{\prime}}
\end{array} \right],	\label{Potential}
\end{eqnarray}
with 
\[
V_{\gamma^{\prime}} = -\frac{(Q_{\chi} g)^2}{4 \pi} \frac{1}{r},  \ \ \ \ \ 
V_{\Phi} = -\frac{2 y^2}{4 \pi} \frac{e^{-m_{\Phi} r}}{r} , 
\]
which contains the contributions from both real and imaginary parts of the $\Phi^{(*)}$. 
Note that $m_{\Phi}$ is the mass of $\Phi$ in the unbroken $U(1)$ symmetry case. Diagonalizing the $V_{\text{d}}^{\prime}$ will simplify each equation in the (\ref{SchroedingerU1Conserving}) to four decoupled equations.

Although $\tilde{V}_{\text{s}}$ has already been diagonalized, we still rotate the basis for further discussions. 
The four corresponding particle states with their potential can then be written as
\begin{eqnarray}
& & \frac{1}{\sqrt{2}} (| \chi \chi \rangle + | \overline{\chi} \overline{\chi} \rangle),~ -V_{\gamma^{\prime}}, \nonumber \\
& & \frac{1}{\sqrt{2}} (| \chi \chi \rangle - | \overline{\chi} \overline{\chi} \rangle) ,~ -V_{\gamma^{\prime}}, \nonumber \\
& & \frac{1}{\sqrt{2}} (| \overline{\chi} \chi \rangle + | \chi \overline{\chi} \rangle),~ V_{\gamma^{\prime}} + V_{\Phi}, \nonumber \\
& & \frac{1}{\sqrt{2}} (| \overline{\chi} \chi \rangle - | \chi \overline{\chi} \rangle), ~ V_{\gamma^{\prime}}-V_{\Phi}. \label{ParticleStateWithPotential}
\end{eqnarray}
It will be convenient to rewrite the above bases in the $|\chi_i \chi_j \rangle$ forms. 
Notice that if we define the 4-spinor with the Weyl spinors defined in (\ref{Spinor2Chi12}),
\begin{eqnarray}
\widetilde{\chi}_1 = \left[
\begin{array}{c}
\chi_1 \\
i \sigma^2 \chi_1^*
\end{array} \right], ~ \widetilde{\chi}_2 = \left[
\begin{array}{c}
i \chi_2 \\
- \sigma^2 \chi_2^*
\end{array} \right],
\end{eqnarray}
we find that the four-spinor $\chi$ and $\chi^C$ can be written in the form of
\begin{eqnarray}
\chi &=& \frac{\widetilde{\chi}_1 - i \widetilde{\chi}_2 }{\sqrt{2}}, \nonumber \\
\chi^C &=& -\frac{\widetilde{\chi}_1 + i \widetilde{\chi}_2}{\sqrt{2}} . \label{Decomposition4Spinor}
\end{eqnarray}
This prompts us to  
\begin{eqnarray}
| \chi \rangle &=& \frac{1}{\sqrt{2}} (| \chi_1 \rangle - i | \chi_2 \rangle), \nonumber \\
| \overline{\chi} \rangle &=& -\frac{1}{\sqrt{2}} (| \chi_1 \rangle + i | \chi_2 \rangle).
\end{eqnarray}
Now we rewrite (\ref{ParticleStateWithPotential}) to be
\begin{eqnarray}
& &\frac{1}{\sqrt{2}} ( | \chi_1 \chi_1 \rangle - | \chi_2 \chi_2 \rangle ),~ -V_{\gamma^{\prime}}, \label{ParticleState12WithPotential1} \\
& &\frac{1}{\sqrt{2}}( | \chi_1 \chi_2 \rangle + | \chi_2 \chi_1 \rangle ),~ -V_{\gamma^{\prime}}, \label{ParticleState12WithPotential2} \\
& &\frac{1}{\sqrt{2}}( | \chi_1 \chi_1 \rangle + | \chi_2 \chi_2 \rangle ),~ V_{\gamma^{\prime}} + V_{\Phi}, \label{ParticleState12WithPotential3} \\
& &\frac{1}{\sqrt{2}}( | \chi_1 \chi_2 \rangle - | \chi_2 \chi_1 \rangle ),~ V_{\gamma^{\prime}} - V_{\Phi},	\label{ParticleState12WithPotential}
\end{eqnarray}
{where the unimportant universal minus signs and phases $i=e^{i \pi}$ in the states of  (\ref{ParticleState12WithPotential2}),  (\ref{ParticleState12WithPotential3}) and (\ref{ParticleState12WithPotential}), have been neglected.} From (\ref{ParticleState12WithPotential}) we can clearly see that only $\frac{1}{\sqrt{2}}( | \chi_1 \chi_1 \rangle + | \chi_2 \chi_2 \rangle )$ feels a completely attractive potential. The interactions in $\frac{1}{\sqrt{2}}( | \chi_1 \chi_2 \rangle + | \chi_2 \chi_1 \rangle )$ and $\frac{1}{\sqrt{2}}( | \chi_1 \chi_1 \rangle - | \chi_2 \chi_2 \rangle )$ are completely repulsive, thus a bound state cannot be formed.

Remember that the total state of a fermion pair should be anti-symmetric when we interchange the two fermionic  components in all parts of the state vector. From (\ref{ParticleState12WithPotential}) we can clearly see the symmetry properties of the wave functions. Therefore, $L+S$ should be even when combined with the $\frac{1}{\sqrt{2}} ( | \chi_1 \chi_1 \rangle \pm | \chi_2 \chi_2 \rangle )$, $\frac{1}{\sqrt{2}}( | \chi_1 \chi_2 \rangle + | \chi_2 \chi_1 \rangle )$, and $L+S$ should be odd when  combined with the $\frac{1}{\sqrt{2}}( | \chi_1 \chi_2 \rangle - | \chi_2 \chi_1 \rangle )$.

\subsection{The case of $U(1) \rightarrow Z_2$ breaking} \label{U1Break2Z2}

If the dark $U(1)$ is spontaneously broken into its $Z_2$ subgroup, $\delta m \neq 0$. 
Then it is convenient to write the Schr\"{o}dinger equation in the $|\chi_i \chi_j \rangle$ basis.     
The total wave function becomes
\begin{eqnarray}
\int d^3\vec{x} ( \psi_{\chi_i \chi_j} (\vec{x}) | \chi_i \chi_j \rangle ) \otimes  |\vec{x}\rangle \otimes |\text{Spin}\rangle. \label{CompleteWaveij}
\end{eqnarray}
If we define
\begin{eqnarray}
\psi_{\text{s}} = \left[ \begin{array}{c}
\psi_{\chi_1 \chi_1} (\vec{x}) \\
\psi_{\chi_2 \chi_2} (\vec{x})
\end{array} \right],~
\psi_{\text{d}} = \left[ \begin{array}{c}
\psi_{\chi_1 \chi_2} (\vec{x}) \\
\psi_{\chi_2 \chi_1} (\vec{x})
\end{array} \right],
\end{eqnarray}
the Schr\"{o}dinger equation can be written as
\begin{eqnarray}
-\frac{\vec{\nabla}^2}{m_{\chi}} \psi_{\text{s}}(\vec{x}) + V_{\text{s}} \psi_{\text{s}}(\vec{x}) &=& E \psi_{\text{s}}(\vec{x}), \label{Schoedingers}\\
-\frac{\vec{\nabla}^2}{m_{\chi}} \psi_{\text{d}}(\vec{x}) + V_{\text{d}} \psi_{\text{d}}(\vec{x}) &=& E \psi_{\text{d}}(\vec{x}), \label{Schoedingerd}	\label{SchroedingerU1Broken}
\end{eqnarray}
where
\begin{eqnarray}
V_{\text{s}} = \left[
\begin{array}{cc}
V_{R} & ( V_{\gamma^{\prime}} + V_{\gamma^{\prime}_\text{L}} ) \\
( V_{\gamma^{\prime}} + V_{\gamma^{\prime}_\text{L}} )  & ( V_{R} + 4 \delta m )
\end{array} \right],~
V_{\text{d}}= - \left[
\begin{array}{cc}
( V_{R} - 2 \delta m ) &  ( V_{\gamma^{\prime}} - V_{\gamma^{\prime}_\text{L}} ) \\
( V_{\gamma^{\prime}} - V_{\gamma^{\prime}_\text{L}} ) &  ( V_{R} -  2 \delta m )
\end{array} \right].		\label{Potential12}
\end{eqnarray}
Here the three potentials are derived by calculating the Fourier transformation of the $\chi_i \chi_j \rightarrow \chi_{i^{\prime}} \chi_{j^{\prime}}$ tree-level amplitudes. The results are 
\begin{eqnarray}
V_{R} & = & -\frac{y^2}{4 \pi} \frac{e^{-m_{R} r}}{r} , \\
V_{\gamma^{\prime}} & = & -\frac{(Q_{\chi} g)^2}{4 \pi} \frac{e^{-m_{\gamma^{\prime}} r}}{r} , \label{PotentialTransverse} \\  V_{\gamma^{\prime}_{\text{L}}} & = &  -\frac{y^2}{4 \pi} \frac{e^{-m_{\gamma^{\prime}} r}}{r} .  \label{PotentialLongitudinal}
\end{eqnarray}
One can diagonalize (\ref{Potential12}) and compare the results with (\ref{ParticleState12WithPotential1}-\ref{ParticleState12WithPotential}) in the $\delta m \rightarrow 0$ limit for a validation. The $V_{\gamma^{\prime}_{\text{L}}} = -\frac{y^2}{4 \pi} \frac{e^{-m_{\gamma^{\prime}} r}}{r}$ originate from the $\frac{i}{q^2-m_{\gamma^{\prime}}^2} \frac{k^{\mu} k^{\nu}}{m_{\gamma^{\prime}}^2}$ term in the $\gamma^{\prime}$ propagator, where $k^{\mu}$ and $k^{\nu}$ finally contract with the on-shell spinors, inducing a $(m_{\chi_1} - m_{\chi_2})^2$ term, which is proportional to $y^2 v_{\Phi}^2$. Eliminating the $v_{\Phi}^2$ with the $m_{\gamma^{\prime}}^{2}$ in the denominator, and $Q_{\Phi}^2 g^2$ cancels with the coupling constants, we finally realize that $V_{\gamma^{\prime}_{\text{L}}}$ is contributed from the longitudinal polarization of $\gamma^{\prime}$. A similar description with a nearly-degenerate multi-component dark matter model is described in detail in Ref.~\cite{WrongskianMethod}.

Another analysis of the $V_{\gamma^{\prime}_{\text{L}}}$ is through the Goldstone equivalent theorem. This term can be understood originating from the exchange of the Goldstone boson $I$. In the non-breaking limit, 
$v_{\Phi} \rightarrow 0$, one can find that $I$-contribution from the $V_{\Phi}$ in the (\ref{ParticleState12WithPotential3}-\ref{ParticleState12WithPotential}) is equivalent to the longitudinal $V_{\gamma^{\prime}_{L}}$ appeared in (\ref{PotentialLongitudinal}). \footnote{ We applied the ``on-shell approximation'' described in Ref.~\cite{Petraki1}. However, for a general $R_\xi$ gauge, the validity of this approximation is a little bit subtle. We illustrate this in the appendix \ref{GaugeIndependence}.}

For the (\ref{Schoedingerd}), the result is the same as the $\delta m=0$ case: this is because, from the structure of 
$V_{\text{d}}$ in the (\ref{Potential12}), we can see the extra $\delta m$ appearing in both of the diagonal elements does not disturb the diagonalizing process of the wave functions. Such a $\delta m$ here only shift the total energy, so the (\ref{Schoedingerd}) can still be decoupled into two independent equations by diagonalizing the $V_{\text{d}}$. 
A standard ``shooting-method'' is applicable for these equations. However, for the (\ref{Schoedingers}), the final wave 
functions will be a mixing between the $|\chi_1 \chi_1 \rangle \pm | \chi_2 \chi_2 \rangle$ basis. 
In the following text, we will address the method for solving these equations in detail.

The general wave functions described by the Eqn.~(\ref{Schoedingers}) are given by
\begin{eqnarray}
\int d^3 \vec{x} (\psi_{\chi_1 \chi_1} (\vec{x}) | \chi_1 \chi_1 \rangle + \psi_{\chi_2 \chi_2} (\vec{x}) | \chi_2 \chi_2 \rangle ) \otimes | \vec{x} \rangle \otimes | \text{Spin} \rangle.
\end{eqnarray}
Let us decompose them into the radial and the angular parts using spherical Harmonics: 
\begin{eqnarray}
\psi_{\chi_1 \chi_1}(\vec{x}) &=& \kappa^{\frac{3}{2}} \left[ \frac{\chi_{1 nl}(\kappa r)}{\kappa r} \right] Y_{l m} (\Omega_{\vec{r}}) \nonumber \\
\psi_{\chi_2 \chi_2}(\vec{x}) &=& \kappa^{\frac{3}{2}} \left[ \frac{\chi_{2 nl}(\kappa r)}{\kappa r} \right] Y_{l m} (\Omega_{\vec{r}}), \label{BoundStateDecomposition}
\end{eqnarray}
where $\kappa = \mu_r \alpha^{\prime}$ is the inverse ``Bohr radius'' and $\mu_r$ is the reduced mass of the $\chi \chi$ system.

The parameter $\alpha^{\prime}$ is some arbitrary reference value which reflects the typical dark interaction strength (see the discussions below). For the value of $\alpha^{\prime}$, it can be arbitrary, and  all the $\alpha^{\prime}$ dependencies will cancel out in the end, leaving all of the physical observables are undisturbed. However, an improper $\alpha^{\prime}$ selection might cause some instabilities in our practical numerical calculations, especially when we are manipulating the exponential  suppression of the $x \rightarrow \infty$ asymptotic condition of the bound states. For this purpose, we recommend $\alpha^{\prime}=\text{Max} \lbrace \frac{(Q_{\chi} g)^2}{4 \pi},~\frac{y^2}{4 \pi} \rbrace$.

Then, the wave equation for the radial part wave function becomes 
\begin{eqnarray}
\chi_{n l}^{\prime \prime}(x) + \left[ -\frac{l(l+1)}{x^2} - \gamma^2 - 2 V_{\text{s}, \alpha^{\prime}} \right] \chi_{n l}(x) = 0, \label{RadialEquation}
\end{eqnarray}
where
\begin{eqnarray}
\chi_{n l}(x) = \left[ \begin{array}{c}
\chi_{1 n l}(x) \\
\chi_{2 n l}(x)
\end{array} \right]
\end{eqnarray}
is the radial wave function vector, and the potential term is given by
\begin{eqnarray}
V_{\text{s}, \alpha^{\prime}} = \left[ \begin{array}{cc}
-\frac{c_1 e^{-\frac{x}{\xi_1}}}{x} & -\frac{(c_2+c_1) e^{-\frac{x}{\xi_2}}}{x} \\
-\frac{(c_2+c_1) e^{-\frac{x}{\xi_2}}}{x} & -\frac{c_1 e^{-\frac{x}{\xi_1}}}{x}+\delta \gamma^2
\end{array} \right]. \label{RadialPotentialS}
\end{eqnarray}
Here $c_1 = \frac{y^2}{4 \pi \alpha^{\prime}}$, $c_2 =  \frac{(Q_{\chi} g)^2}{4 \pi \alpha^{\prime}}$ are the relative interaction strength compared with the $\alpha^{\prime}$, $\xi_1 = \frac{\kappa}{m_{R}}$ and $\xi_2 = \frac{\kappa}{m_{\gamma^{\prime}}}$ are the characteristic length scales  of the Yukawa potentials
generated by dark Higgs $R$ and the dark photon $\gamma^{'}$, respectively. 
The parameter $\gamma$  is related with the energy eigenvalue in the unit of the ``Bohr energy'', 
and is defined as
\begin{eqnarray}
\gamma = \frac{\sqrt{-2 \mu_r E}}{\kappa},
\end{eqnarray}
where $E$ is the  energy eigenvalue of the Schr\"{o}dinger equation. Finally we introduce another new parameter $\delta \gamma^2$, which is the reduced $2 \delta m$: 
\begin{eqnarray}
\delta \gamma^2 = \frac{4 \delta m \mu_r}{\kappa^2}.
\end{eqnarray}

\subsection{Solving the Schr\"{o}dinger equations for the bound states}

Solving the Schr\"{o}dinger  equations for the bound states is based upon the so-called  ``shooting method''. 
The boundary conditions at the $x \rightarrow \infty$ are replaced by some finite values. Here we adopt $\chi_{n l} (x=12)= 0$. Without loss of generality, we assume $\delta \gamma > 0$. Notice that if $x$ is sufficiently large, the asymptotic behaviors of the $\chi_{1nl}$  and $\chi_{2nl}$ become
\begin{eqnarray}
\left\lbrace \begin{array}{l}
\chi_{1nl}(x \rightarrow \infty) \sim e^{-\gamma x} \\
\chi_{2nl}(x \rightarrow \infty) \sim e^{-\sqrt{\gamma^2+\delta \gamma^2} x}.
\end{array} \right.
\end{eqnarray}
This means that $\chi_{2 n l}$ drops faster than the $\chi_{1 n l}$ in the $x \rightarrow \infty$ condition, so a universal infinite boundary condition $\chi_{n l} (x=12) =0$ will cause the numerical instability in the $\chi_{2 n l}$ calculations. Therefore, we need to reduce the $\chi_{2 n l}$ boundary area. Observing the structure of (\ref{RadialPotentialS}), we realize that when the absolute value of the off-diagonal element $|\frac{(c_2 + c_1) e^{-\frac{x}{\xi_2}}}{x}|$ is much smaller than $|\delta \gamma^2|$, $\chi_{1nl}$ and $\chi_{2nl}$ will nearly decouple and evolve independently.  Then if $|\frac{c_1 e^{-\frac{x}{\xi_1}}}{x}| \ll \delta |\gamma^2|$, $\chi_{2 n l} \sim e^{-\sqrt{\gamma^2+\delta \gamma^2} x }$ which vanishes quickly.  The boundary condition on $\chi_{2nl}$ should be determined in the intersection of these two ranges. Noticing that in our following numerical calculations, we adopt $c_1 \leq c_2$, and $\xi_1 \sim \xi_2$ by the order of magnitudes, after various trials with the criterion of a smoothly-shaped wave function, we finally choose to solve the following simple equation
\begin{eqnarray}
\frac{c_2 e^{-\frac{x_0^{\prime}}{\xi_2}}}{x_0^{\prime}} = \frac{\delta \gamma^2}{20}.
\end{eqnarray}
to acquire $x_0^{\prime} > 0$ in our following calculations. Obtain $x_0 = \text{Min} \lbrace x_0^{\prime},~ 12\rbrace$, fix some $\gamma^2$, and adopt the initial condition
\begin{eqnarray}
\left\lbrace \begin{array}{c}
\chi_{1 n l}(x\rightarrow 0) = x^{l+1} \\
\chi_{2 n l}(x\rightarrow 0) = A x^{l+1} 
\end{array} \right.	\label{BoundStateBoundary}
\end{eqnarray}
to use some numerical method to solve the (\ref{RadialEquation}) from $x \rightarrow 0$ to $x_0$. 
Then we can find the appropriate $A$ for $\chi_{2 n l}(x_0) = 0$, delete all the terms involving $\chi_{2 n l}(x)$, and solve the $\chi_{1 n l}(x)$ equation.  We continue to solve the $\chi_{1 n l}(x)$ within the range $[x_0,~12]$. Changing different $\gamma^2$ and repeating the above process, we can finally reach $\chi_{1 n l}(12) = 0$ and determine the eigenvalue $\gamma^2$.

\begin{figure} 
\includegraphics[width=3in]{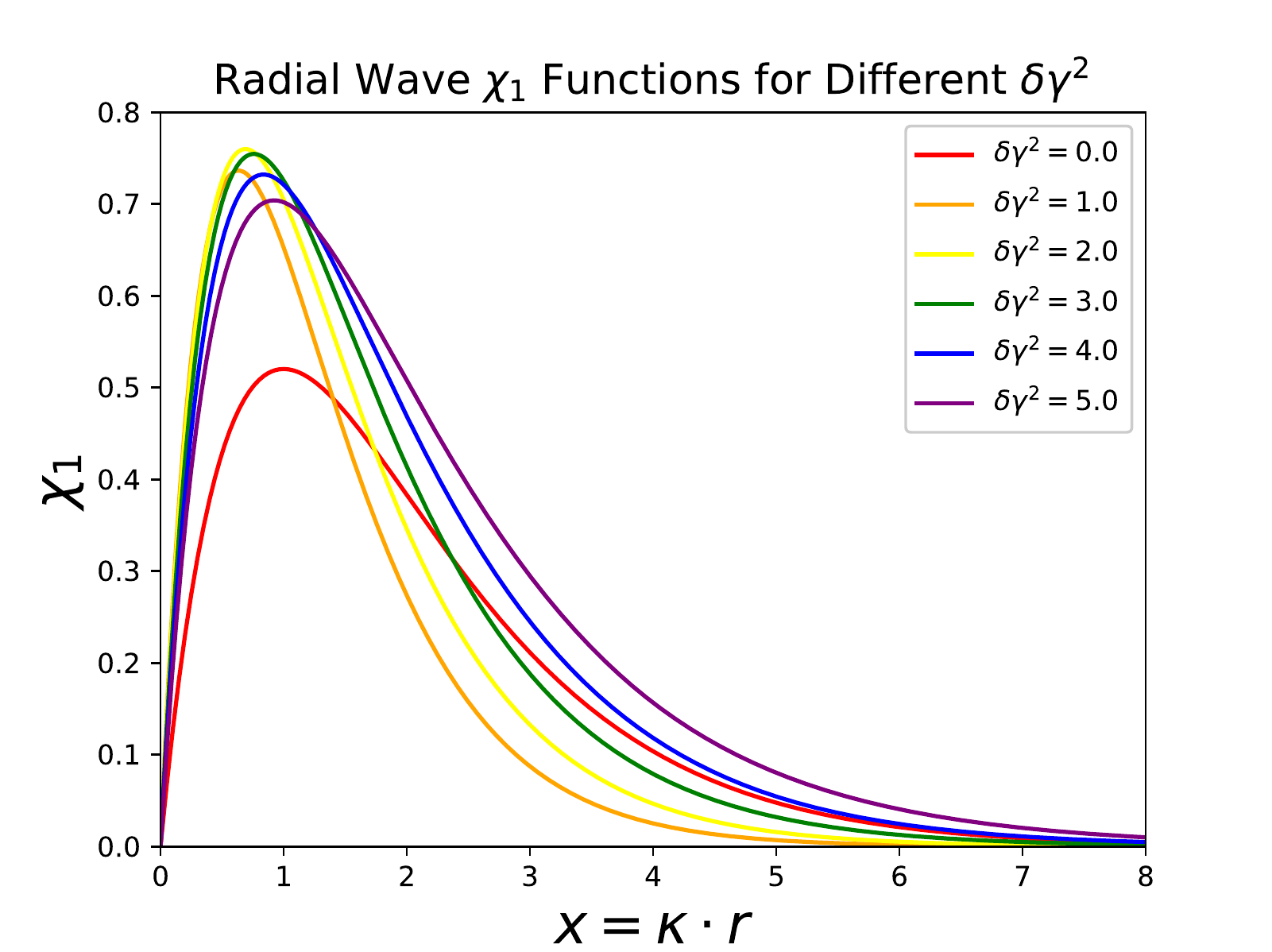}
\includegraphics[width=3in]{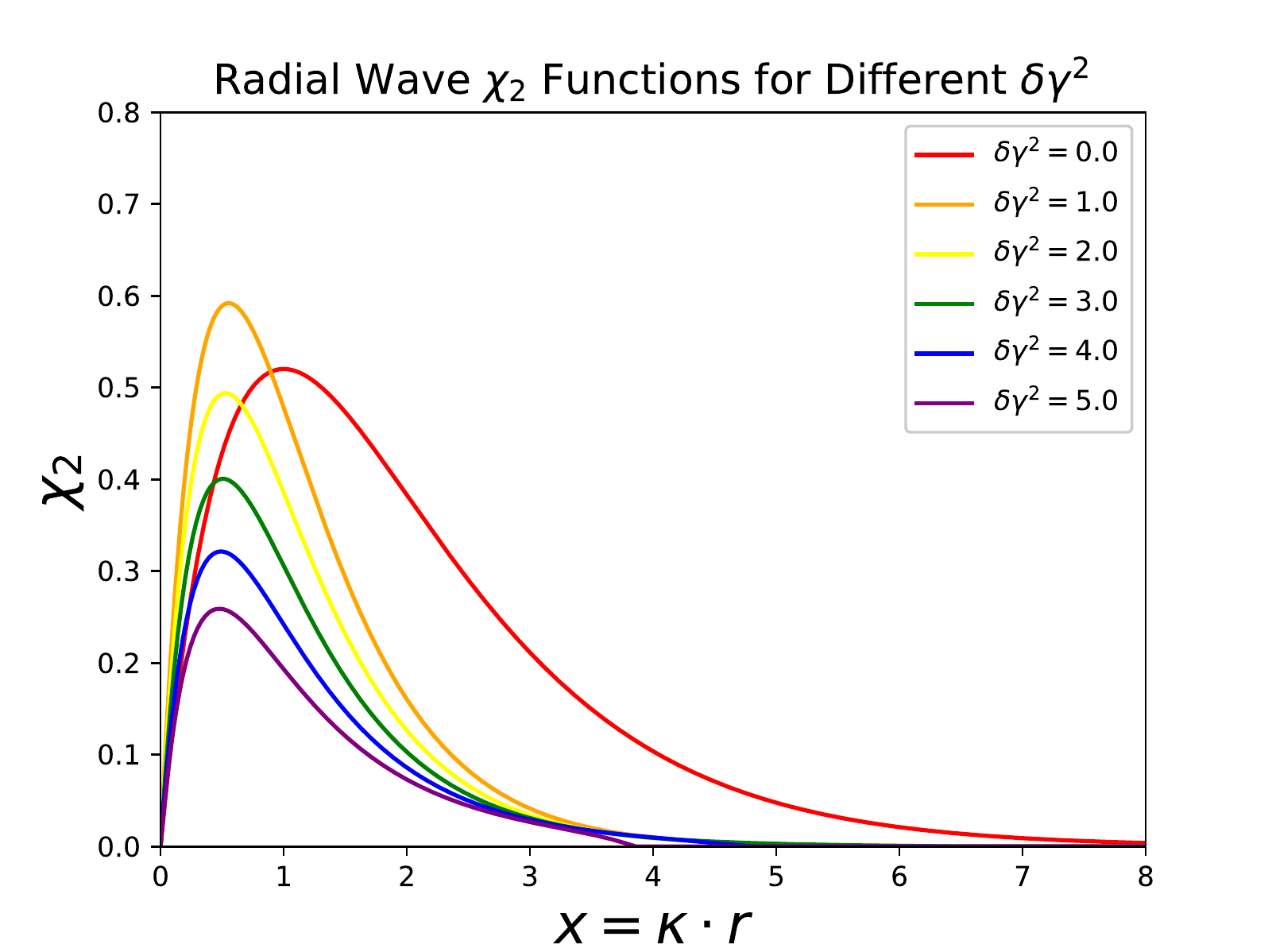}
\caption{Wave functions of the ground state for different $\delta \gamma^2$. Here we adopt $c_1 = 0.35$, $c_2=1$, $\xi_1 = 200$, $\xi_2 = 100$. We can see clearly that the $\chi_2$ reduces as the $\delta \gamma^2$ accumulates. Here we only plot the $A>0$ case, and the wave functions are normalized.} \label{WaveFunctionSample}
\end{figure}

Let us note that when we determine the $A$ in the first step, there might be two solutions. One is $A>0$, and the other is $A<0$. Both of them are possible and can give different $\gamma^2$'s in the final step. In the $\delta \gamma \rightarrow 0$ limit, these two solutions degenerate to the $\frac{1}{\sqrt{2}} | \chi_1 \chi_1 \rangle + | \chi_2 \chi_2 \rangle$ and $\frac{1}{\sqrt{2}} | \chi_1 \chi_1 \rangle - |  \chi_2 \chi_2 \rangle$ states described in the (\ref{ParticleState12WithPotential2}, \ref{ParticleState12WithPotential}). As the $\delta \gamma^2$ accumulates, the wave function will depart from the (\ref{ParticleState12WithPotential2}, \ref{ParticleState12WithPotential}). This can be clearly seen from Fig.~\ref{WaveFunctionSample}.

\begin{figure}
\begin{center}
\includegraphics[width=4.0in]{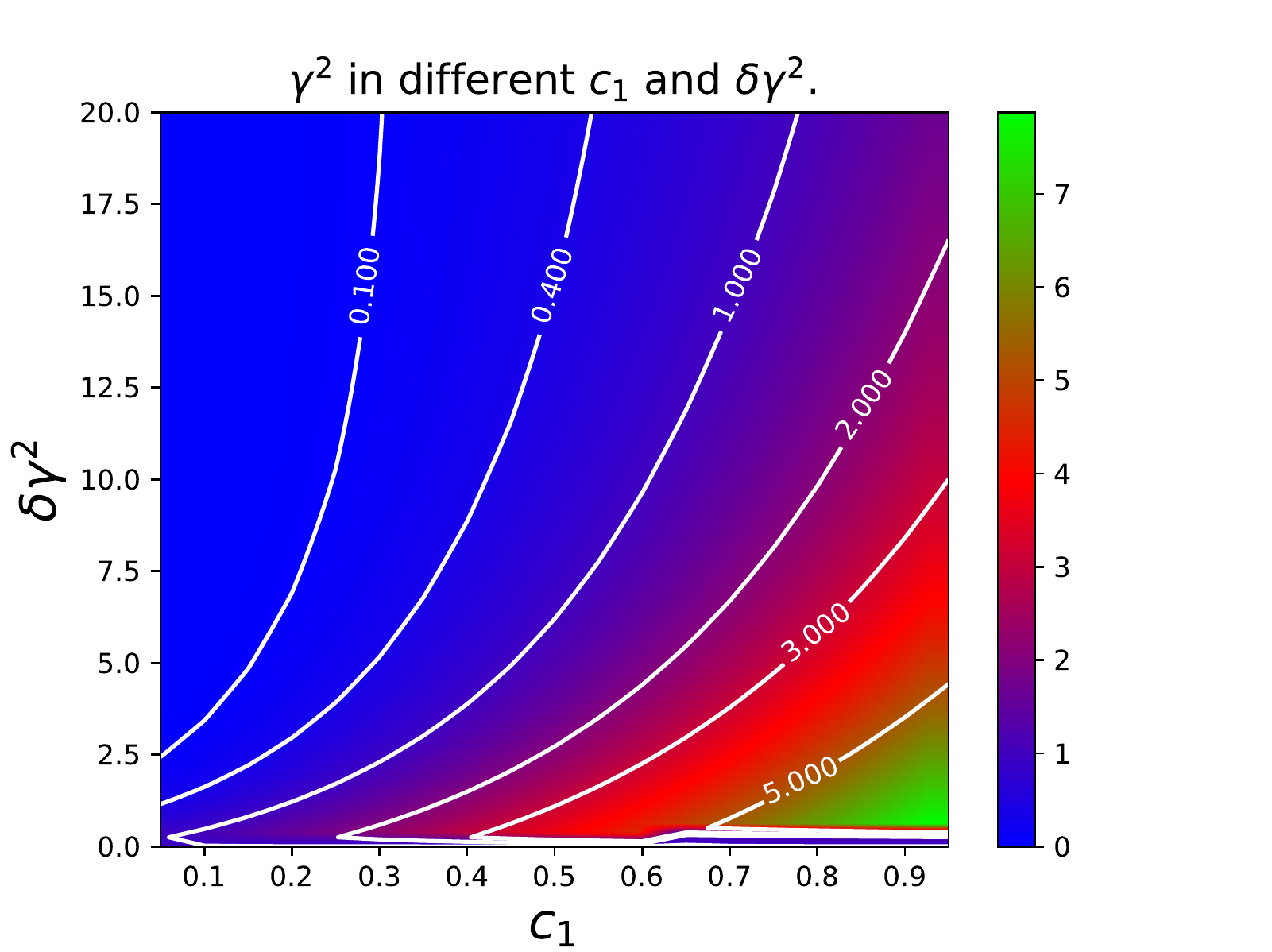}
\end{center}
\caption{$\gamma^2$, which indicates the bound-energy, versus different $c_1$ and $\delta \gamma^2$. Here $c_2$ is fixed to be 1, and  $\xi_1 = 200$, $\xi_2 = 100$.} \label{BoundEnergy}
\end{figure}

As the $\delta \gamma^2$ accumulates, the $\chi_2 \chi_2$ elements will be reduced in the ground-state wave functions, so the system will become more similar to the one-component situations. In Fig~\ref{BoundEnergy}, we can see that as $\delta \gamma^2$ increases, the bound energy parameter $\gamma^2$ approaches 0. This indicates that the $\chi_2$ decouples in the large $\delta \gamma^2$ limit.

For the completeness of this section, we point out that (\ref{Schoedingerd}) can also be decomposed into 
the combinations of the angular and radial components following the similar processes.  The radial functions are 
similar to (\ref{RadialEquation}), and the potential term $V_{\text{s}, \alpha^{\prime}}$ should be replaced by 
\begin{eqnarray}
V_{\text{d}, \alpha^{\prime}} = \left[ \begin{array}{cc}
\frac{c_1 e^{-\frac{x}{\xi_1}}}{x} & \frac{(c_2-c_1) e^{-\frac{x}{\xi_2}}}{x} \\
\frac{(c_2-c_1) e^{-\frac{x}{\xi_2}}}{x} & \frac{c_1 e^{-\frac{x}{\xi_1}}}{x}
\end{array} \right],	\label{RadialPotentialD}
\end{eqnarray}
where the universal energy shift $\delta m$ terms are removed. One can still follow the steps described above, or beforehand diagonalize (\ref{RadialPotentialD}) to solve the wave functions.

\begin{table}\centering
\begin{tabular}{|c|c|c|c|}
\hline
$| \chi_i \chi_j \rangle $ property & Sign of $A$ & Sign of $L+S$ & $n^{2 s+1} l_J ^{\text{s}/\text{d} \pm}$ \\
\hline
$| \chi_1 \chi_1 \rangle \leftrightarrow | \chi_2 \chi_2 \rangle$ & $+$ & Even & $1^1 \text{S}_0^{\text{s} +}$,  $1^3 \text{P}_{0, 1, 2}^{\text{s} +}$, $2^1 \text{S}_0^{\text{s} +}$\\
\hline
$| \chi_1 \chi_1 \rangle \leftrightarrow | \chi_2 \chi_2 \rangle$ & $-$ & Even &  $1^1 \text{S}_0^{\text{s} -}$,  $1^3 \text{P}_{0, 1, 2}^{\text{s} -}$, $2^1 \text{S}_0^{\text{s} -}$ \\
\hline
$| \chi_1 \chi_2 \rangle \leftrightarrow | \chi_2 \chi_1 \rangle$ & $-$ & Odd & $1^3 \text{S}_1^{\text{d} -}$,  $1^1 \text{P}_{1}^{\text{d} -}$, $2^3 \text{S}_1^{\text{d} -}$\\
\hline
$| \chi_1 \chi_2 \rangle \leftrightarrow | \chi_2 \chi_1 \rangle$ & $+$  & Even & Bound state does not exist. \\
\hline
\end{tabular}
\caption{Quantum characters of different states} \label{Classification}
\end{table}

The above discussions mainly focus on the spatial part of the wave function. The total wave function of a fermionic pair expressed by (\ref{TotalWaveOriginal}, \ref{CompleteWaveij}) should be anti-symmetric. One might be familiar with the $L+S$ being odd or even for an identical particle system, however, here, we need to take into account the ``flavour'' wave function $|\chi_i \chi_j \rangle$. For example, if the ``flavour'' part is given by $\frac{1}{\sqrt{2}}(|\chi_1 \chi_2 \rangle - |\chi_2 \chi_1\rangle)$ which is anti-symmetric, we need an odd $L+S$, which is corresponding for a symmetric configuration-spin wave function, so the totally anti-symmetric wave function is acquired. Combining all the descriptions above, and considering the anti-symmetry character of the fermion pairs, we can then classify the bound states with the quantum numbers characterized in Tab.~\ref{Classification}. We had extended the traditional spectroscopic notation $n^{2 s+1} l_J$ symbol to $n^{2 s+1} l_J^{\text{s}/\text{d}}\pm$, where ``s'' or ``d'' indicates the ``same'' or ``different'' in the $|\chi_i \chi_j \rangle$ wave function, and $\pm$ indicates the sign of $A$ at the origin of the wave functions. The bound state does not exist in the last line of Tab.~\ref{Classification}, which has been described in the previous discussions of (\ref{ParticleState12WithPotential2}). For the second line of Tab.~\ref{Classification}, similar to the (\ref{ParticleState12WithPotential1}), usually the bound state does not exist. However, notice that because $m_{\gamma^{\prime}} \neq m_R$, in some particular parameter space, (\ref{RadialPotentialS}) can induce some kind of ``molecule-like'' potentials. In this paper, we only list the quantum number possibilities of the bound state in the second line of Tab.~\ref{Classification}. However such ``molecule-like'' weird bound states are omitted in the following discussions for simplicity. This situation is left for our future researches.

In the following of this paper we are going to calculate the contributions from these bound states listed in Tab.~\ref{Classification} to the freeze-out processes.

\section{Bound State Formation Cross Section and Dissociation Rate} \label{BSFProcesses}

\subsection{Calculation of Bound State Formation Cross Section}
The dark matter bound states are formed by the scattering of two free dark matter particles, with emission of $\gamma^{\prime}$ or $R$  that takes the extra energy away from the bound system. Now we are going to calculate the transition amplitudes. Our discussions and derivations are based on the symbols in Ref.~\cite{Petraki1, Petraki2}. Although Ref.~\cite{Petraki1, Petraki2} did not consider the multi-component wave functions, we can just extend the ``overlap'' integrals $\mathcal{I}$, $\vec{\mathcal{J}}$  and $\mathcal{K}$ in the (2.7a-c) of Ref.~\cite{Petraki2} to the two-component case rather than deriving them starting from the Dyson-Schwinger equations. We omit the radiation from the interchanging mediators due to the reason described in Appendix \ref{Reason}.  

\begin{figure}[b]
\includegraphics[width=0.49\textwidth]{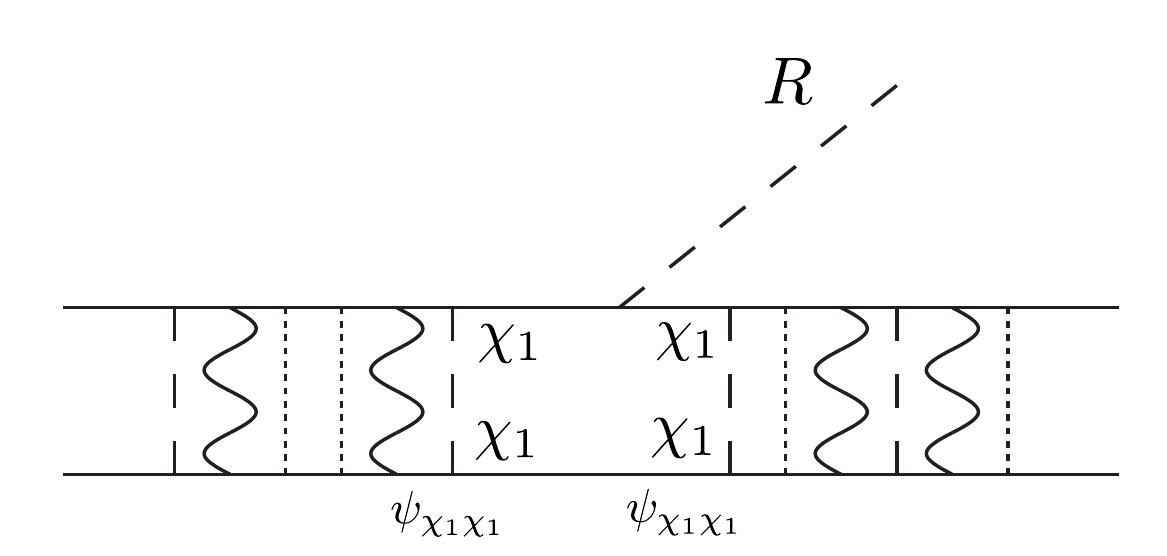}
\includegraphics[width=0.49\textwidth]{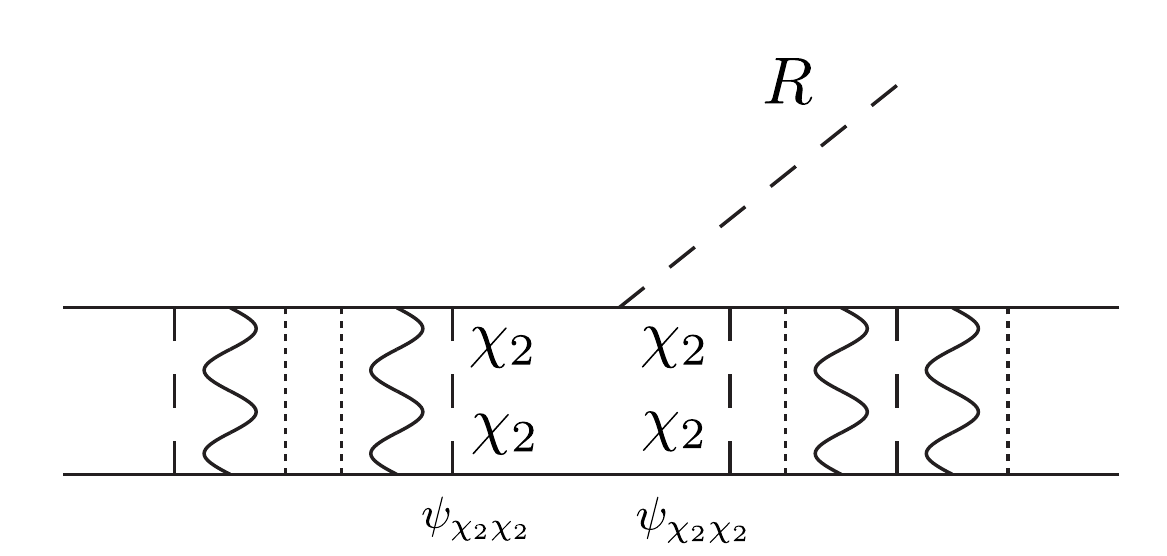}
\includegraphics[width=0.49\textwidth]{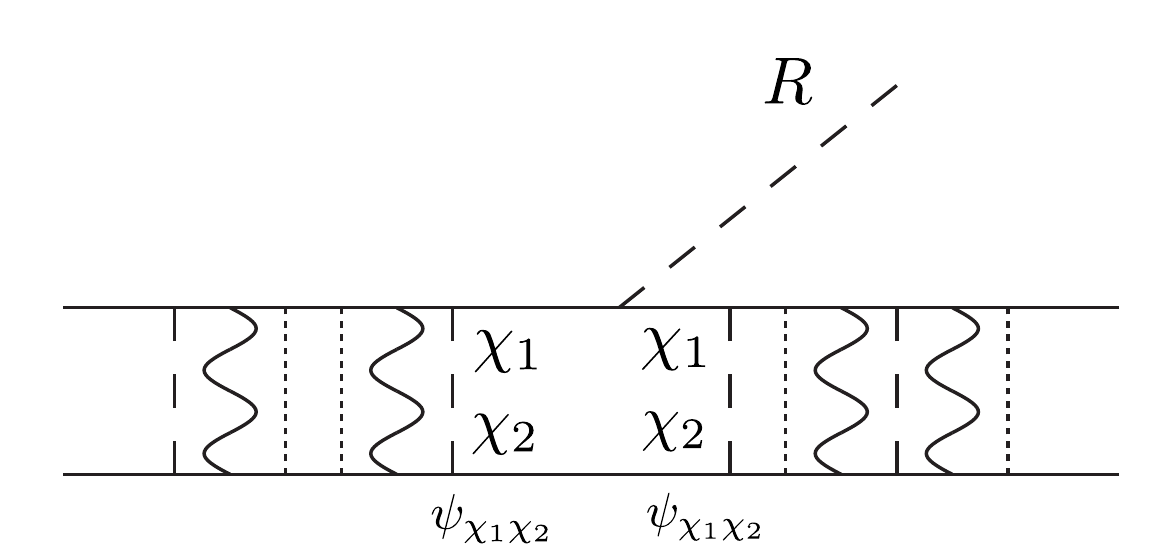}
	\includegraphics[width=0.49\textwidth]{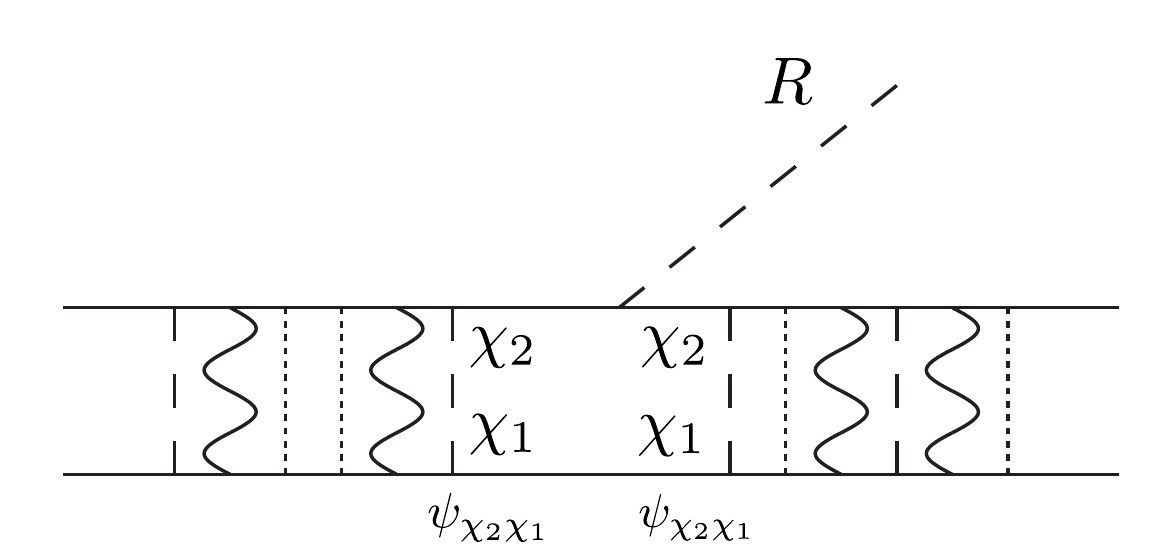}
\caption{Sketch diagrams emitting a $R$ scalar boson. Notice that since the $\chi_1 \chi_1 R$ and $\chi_2 \chi_2 R$ couplings take different sighs, so the $\chi_{1 l_2 k i}^*(x) \chi_{1 n l_1}(x)$ and $\chi_{2 l_2 k i}^*(x) \chi_{2 n l_1}(x)$ should also take the opposite sign.} \label{EmittingR}
\end{figure}

\begin{figure}
\includegraphics[width=0.49\textwidth]{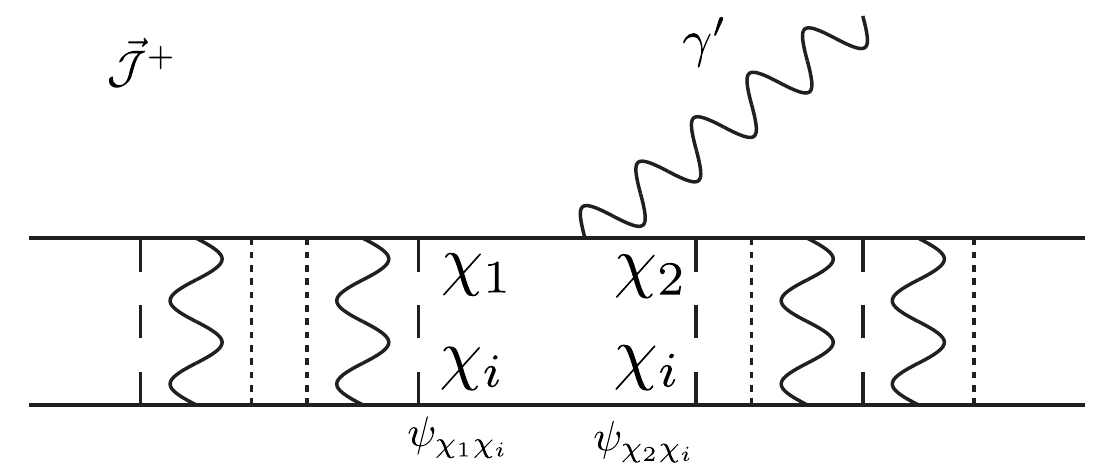}
\includegraphics[width=0.49\textwidth]{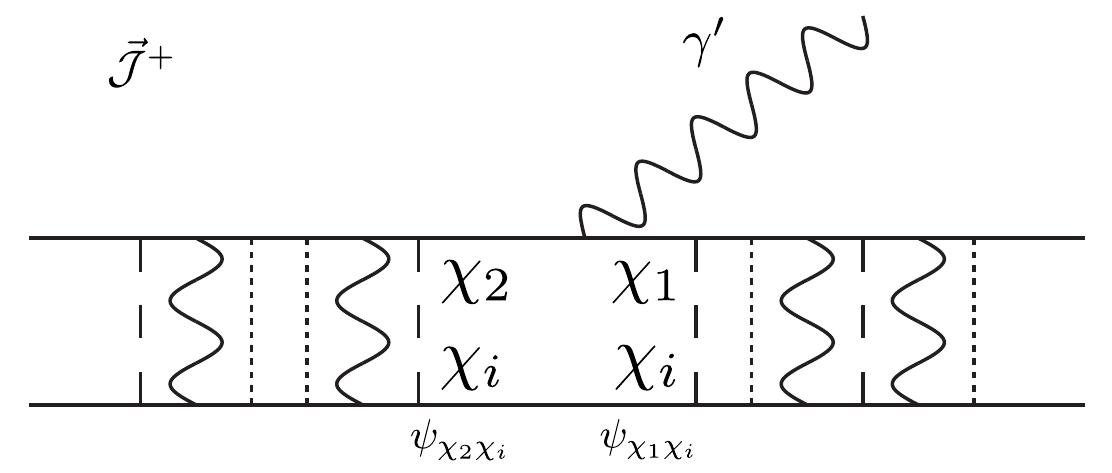}
\includegraphics[width=0.49\textwidth]{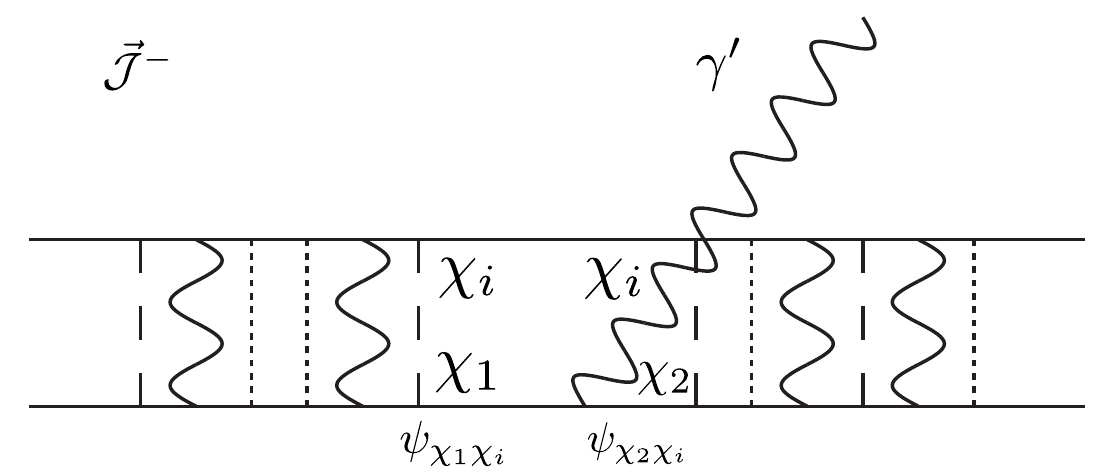}
\includegraphics[width=0.49\textwidth]{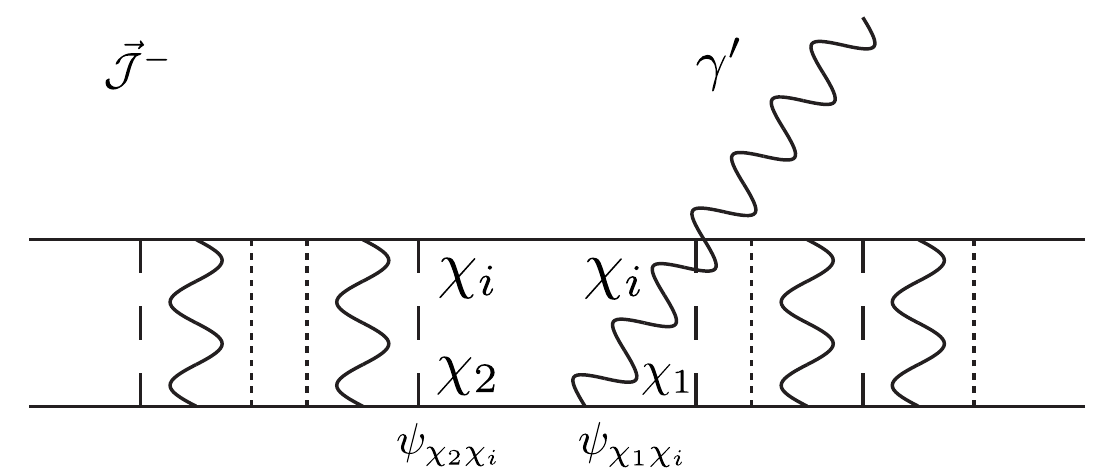}
\caption{Sketch diagrams emitting a $\gamma^{\prime}$. All the coupling signs are the same, but $\vec{\mathcal{J}}^{+}$ will induce a cross relation between the components of the $\psi_{\text{s}}$ and $\psi_{\text{d}}$} \label{EmittingGammaPrime}
\end{figure}

Below let us first list the integrals required in this paper: 
\begin{eqnarray}
\mathcal{I}_{\text{s}, \vec{k},nlm} (\vec{b}) &=& \int d^3 \vec{r} \psi^*_{\text{s}, nlm}(\vec{r}) \sigma^3 \phi_{\text{s}, \vec{k} }(\vec{r}) e^{-i \vec{b} \cdot \vec{r}}, \label{Is} \\
\mathcal{I}_{\text{d}, \vec{k},nlm} (\vec{b}) &=& \int d^3 \vec{r} \psi^*_{\text{d}, nlm}(\vec{r}) \sigma^3 \phi_{\text{d}, \vec{k} }(\vec{r}) e^{-i \vec{b} \cdot \vec{r}}, \label{Id} \\
\mathcal{I}_{\text{s $\rightarrow$ d}, \vec{k},nlm}^+ (\vec{b}) &=& \int d^3 \vec{r} \psi^*_{\text{d}, nlm}(\vec{r}) (\sigma^1) \phi_{\text{s}, \vec{k} }(\vec{r}) e^{-i \vec{b} \cdot \vec{r}}, \label{Isdp} \\
\mathcal{I}_{\text{d $\rightarrow$ s}, \vec{k},nlm}^+ (\vec{b}) &=& \int d^3 \vec{r} \psi^*_{\text{s}, nlm}(\vec{r}) (\sigma^1) \phi_{\text{d}, \vec{k} }(\vec{r}) e^{-i \vec{b} \cdot \vec{r}}, \label{Idsp} \\
\mathcal{I}_{\text{s $\rightarrow$ d}, \vec{k},nlm}^- (\vec{b}) &=& \int d^3 \vec{r} \psi^*_{\text{d}, nlm}(\vec{r}) \phi_{\text{s}, \vec{k} }(\vec{r}) e^{-i \vec{b} \cdot \vec{r}}, \label{Isdm} \\
\mathcal{I}_{\text{d $\rightarrow$ s}, \vec{k},nlm}^- (\vec{b}) &=& \int d^3 \vec{r} \psi^*_{\text{s}, nlm}(\vec{r}) \phi_{\text{d}, \vec{k} }(\vec{r}) e^{-i \vec{b} \cdot \vec{r}}, \label{Idsm} \\
\vec{\mathcal{J}}_{\text{s} \rightarrow \text{d}, \vec{k},nlm}^{+} (\vec{b}) &=& i \int d^3 \vec{r} \vec{\nabla}[\psi_{\text{d}, nlm}(\vec{r})] (i \sigma^2) \phi_{\text{s}, \vec{k} }(\vec{r}) e^{-i \vec{b} \cdot \vec{r}}, \label{Jpsd} \\
\vec{\mathcal{J}}_{\text{d} \rightarrow \text{s}, \vec{k},nlm}^{+} (\vec{b}) &=& i \int d^3 \vec{r} \vec{\nabla}[\psi_{\text{s}, nlm}(\vec{r})] (i \sigma^2) \phi_{\text{d}, \vec{k} }(\vec{r}) e^{-i \vec{b} \cdot \vec{r}}, \label{Jpds} \\
\vec{\mathcal{J}}_{\text{s} \rightarrow \text{d}, \vec{k},nlm}^{-} (\vec{b}) &=& i \int d^3 \vec{r} \vec{\nabla}[\psi_{\text{d}, nlm}(\vec{r})] \sigma^3 \phi_{\text{s}, \vec{k} }(\vec{r}) e^{-i \vec{b} \cdot \vec{r}}, \label{Jmsd} \\
\vec{\mathcal{J}}_{\text{d} \rightarrow \text{s}, \vec{k},nlm}^{-} (\vec{b}) &=& i \int d^3 \vec{r} \vec{\nabla}[\psi_{\text{s}, nlm}(\vec{r})] \sigma^3 \phi_{\text{d}, \vec{k} }(\vec{r}) e^{-i \vec{b} \cdot \vec{r}}, \label{Jmds} \\
\mathcal{K}_{\text{s}, \vec{k},nlm} (\vec{b}) &=& -\int d^3 \vec{r} \vec{\nabla}^2 [\psi^*_{\text{s}, nlm}(\vec{r})] \sigma^3 \phi_{\text{s}, \vec{k} }(\vec{r}) e^{-i \vec{b} \cdot \vec{r}}, \label{Ks} \\
\mathcal{K}_{\text{d}, \vec{k},nlm} (\vec{b}) &=& -\int d^3 \vec{r} \vec{\nabla}^2 [\psi^*_{\text{d}, nlm}(\vec{r})] \sigma^3 \phi_{\text{d}, \vec{k} }(\vec{r}) e^{-i \vec{b} \cdot \vec{r}}. \label{Kd}
\end{eqnarray}
Here $\mathcal{I}$ and $\mathcal{K}$ are related to the $R-$emission processes, while 
$\vec{\mathcal{J}}$ corresponds to the $\gamma^{\prime}-$emission process. $\vec{b}$ is the variable for these functions, and will be set to something like $\frac{\vec{p}_{\gamma^{\prime}, I, R}}{2}$ in our later discussions. $\psi_{(sd),nlm}(\vec{r})$ is the two-component wave function for the bound state with the notations $n^{2 s+1} l_J^{\rm{s}/\rm{d}}$. and $\phi_{\rm{s}/\rm{d}, \vec{k}}(\vec{r})$ is the two-component ``same/different particle species'' wave function for the scattering state with the (unreduced) initial relative momentum $\vec{k}$. In the following texts, we will express all these overlap integrals with the radial wave functions for further calculations. For the two-component dark matter model, coupling constants might be different for each component. Therefore, for notational convenience, we introduce $2\times 2$ Pauli spin matrices $\sigma^i$ in (\ref{Is})--(\ref{Kd}) to connect the two doublet wave functions, although these wave functions have no direct relationship with any $SU(2)$ group. The reason for us to adopt 
$\sigma^3$ in the $\mathcal{I}$ and $\mathcal{K}$ has been sketched in the Fig~\ref{EmittingR}. Notice that the couplings of $\chi_1-\chi_1-R$ and $\chi_2-\chi_2-R$ take opposite signs, so that $\psi_{\chi_1 \chi_1} \leftrightarrow \psi_{\chi_1 \chi_1}$ and 
$\psi_{\chi_2 \chi_2} \leftrightarrow \psi_{\chi_2 \chi_2}$, or $\psi_{\chi_1 \chi_2} \leftrightarrow 
\psi_{\chi_1 \chi_2}$ and $\psi_{\chi_2 \chi_1} \leftrightarrow \psi_{\chi_2 \chi_1}$ contributions are opposite. In Fig.~\ref{EmittingGammaPrime}, we can see that the $\vec{\mathcal{J}}^{+}$ connects 
the crossing components of the $\left[ \begin{array}{c} \psi_{\chi_1 \chi_1} \\ \psi_{\chi_2 \chi_2} 
\end{array} \right]$ and $\left[ \begin{array}{c} \psi_{\chi_1 \chi_2} \\ \psi_{\chi_2 \chi_1} \end{array} 
\right]$ multiplets if we take $i=1$ or $2$ into Fig.~\ref{EmittingGammaPrime}. This is the reason 
why the  $\sigma^1$ and $\sigma^2$ appear in the (\ref{Is}-\ref{Kd}).

Equipped with all these overlap integrals, we are now ready to calculate the transition amplitudes. One can directly follow the processes to derive (5.21), (5.54) in Ref.~\cite{Petraki1}, and just extend all the wave functions in two-component case to acquire the  transition amplitudes for the bound state formation. In the rest frame of the final bound state system, they are given by
\begin{eqnarray}
\mathcal{M}_{\text{s or d}, \vec{k} \rightarrow nlm+R} \simeq &-& y M \sqrt{2 \mu_r} [ \mathcal{I}_{\text{s or d}, \vec{k},nlm} (\frac{\vec{p}_{R}}{2}) + \mathcal{I}_{\text{s or d}, \vec{k},nlm} (\frac{\vec{p}_{R}}{2}) \nonumber \\
& & \left.  + \frac{ \mathcal{K}_{\text{s or d}, \vec{k},nlm} (\frac{\vec{p}_{R}}{2}) + \mathcal{K}_{\text{s or d}, \vec{k},nlm} (\frac{\vec{p}_{R}}{2}) }{2 M \mu_r} \right],  \label{MR}\\
\mathcal{M}_{\text{s$\rightarrow$d}, \vec{k} \rightarrow nlm+\gamma^{\prime}}^j \simeq &-& 2 Q_{\chi} g \sqrt{2 \mu_r} \left[ 2 \mathcal{J}^{+ j}_{\text{s$\rightarrow$d}, \vec{k}, nlm}(\frac{\vec{p}_{\gamma^{\prime}}}{2}) - 2 \mathcal{J}^{- j}_{\text{s$\rightarrow$d}, \vec{k}, nlm}(\frac{\vec{p}_{\gamma^{\prime}}}{2}) \right], \label{MGammaSD} \\ 
\mathcal{M}_{\text{d$\rightarrow$s}, \vec{k} \rightarrow nlm+\gamma^{\prime}}^j \simeq &-& 2 Q_{\chi} g \sqrt{2 \mu_r} \left[ 2 \mathcal{J}^{+ j}_{\text{d$\rightarrow$s}, \vec{k}, nlm}(\frac{\vec{p}_{\gamma^{\prime}}}{2}) - 2 \mathcal{J}^{- j}_{\text{d$\rightarrow$s}, \vec{k}, nlm}(\frac{\vec{p}_{\gamma^{\prime}}}{2}) \right], \label{MGammaDS} 
\end{eqnarray}
where $\vec{p}_{R, \gamma^{\prime}}$ are the momentums of the emitted $R$, $\gamma^{\prime}$ particles respectively, 
and $M$ is the total mass of the two-body system, $\mu_r$ is again the center of mass reduced mass. Since the two majorana particles are nearly degenerate, 
we adopt $M=2 m_{\chi}$, and $\mu_r = \frac{m_{\chi}}{2}$,  neglecting $\delta m$ in the reduced mass  $\mu_r$.  

With these amplitudes, we can use
\begin{eqnarray}
v_{\text{rel}} \frac{d \sigma^{\lbrace nlm \rbrace}_{\text{BSF}}}{d \Omega} = \frac{|\vec{P}_{\phi}|}{64 \pi^2 M^2 \mu_r} |\mathcal{M}_{\vec{k} \rightarrow nlm} |^2 \label{CrossSectionV}
\end{eqnarray}
to calculate the differential and the total cross sections times velocity. Here $\vec{P}_{\phi}$ indicates $\vec{p}_{R}$ or $\vec{p}_{\gamma^{\prime}}$. The calculation of the amplitude for $\vec{k} \rightarrow nlm + R$ is  straighforward: just take (\ref{MR}) directly into  (\ref{CrossSectionV}). For the cross section of the process $\vec{k} \rightarrow nlm + \gamma^{\prime}$,  we have to modify  Eqn.~(3.3) in Ref.~\cite{Petraki2} because the Ward identity $p_{\gamma^{\prime}}^{\mu} \mathcal{M}_{\mu} = 0$ becomes invalid in our case. It should be replaced with \footnote{See Appendix \ref{WTProof} for detailed discussions.}
\begin{eqnarray}
\mathcal{M}_{\mu} p_{\gamma^{\prime}}^{\mu} = \Delta m \mathcal{M}_{\text{GS}}, \label{WDIdentity}
\end{eqnarray}
where $\Delta m$ is the mass difference between the previous and latter fermion species in the F-F-V vertex times the sign of the corresponding gauge coupling. Simple calculations by enumerating all the possible vertices in Fig.~\ref{EmittingGammaPrime} give $\Delta m = 2 \delta m = 2 y v_{\Phi}$. $\mathcal{M}_{\text{GS}}$ is the amplitude of changing the emitting $\gamma^{\prime}$ into the Goldstone boson. It is calculated to be
\begin{eqnarray}
& & \mathcal{M}_{\text{GS, s$\rightarrow$d, or d$\rightarrow$s}, \vec{k} \rightarrow nlm + \gamma^{\prime}} \nonumber \\
& = & 2 (Q_{\chi} g)  \sqrt{2 \mu_r} (M) \left[ \mathcal{I}_{\text{s$\rightarrow$d, or d$\rightarrow$s}, \vec{k},nlm}^+ (\frac{\vec{p}_{\gamma^{\prime}}}{2}) + \mathcal{I}_{\text{s$\rightarrow$d, or d$\rightarrow$s}, \vec{k},nlm}^- (\frac{\vec{p}_{\gamma^{\prime}}}{2}) \right]. \label{GoldStoneEmission}
\end{eqnarray}
Therefore, the
\begin{eqnarray}
\mathcal{M}^0 = \frac{p_{\gamma^{\prime} i} \mathcal{M}^i + \Delta m \mathcal{M}_{\text{GS}}}{p_{\gamma^{\prime}}^0}, \label{M0Equation}
\end{eqnarray}
so finally,
\begin{eqnarray}
& & \sum_{\epsilon} | \mathcal{M}_{\vec{k} \rightarrow nlm} |^2 \nonumber \\
&=& -\left( g_{\mu \nu} - \frac{p_{\gamma^{\prime}, \mu} p_{\gamma^{\prime}, \nu}}{m_{\gamma^{\prime}}^2} \right) M^{\mu}_{\vec{k} \rightarrow nlm} M^{\nu *}_{\vec{k} \rightarrow nlm} \nonumber \\
&=& M^{j}_{\vec{k} \rightarrow nlm} M^{j *}_{\vec{k} \rightarrow nlm} - \frac{|p_{\gamma^{\prime}}^j \mathcal{M}^j_{\vec{k} \rightarrow nlm} + \Delta m \mathcal{M}_{\text{GS} \vec{k} \rightarrow nlm}|^2}{p_{\gamma^{\prime}}^2 + m_{\gamma^{\prime}}^2} + \frac{\Delta m^2}{m_{\gamma^{\prime}}^2} |\mathcal{M}_{\text{GS} \vec{k} \rightarrow nlm}|^2. \label{EmissionVectorAmp}
\end{eqnarray}
Notice that $\frac{(\Delta m)^2}{(m_{\gamma^{\prime}})^2} = \frac{4 y^2}{Q_{\chi}^2 g^2}$, and the $Q_{\chi}^2 g^2$ will be cancelled with the coupling constants in (\ref{GoldStoneEmission}). Therefore, the third term in (\ref{EmissionVectorAmp}) is actually the Goldstone-equivalent term.

For all of these overlap integrals, they should be treated differently in two kinematic regimes. When $\mu v \gtrsim \kappa$, Born approximation is valid, and many higher angular momentum partial waves participate the process. And the distortion of the scattering state wave functions compared with the plane wave functions is not that severe, so we can use plane wave to approximate for the scattering wave functions to preserve the higher angular momentum partial wave contributions. Furthermore, when $\mu v^2 \lesssim \kappa$, the momentum $p_{\gamma^{\prime}/R} \sim$ Max $\lbrace \mu v^2, \mu \alpha^{\prime 2} \rbrace$ of the radiated $\gamma^{\prime}/R$ becomes much smaller than the $\kappa$. 
Therefore its higher angular momentum partial wave contributions are suppressed due to the small value of $j_l(\frac{p_{\gamma^{\prime}/R} x}{2 \kappa})$ when $x \lesssim \frac{1}{\alpha^{\prime}}$. Also the distortions of the scattering wave functions become significant, so we use the lowest partial waves to calculate the boundstate formation processes. In practical calculations, usually the two ranges $\mu v \gtrsim \kappa$ and $\mu v^2 \lesssim \kappa$ have enough overlaps for one {to} determine a boundary point to shift smoothly between these two methods. Practically, after some trials on some benchmark points, and notice that $p_{\gamma^{\prime}/R} \sim \text{max} \lbrace \mu v^2, \mu \alpha^{\prime 2} \rbrace$, we adopt $\frac{p_{\gamma^{\prime}/R}}{2 \kappa} = 0.8$ 
as the boundary between  these two methods. 

In the plane wave approximation (or the zeroth order Born approximation), the plane wave function $\left[ \begin{array}{c} e^{-i \vec{k} \cdot \vec{r}} \\ 0 \end{array} \right]$ or $\left[ \begin{array}{c} 0 \\ e^{-i \vec{k} \cdot \vec{r}} \end{array} \right]$ is applied to estimate $\phi_{\vec{k}}$ in different initial states. Let us define $\vec{b}^{\prime}= \vec{b} + \vec{k}$, where $\vec{b}$ is still the formal variable to be replaced by $\frac{\vec{p}_{\gamma^{\prime}/R}}{2}$ in the actual following calculations. Notice that one can express a plane wave in terms of the spherical Bessel functions $j_l $ and Legendre polynomials: 
\begin{eqnarray}
e^{-i \vec{b}^{\prime} \cdot \vec{r}} = \sum_{l=0}^{\infty} (2 l + 1)(-i)^l j_l (b^{\prime} r) P_l(\hat{\mathbf{b}}^{\prime} \cdot \hat{\mathbf{r}}) . \label{PlaneExpansion}
\end{eqnarray}
Then, after some expansions and contractions of the integrations, we acquire
\begin{eqnarray}
\int d^3 \vec{r} \psi_{i, \text{s or d},nlm}^*( \vec{r} ) e^{-i \vec{b}^{\prime} \cdot \vec{r}} = (-i)^l \frac{4 \pi}{\kappa^{\frac{3}{2}}} \int_0^{\infty} dz z Y_{lm}^*(\vec{\Omega}_{\vec{b}^{\prime}}) \chi_{i nl}(z) j_l(\frac{b^{\prime} z}{\kappa}), \label{BornApproximation}
\end{eqnarray}
where $\psi_{i, \text{s or d},nlm}^*( \vec{r} )$ or $\chi_{i nl}(z)$ are the $i$-th element of the two-component wave function $\psi_{\text{s or d},nlm}^*( \vec{r} )$ or $\chi_{nl}(z)$ (i=1, 2). This can be used to estimate (\ref{Is})--(\ref{Idsm}). For the (\ref{Ks})--(\ref{Kd}), we can use the schr\"{o}dinger equation to eliminate the $\vec{\nabla}^2$ in $\vec{\nabla}^2 \left[\psi_{\text{s or d}, nlm}(\vec{r}) \right]$, then again apply (\ref{BornApproximation}). For the (\ref{Jpsd})--(\ref{Jmds}), we can use the partial integration method to cast the $\vec{\nabla}$ to in front of $e^{-i \vec{b}^{\prime} \cdot \vec{r}}$ to extract a $\vec{b}^{\prime}$ factor, and then apply (\ref{BornApproximation}) to calculate the remained part.

For the partial wave method, we need to solve (\ref{Schoedingers}) and (\ref{Schoedingerd}) for the complete 
$\phi_{\rm{s}/\rm{d}, \vec{k}}$. As in (\ref{BoundStateDecomposition}), we extract the radial wave function 
$\chi_{(\rm{s}/\rm{d}), \vec{k}}$
\begin{eqnarray}
\phi_{(\text{s}/\text{d}), \vec{k}}(\vec{x}) &=& \kappa^{\frac{3}{2}} \left[ \frac{\chi_{(\text{s}/\text{d}), lk}(\kappa r)}{\kappa r} \right] Y_{l m} (\Omega_{\vec{r}}).
\end{eqnarray}
In the following discussions, we only describe the ``same-flavour'' case in the non-zero $\delta \gamma^2$ case. The ``different-flavour'' case can be directly inferred by simply setting $\delta \gamma^2 = 0$. Therefore, we omit the $(\text{s}/\text{d})$ subscripts for brevity in the rest of this subsection. 

Starting from the $x \rightarrow 0$ boundary, Eq. (\ref{BoundStateBoundary}) need to be modified to
\begin{eqnarray}
\left\lbrace \begin{array}{c}
\chi_{1 l k}(x\rightarrow 0) = A_1 x^{l+1} \\
\chi_{2 l k}(x\rightarrow 0) = A_2 x^{l+1}
\end{array} \right.,	\label{ScatteringStateBoundary}
\end{eqnarray}
where $\chi_{i k}$ indicates the two-component radial wave functions 
\begin{eqnarray}
\chi_{l k}(x) = \left[ \begin{array}{c}
\chi_{1 l k}(x) \\
\chi_{2 l k}(x)
\end{array} \right] ,
\end{eqnarray}
satisfying (\ref{RadialEquation}), (\ref{RadialPotentialS}), (\ref{RadialPotentialD}). 

In the $-\gamma^2 > \delta \gamma^2$ case, the threshold for the inelastic scattering $\chi_1 \chi_1 \rightarrow \chi_2 \chi_2$ opens. There are two linearly independent solutions to (\ref{RadialEquation}) with potential (\ref{RadialPotentialS}), (\ref{RadialPotentialD}): $A_1=0$, $A_2 \neq 0$ and $A_1 \neq 0$, $A_2=0$. With these two initial conditions, we can solve (\ref{RadialEquation}) in the $x \rightarrow \infty$ limit. Then we adjust the absolute value of $A_1$ and $A_2$, and finally acquire the asymptotic form  of $\chi_{l i}^{(1), (2)}$ that correspond to the two initial conditions: 
\begin{eqnarray}
& &
\left\lbrace \begin{array}{l}
\chi_{1 l k}^{(1)}(x \rightarrow \infty)  \rightarrow \sin(k_1 x + \delta_{l 1}) \\
\chi_{2 l k}^{(1)}(x \rightarrow \infty)  \rightarrow t \sin(k_2 x + \delta_{l 2})
\end{array}\right., \text{~~~for the $A_2 = 0$ boundary condition}, \nonumber \\
& &
\left\lbrace \begin{array}{l}
\chi_{1 l k}^{(2)}(x \rightarrow \infty)  \rightarrow t^{\prime} \sin(k_1 x + \delta_{l 1}^{\prime}) \\
\chi_{2 l k}^{(2)}(x \rightarrow \infty)  \rightarrow  \sin(k_2 x + \delta_{l 2}^{\prime})
\end{array}\right., \text{~~~for the $A_1 = 0$ boundary condition}, \label{TwoSolutions}
\end{eqnarray}
where $k_1$ and $k_2$ are the dimensionless relative momentum of the two particle pairs reduced by $\kappa$. Recombining the two solutions (\ref{TwoSolutions}), we acquire the general solution of Eqn.~(\ref{RadialEquation})
\begin{eqnarray}
A_l \chi_{l k}^{(1)}(x) + B_l \chi_{l k}^{(2)}(x). \label{ScatteringForm}
\end{eqnarray}
The sine functions can be decomposed into exponential incoming and outgoing wave functions 
\[
\sin (k x + \delta) = \frac{e^{i (k x + \delta)} - e^{-i (k x + \delta)}}{2 i} .
\] 
If we want a pure $\chi_1 \chi_1$ initial state, we acquire
\begin{eqnarray}
\left\lbrace \begin{array}{l}
A_{l 1}  e^{-i \delta_{l1}} + B_{l 1}  t^{\prime} e^{-i \delta_{l1}^{\prime}} = i \frac{1}{k_1} \\
A_{l 1}  t e^{-i \delta_{l2}} + B_{l 1}  e^{-i \delta_{l2}^{\prime}} = 0
\end{array}\right.. \label{ScatteringFormAsymptotic1}
\end{eqnarray}
This is the no-incoming wave function condition for   $\chi_2 \chi_2$ pair, and the incoming wave function 
of $\chi_1 \chi_1$ is normalized. The solution for $A_l$ and $B_l$ is
\begin{eqnarray}
\left\lbrace
\begin{array}{l}
A_{l 1} = \frac{i e^{i \delta_{l 1} + i \delta_{l 1}^{\prime} + i \delta_{l 2}}}{e^{i \delta_{l 1}^{\prime} + i \delta_{l 2}} - e^{i \delta_{l 1} + i \delta_{l 2}^{\prime}} t t^{\prime}} \frac{1}{k_1} \\
B_{l 1}  = \frac{i e^{i \delta_{l 1} + i \delta_{l 1}^{\prime} + i \delta_{l 2}^{\prime} } t }{-e^{i \delta_{l 1}^{\prime} + i \delta_{l 2}} + e^{i \delta_{l 1} + i \delta_{l 2}^{\prime}} t t^{\prime}} \frac{1}{k_1}
\end{array}\right..	\label{Solution1}
\end{eqnarray}
For the pure $\chi_2 \chi_2$ initial state, we have
\begin{eqnarray}
\left\lbrace \begin{array}{l}
A_{l 2} e^{-i \delta_{l1}} + B_{l 2} t^{\prime} e^{-i \delta_{l1}^{\prime}} = 0 \\
A_{l 2} t e^{-i \delta_{l2}} + B_{l 2} e^{-i \delta_{l2}^{\prime}} = i \frac{1}{k_2}
\end{array}\right.. \label{ScatteringFormAsymptotic2}
\end{eqnarray}
The corresponding solutions are
\begin{eqnarray}
\left\lbrace
\begin{array}{l}
A_{l 2} = \frac{i e^{i \delta_{l 1} + i \delta_{l 2} + i \delta_{l 2}^{\prime}} t^{\prime} }{-e^{i \delta_{l 1}^{\prime} + i \delta_{l 2}} + e^{i \delta_{l 1} + i \delta_{l 2}^{\prime}} t t^{\prime}} \frac{1}{k_2} \\
B_{l 2} = -\frac{i e^{i \delta_{l 1}^{\prime} + i \delta_{l 2} + i \delta_{l 2}^{\prime} } }{-e^{i \delta_{l 1}^{\prime} + i \delta_{l 2}} + e^{i \delta_{l 1} + i \delta_{l 2}^{\prime}} t t^{\prime}} \frac{1}{k_2}
\end{array}\right..	\label{Solution2}
\end{eqnarray}
With (\ref{Solution1}) and (\ref{Solution2}), we can also calculate the $\chi_1 \chi_1 \leftrightarrow \chi_2 \chi_2$ cross sections, and the $\chi_i \chi_i$ self-scattering cross sections. However, in this paper, we shall only utilize
$A_{l 1,2} \chi_{l i}^{(1)}(x) + B_{l 1,2} \chi_{l i}^{(2)}(x)$ in order to calculate the wave functions' ``overlap'' between the 
$\chi_1 \chi_1$ or $\chi_2 \chi_2$ initial states and the bound states.

In the $0< -\gamma^2 < \delta \gamma^2$ case, the $\chi_1 \chi_1 \rightarrow \chi_2 \chi_2$ process is kinematically closed. Only self-scattering $\chi_1 \chi_1 \rightarrow\chi_1 \chi_1$ is available.   In the language of wave functions,
\begin{eqnarray}
& &
\left\lbrace \begin{array}{l}
\chi_{1 l k}^{(1)}(x \rightarrow \infty)  \rightarrow \frac{\kappa}{k_1} \sin(k_1 x + \delta_{l 1}) \\
\chi_{2 l k}^{(1)}(x \rightarrow \infty)  \propto e^{-\sqrt{\delta \gamma^2 + \gamma^2} x}
\end{array}\right.. \nonumber  \label{BelowThresholdSolution}
\end{eqnarray}
We need to adjust $A_1$ and $A_2$ in Eq. (\ref{ScatteringForm}) properly to acquire this boundary 
condition. In this case, neither $\chi_{1 l k}^{(2)}(x \rightarrow \infty)$ nor 
$\chi_{2 l k}^{(2)}(x \rightarrow \infty)$ is available since there is no asymptotic on-shell $\chi_2 \chi_2$ pair on this energy.

In summary, we denote
\begin{eqnarray}
\chi_{l k 1} (x)= A_{l 1} \chi_{l k}^{(1)}(x) + B_{l 1} \chi_{l k}^{(2)}(x) \nonumber \\
\chi_{l k 2} (x)= A_{l 2} \chi_{l k}^{(1)}(x) + B_{l 2} \chi_{l k}^{(2)}(x) \label{DifferentInitial}
\end{eqnarray}
in the $-\gamma^2 \geq \delta \gamma^2$ case, indicating $\chi_1 \chi_1$ and $\chi_2 \chi_2$ initial states,  
respectively. On the other hand, for the $-\gamma^2 < \delta \gamma^2$ case, we would have
\begin{eqnarray}
\chi_{l k 1} (x)=\chi_{l k}^{(1)}(x)
\end{eqnarray}
for the $\chi_1 \chi_1$ initial states only. For the $\chi_1 \chi_2$ or $\chi_2 \chi_1$ initial states, 
(\ref{DifferentInitial}) is still valid.

Following Ref.~\cite{Petraki2} and expanding $e^{-i \vec{b} \cdot \vec{r}}$ into partial waves, 
we acquire the following overlap integrals for the lowest orders in the ``spherical Bessel function indices''
\begin{eqnarray}
\mathcal{I}^t_{\vec{k}, n00}(\vec{b}) &=& \left( \frac{4 \pi}{\kappa} \right)^{\frac{3}{2}} \frac{1}{4 \pi} \int_0^{\infty} dx \chi_{n, l=0}^*(x) A \chi_{|\vec{k}|, l=0}(x)  j_0 \left( \frac{b x}{\kappa} \right) \nonumber \\
&-& \sqrt{\frac{4 \pi}{\kappa^3}} i P_1(\hat{\mathbf{k}} \cdot \hat{\mathbf{b}}) \int_0^{\infty} dx \chi_{n, l=0}^*(x) A \chi_{|\vec{k}|, l=1}(x) 3 j_1\left( \frac{b x}{k} \right) , \label{IDetail1} \\
\mathcal{I}^t_{\vec{k}, l=0 \rightarrow n10}(\vec{b}) &=& -\left( \frac{4 \pi}{\kappa} \right)^{\frac{3}{2}} i \frac{1}{4 \pi} \int_0^{\infty} dx \chi_{n, l=0}^*(x) A \chi_{|\vec{k}|, l=1}(x) 3 j_1 \left( \frac{b x}{\kappa} \right), \label{IDetail2} \\
\mathcal{I}^t_{\vec{k}, l=0 \rightarrow n11}(\vec{b}) &=& -\left( \frac{4 \pi}{\kappa} \right)^{\frac{3}{2}} i \frac{-1}{4 \pi} \sqrt{\frac{3}{2}} \sin \theta e^{-i \phi} \int_0^{\infty} dx \chi_{n, l=0}^*(x) A \chi_{|\vec{k}|, l=1}(x) 3 j_1 \left( \frac{b x}{\kappa} \right), \label{IDetail3} \\
\mathcal{I}^t_{\vec{k}, l=1 \rightarrow n10}(\vec{b}) &=& -\left( \frac{4 \pi}{\kappa} \right)^{\frac{3}{2}} i \frac{\sqrt{3}}{4 \pi}  \cos \theta \int_0^{\infty} dx \chi_{n, l=0}^*(x) A \chi_{|\vec{k}|, l=1}(x) j_0 \left( \frac{b x}{\kappa} \right), \label{IDetail4} \\
\mathcal{I}^t_{\vec{k}, l=2 \rightarrow n10}(\vec{b}) &=& -\left( \frac{4 \pi}{\kappa} \right)^{\frac{3}{2}} i \frac{\sqrt{3}}{2 \pi}  \cos \theta \int_0^{\infty} dx \chi_{n, l=0}^*(x) A \chi_{|\vec{k}|, l=1}(x) 3 j_1 \left( \frac{b x}{\kappa} \right), \label{IDetail5} \\
\mathcal{I}^t_{\vec{k}, l=2 \rightarrow n11}(\vec{b}) &=& -\left( \frac{4 \pi}{\kappa} \right)^{\frac{3}{2}} i \frac{\sqrt{15}}{4 \pi}  \sin \theta e^{-i \phi} \int_0^{\infty} dx \chi_{n, l=0}^*(x) A \chi_{|\vec{k}|, l=1}(x) 3 j_1 \left( \frac{b x}{\kappa} \right), \label{IDetail6}
\end{eqnarray}
where $\mathcal{I}^t_{\dots}$ indicates the corresponding partial wave expansions for all the overlap integrals, (\ref{Is})--(\ref{Idsm}), with the upper symbol $t$ being either empty or $\pm$.  And $A$ is the corresponding $\sigma^3$ or $i \sigma^2$ for each integrals. Finally $\chi_{n, l}(x)$ and  $\chi_{|\vec{k}|, l}(x)$  are the 2-dimensional radial wave functions for the corresponding bound states or scattering states.

Compared with the Ref.~\cite{Petraki2}, we keep the spherical Bessel function $j_n\left(\frac{b x}{\kappa} \right)$ without expanding them to the $\left(\frac{b}{\kappa}\right)^{n}$ series for more precise calculations. The indices in the $j_n\left(\frac{b x}{\kappa} \right)$ are similar to the indices of the $\left(\frac{b}{\kappa}\right)^{n}$ in the order of magnitude approximations. Unlike in Ref.~\cite{Petraki2}, the appearance of $A = \sigma^{0,1,2,3}$, and the non-orthogonal wave functions in $\mathcal{I}$ functions leave us non-zero values for the overlap integrals, $\int_0^{\infty} dx \chi_{n l}^* (x) A \chi_{|\vec{k}|, l_I} (x)$ in the lowest order of $\frac{b}{\kappa}$,   so the $\left( \frac{b}{\kappa} \right)^0$ order term should be calculated. However, in some cases, the $\left( \frac{b}{\kappa} \right)^0$ order term of the $\mathcal{I}$ expansions is still rather small:  
for example, in the $\delta \gamma^2 \rightarrow \infty$ case. Therefore, a $\left( \frac{b}{\kappa} \right)^1$ term is still needed, and we have expanded the $\mathcal{I}^t_{k, n00}(\vec{b})$, $\mathcal{I}^t_{k, n11}(\vec{b})$, $\mathcal{I}^t_{k, n10}(\vec{b})$ up to this level.

For the $\mathcal{K}$'s, let us remember that the $\vec{\nabla}^2$ in the 
(\ref{Ks})--(\ref{Kd}) can be replaced by other parts of the schr\"{o}dinger equation in (\ref{Schoedingers})--(\ref{Schoedingerd}).  Then it is easy to insert these terms in (\ref{IDetail1})--(\ref{IDetail6}) for us to acquire the corresponding $\mathcal{K}_{\cdots}$ expansion terms.

Finally, we show the leading order expressions for $\vec{\mathcal{J}}$ functions.
\begin{eqnarray}
\vec{\mathcal{J}}^{\pm}_{\vec{k}, n00}(\vec{b}) &\simeq& -\hat{\mathbf{k}} \sqrt{\frac{4 \pi}{\kappa}} \int_0^{\infty} dx \left[ \frac{d \chi_{n, l=0}^*(x)}{d x} - \frac{\chi_{n, l=0}^* (x)}{x} \right] A \chi_{|\vec{k}|, l=1}(x) \frac{3 j_1 \left( \frac{b x}{\kappa} \right)}{\left( \frac{b x}{\kappa} \right)}, \\
\vec{\mathcal{J}^{\pm}}_{\vec{k}, n10}(\vec{b}) &\simeq& i \sqrt{\frac{12 \pi}{\kappa}} \left\lbrace ( \hat{\mathbf{k}} \cos \theta_{\vec{k}} - \frac{\hat{\mathbf{e}}_z}{3}) \int_0^{\infty} dx \left[ \chi^{\prime}_{n, l=1}(x) - \frac{2 \chi_{n, l=1}(x)}{x}\right]^* A \chi_{|\vec{k}|, l=2}(x) j_0 \left( \frac{b x}{\kappa} \right) \right. \nonumber \\
& & \left. + \frac{\hat{\mathbf{e}}_z}{3} \int_0^{\infty} dx \left[ \chi^{\prime}_{n, l=1}(x) + \frac{\chi_{n, l=1}(x)}{x}\right]^* \chi_{|\vec{k}|, l=0}(x) j_0 \left( \frac{b x}{\kappa} \right) \right\rbrace, \\
\vec{\mathcal{J}}^{\pm}_{\vec{k}, n11}(\vec{b}) &\simeq& -i \sqrt{\frac{6 \pi}{\kappa}} \left\lbrace ( \hat{\mathbf{k}} \sin \theta_{\vec{k}} e^{i \phi_{\vec{k}}} - \frac{\hat{\mathbf{e}}_x + i \hat{\mathbf{e}}_y}{3}) \int_0^{\infty} dx \left[ \chi^{\prime}_{n, l=1}(x) - \frac{2 \chi_{n, l=1}(x)}{x}\right]^* A \chi_{|\vec{k}|, l=2}(x) j_0 \left( \frac{b x}{\kappa} \right) \right. \nonumber \\
& & \left. + \frac{\hat{\mathbf{e}}_x+ i \hat{\mathbf{e}}_y}{3} \int_0^{\infty} dx \left[ \chi^{\prime}_{n, l=1}(x) + \frac{\chi_{n, l=1}(x)}{x}\right]^* A \chi_{|\vec{k}|, l=0}(x) j_0 \left( \frac{b x}{\kappa} \right) \right\rbrace, \\
\vec{\mathcal{J}}^{\pm}_{\vec{k}, n1(-1)}(\vec{b}) &=& -\vec{J}_{\vec{k}, n11}(\vec{b}).
\end{eqnarray}
Here, again $A$ is the $i \sigma^2$ or $\sigma^3$ corresponding to the $+$ or $-$ upper indices.

\subsection{Pair Annihilations Inside the Bound State Particles}

Once a dark matter bound state is formed, it might dissociate into two free dark matter particle again after scattering with a dark photon or dark Higgs boson within the surrounding plasma. During the dark matter freeze-out processes, the bound state formation processes are competing with the bound state dissociation processes. Only when the bound states decay rapidly enough before their dissociation, can the bound state formation effectively contribute to the total annihilation processes \cite{EllisFengLuo}. The dark matter dissociation rate can be evaluated conveniently by Eq. (\ref{DissociationFormula}) in the next subsection 4.3.  The annihilation of the two components inside the bound state particle is the main contribution to the bound state decay processes. In this subsection, we concentrate at the algorithm to calculate the such decay widths. 

To calculate the decay width, we need to calculate the squared annihilation amplitude of the particle pair in a particular wave function of relative motion, which is proportional to the $\psi(0)^2$ $\rightarrow$ $|\psi(0)|^2$ in the s-wave, and proportional to the $\psi^{\prime}(0)^2$ $\rightarrow$ $| \psi^{\prime}(0) |^2$ in the p-wave situation. However, it is more 
convenient to calculate the perturbative squared annihilation amplitude of a plane wave. 
For example, if we calculate the squared annihilation amplitude of two particles with the momentum $(E_1, 0, 0, p)$ and $(E_2, 0, 0, -p)$ travelling along the $\pm z$ direction, and the total momentum of the particle pair is zero, we write down the phase space integrated square amplitude in the form of
\begin{eqnarray}
\int \sum_{i} |\mathcal{M}_{zi} (E_1, E_2, p, \theta, \phi)|^2 \sin \theta d \theta d \phi, \label{MSquareForm}
\end{eqnarray}
where $\theta$, $\phi$ are the final state phase space angles, and $i$ indicates all of the polarization indices in the final states. Then, we need to extract the s-wave and p-wave contributions from the (\ref{MSquareForm}). This is done by ``somehow'' expanding the (\ref{MSquareForm}) and find the coefficient of the $v^0$ and $v^2$ terms, where $v$ is the relative velocity of the two particles, and can be expressed by
\begin{eqnarray}
v = \frac{p}{E_1}+\frac{p}{E_2}.
\end{eqnarray}
Notice that the $v$ dependence of $E_{1,2}$ should not expanded, which is the meaning of the ``somehow'' that we have noted. Then we can make the expansion
\begin{eqnarray}
\mathcal{M}_{zi}(E_1, E_2, p, \theta, \phi) \approx \mathcal{M}^0_{zi} (E_1, E_2, p, \theta, \phi) +  \mathcal{M}^1_{zi} (E_1, E_2, p, \theta, \phi) v,  \label{MFormV}
\end{eqnarray}
and take (\ref{MFormV}) into (\ref{MSquareForm}) to find out the $v^0$ and $v^2$ terms to acquire the s-wave and 
p-wave contributions to the amplitudes. We then need to multiply the $v^0$ term with $|\psi(0)|^2$, 
and replace $v^2$ term with $\frac{|\psi^{\prime}(0)|^2}{\mu_r^2}$  to acquire the decay amplitudes.  
\footnote{Practically these should be $|\psi_{\chi_1 \chi_1}(0) \pm \psi_{\chi_2 \chi_2}(0)|^2$, $|\psi_{\chi_1 \chi_2}(0) \pm \psi_{\chi_2 \chi_1}(0)|^2$, $\frac{|\psi^{\prime}_{\chi_1 \chi_1}(0) \pm \psi^{\prime}_{\chi_2 \chi_2}(0)|^2}{\mu_r^2}$, or $\frac{|\psi^{\prime}_{\chi_1 \chi_2}(0) \pm \psi^{\prime}_{\chi_2 \chi_1}(0)|^2}{\mu_r^2}$, depending on the final states}

However, we should note that $E_1$ and $E_2$ also depend on $v$. Directly expanding on $E_1$ and $E_2$ will also contribute to  $v^2$ terms, which are not the real $p-$wave contributions however. Notice that we should not expand $E_1$ and $E_2$ on $v$ during our separation of the s- and p-wave contributions.

From the (\ref{PlaneExpansion}) we can see that the plane wave along the z-direction only contains the $l_z=0$ element in the $l=1$ partial wave. We will also need the $l_x=0$ or $l_y=0$ annihilation amplitude to calculate the decay of the states in different total angular momentum $j$'s. This is done by calculating $\mathcal{M}_{xi}(E_1, E_2, p, \theta, \phi)$ and $\mathcal{M}_{yi}(E_1, E_2, p, \theta, \phi)$, which are corresponding to the annihilation of two particles travelling along the $\pm x$ and $\pm y$ directions, with the total momentum of the two particle system being zero.

We also need to calculate the amplitudes for various total spins of the initial states. The amplitudes can be generically written as, 
\begin{eqnarray}
\mathcal{M} &=& \overline{u} \dots v\text{, or }\overline{v} \dots u. \nonumber \\
&=& Tr[v \overline{u} \dots]\text{, or }Tr[u \overline{v} \dots].
\end{eqnarray}
where $u$ and $v$ are the four-spinors of the initial states. Now $v(p_1) \overline{u}(p_2)$, or $u(p_1) \overline{v}(p_2)$ can be written in the form of
\begin{eqnarray}
v(p_1) \overline{u}(p_2) =  \left[ \begin{array}{cc}
\sqrt{\sigma \cdot p_1} X (\sqrt{\sigma \cdot p_2})^{\dagger} & \sqrt{\sigma \cdot p_1} X (\sqrt{\overline{\sigma} \cdot p_2})^{\dagger} \\
-\sqrt{\overline{\sigma} \cdot p_1} X (\sqrt{\sigma \cdot p_2})^{\dagger} & -\sqrt{\overline{\sigma} \cdot p_1} X (\sqrt{\overline{\sigma} \cdot p_2})^{\dagger}
\end{array} \right] \gamma^0 \nonumber \\
u(p_1) \overline{v}(p_2) =\left[ \begin{array}{cc}
\sqrt{\sigma \cdot p_1} X (\sqrt{\sigma \cdot p_2})^{\dagger} & -\sqrt{\sigma \cdot p_1} X (\sqrt{\overline{\sigma} \cdot p_2})^{\dagger} \\
\sqrt{\overline{\sigma} \cdot p_1} X (\sqrt{\sigma \cdot p_2})^{\dagger} & -\sqrt{\overline{\sigma} \cdot p_1} X (\sqrt{\overline{\sigma} \cdot p_2})^{\dagger}
\end{array} \right] \gamma^0,
\end{eqnarray}
where $X$ indicates the total spin of the particle pair. For the $S=0$ case, the $X = \frac{\sigma^0}{\sqrt{2}}$. For the $S=1$ case, the $X = \frac{\sigma^{1,2,3}}{\sqrt{2}}$, which are corresponding to the $S_x=0$, $S_y=0$, and $S_z=0$ respectively. 

Now we are ready to calculate the annihilation amplitude for all the combinations of different $|L L_z S S_z\rangle$. Notice that in the $L=1$, $S=1$ case,
\begin{eqnarray}
|J=2, J_z=0 \rangle &=& -\frac{1}{\sqrt{6}}(|L_x=0, S_x=0 \rangle + |L_y=0, S_y=0 \rangle - 2 |L_y=0, S_y=0 \rangle), \nonumber \\
|J=1, J_z=0 \rangle &=& \frac{i}{\sqrt{2}}(|L_x=0, S_y=0 \rangle - |L_y=0, S_x=0 \rangle \rangle), \nonumber \\
|J=0, J_z=0 \rangle &=& -\frac{1}{\sqrt{3}}(|L_x=0, S_x=0 \rangle + |L_y=0, S_y=0 \rangle + |L_y=0, S_y=0 \rangle),
\end{eqnarray}
we can follow this to recombine the amplitudes to calculate the decay width of different $|J J_Z \rangle$ states.

Not all of the bound states can have two-body decay channels. Some are prohibited by the he C/P-parity and the dark $U(1)$ charge conservation laws. This will be discussed in detail in Appendix \ref{QuantumNumbers}.  Let us apply the Goldstone-equivalence theorem, and we ignore the masses of the final state scalar particles since $m_{\gamma^{\prime}} \ll m_{\chi_{1,2}}$ and $m_R \ll m_{\chi_{1,2}}$, and the longitudinal polarization of the $\gamma^{\prime}$ is replaced with the Goldstone mode $I$. It is then convenient to analyze the two-body decay channels in the basis of $\chi$, $\overline{\chi}$ before the spontaneously symmetry breaking, and then rotate to the $\chi_{1,2}$ basis for calculations. The result is shown in Table~\ref{TwoBodyDecay}. The integrated amplitude of all the possible two-body final states are listed in Table~\ref{TwoBodyM2}.

\begin{table}
\begin{tabular}{|c|c|c|c|c|}
\hline
$\overline{\chi}$/$\chi$ states & $\chi_{1,2}$ states & $^{2S+1}L_J$ & $ J^{CP}$ & two-body decay channels \\
\hline
$\overline{\chi} \chi$/$\chi \overline{\chi}$ & $| \chi_1 \chi_1 \rangle + | \chi_2 \chi_2 \rangle$ & 
$^1S_0$ & $ 0^{+-}$ & $\gamma^{\prime} \gamma^{\prime}$ \\
\hline
 & $| \chi_1 \chi_2 \rangle - | \chi_2 \chi_1 \rangle$ & $^1P_1 $ & $1^{- -}$ & $I R$ \\
\hline
 & & $^3S_1$ & $1^{-+}$ & No two-body decay channel \\
 \hline
 & $| \chi_1 \chi_1 \rangle + | \chi_2 \chi_2 \rangle$ & $^3P_0$ & $0^{+ +}$ & $R R$, $I I$, $\gamma^{\prime} \gamma^{\prime}$ \\
\hline
 & & $^3P_1$  & $1^{+ +}$ & No two-body decay channel \\
\hline
 & & $^3P_2$ & $2^{+ +}$ & $R R$, $I I$, $\gamma^{\prime} \gamma^{\prime}$ \\
 \hline
 \hline
 \hline
$\chi \chi$/$\overline{\chi}  \overline{\chi}$ & $| \chi_1 \chi_1 \rangle - | \chi_2 \chi_2 \rangle$ & 
$^1S_0$ & $0^{+ +}$ & No two-body decay channel \\
\hline
 & & $^3P_0$ & $0^{+ -}$ & No two-body decay channel \\
\hline
 & $| \chi_1 \chi_1 \rangle - | \chi_2 \chi_2 \rangle$ & $^3P_{1(2)}$ & $1 (2)^{+ -}$ & 
 $R \gamma^{\prime}$ \\
\hline
  & $| \chi_1 \chi_2 \rangle +  | \chi_2 \chi_1 \rangle$ & $^3P_{1(2)}$  & $1 (2)^{- -}$ & 
  $I \gamma^{\prime}$ \\
\hline
\end{tabular}
\caption{Two-body decay channels. The $^{2S+1}L_J$ and $J^{CP}$ are the usual spectroscopy 
notations,  indicating the quantum numbers of the total spin $S$, orbital angular momentum $L$, 
total angular momentum $J$, $C$-parity, and parity under space inversion.} \label{TwoBodyDecay}
\end{table}
\begin{table}
\begin{tabular}{|c|c|c|c|}
\hline
Channel & $^{2S+1}L_J , C$ & Kernel & Coefficient \\
\hline
$\gamma^{'} \gamma^{'}$ & $^1S_0 , +$ & $\frac{512 E^2 (E+m_{\chi})^2 \pi}{(E^2+m_{\chi}^2)^2}$ & $2 \left( \frac{g^{\prime}}{2} \right)^4 \left| \psi_{\chi_1 \chi_1}(0)+\psi_{\chi_2 \chi_2}(0) \right|^2$ \\
\hline
 & $ ^3P_0 , +$  & $\frac{128 E^2 (E-5 m_{\chi})^2 \pi}{3 (E^2 + m_{\chi}^2)^2}$ & $2 \left( \frac{g^{\prime}}{2} \right)^4 \frac{1}{m_{\chi}^2} \left| \psi_{\chi_1 \chi_1}^{\prime}(0) + \psi_{\chi_2 \chi_2}^{\prime}(0) \right|^2$ \\
\hline
  & $^3P_2 , +$ & $\frac{1792}{15} \frac{E^2 (E+m_{\chi})^2 \pi}{(E^2 + m_{\chi}^2)^2}$ & $2 \left( \frac{g^{\prime}}{2} \right)^4 \frac{1}{m_{\chi}^2} \left| \psi_{\chi_1 \chi_1}^{\prime}(0) + \psi_{\chi_2 \chi_2}^{\prime}(0) \right|^2$ \\
\hline
\hline
$R \gamma^{'}$ & $^3P_1 , -$ & $\frac{256 E^4 m_{\chi}^2 (E+m_{\chi})^2 \pi}{3 (E^2 + m_{\chi}^2)^4}$ & $2 \left( \frac{g^{\prime}}{2} \frac{y}{2} \right)^2 \frac{1}{m_{\chi}^2} \left| \psi_{\chi_1 \chi_2}^{\prime}(0) + \psi_{\chi_2 \chi_1}^{\prime}(0) \right|^2$ \\
\hline
 & $^3P_2 , -$& $\frac{256 E^4 m_{\chi}^2 (E+m_{\chi})^2 \pi}{5 (E^2 + m_{\chi}^2)^4}$ & $2 \left( \frac{g^{\prime}}{2} \frac{y}{2}  \right)^2 \frac{1}{m_{\chi}^2} \left| \psi_{\chi_1 \chi_2}^{\prime}(0) + \psi_{\chi_2 \chi_1}^{\prime}(0) \right|^2$ \\
\hline
\hline
$I \gamma^{'}$ & $^3P_1 , +$ & $\frac{256 E^4 m_{\chi}^2 (E+m_{\chi})^2 \pi}{3 (E^2 + m_{\chi}^2)^4}$ & $2  \left( \frac{g^{\prime}}{2} \frac{y}{2}  \right)^2 \frac{1}{m_{\chi}^2} \left| \psi_{\chi_1 \chi_1}^{\prime}(0) - \psi_{\chi_2 \chi_2}^{\prime}(0) \right|^2$ \\
\hline
 & $^3P_2 , +$ & $\frac{256 E^4 m_{\chi}^2 (E+m_{\chi})^2 \pi}{5 (E^2 + m_{\chi}^2)^4}$ & $2 \left( \frac{g^{\prime}}{2} \frac{y}{2}  \right)^2 \frac{1}{m_{\chi}^2} \left| \psi_{\chi_1 \chi_1}^{\prime}(0) - \psi_{\chi_2 \chi_2}^{\prime}(0) \right|^2$ \\
\hline
\hline
$RR$ & $^3P_0 , +$ & $\frac{64 E^2 (E^3 + 7 E^2 m_{\chi} + 3 E m_{\chi}^2 + 9 m_{\chi}^3)^2 \pi}{3 (E^2+m_{\chi}^2)^4}$ & $2 \left(  \frac{g^{\prime}}{2}  \right)^4 \frac{1}{m_{\chi}^2} \left| \psi_{\chi_1 \chi_1}^{\prime}(0) + \psi_{\chi_2 \chi_2}^{\prime}(0) \right|^2$ \\
\hline
 & $^3P_2 , +$ & $\frac{512 E^6 (E+m_{\chi})^2 \pi}{15 (E^2+m_{\chi}^2)^4}$ & $2 \left(  \frac{y}{2}  \right)^4 \frac{1}{m_{\chi}^2} \left| \psi_{\chi_1 \chi_1}^{\prime}(0) + \psi_{\chi_2 \chi_2}^{\prime}(0) \right|^2$ \\
\hline
\hline
$IR$ & $^3S_1 , -$ & $\frac{16 (E+m_{\chi})^2 \left[ E^2 (-4+\frac{4 Q_{\chi}^2 g^2}{y^2}) + \frac{4 Q_{\chi}^2 g^2}{y^2} m_{\chi}^2 \right]^2  \pi}{3 E^2 (E^2+m_{\chi}^2)^2}$ & $2 \left(  \frac{y}{2}  \right)^4 \left| \psi_{\chi_1 \chi_2}(0) + \psi_{\chi_2 \chi_1}(0) \right|^2$ \\
\hline
\hline
$II$ & $^3P_0 , +$ & $\frac{64 E^2 (E^3 + 7 E^2 m_{\chi} + 3 E m_{\chi}^2 + 9 m_{\chi}^3)^2 \pi}{3 (E^2+m_{\chi}^2)^4}$ & $2 \left(  \frac{y}{2}  \right)^4 \frac{1}{m_{\chi}^2} \left| \psi_{\chi_1 \chi_1}^{\prime}(0) + \psi_{\chi_2 \chi_2}^{\prime}(0) \right|^2$ \\
\hline
 & $^3P_2 , +$ & $\frac{512 E^6 (E+m_{\chi})^2 \pi}{15 (E^2+m_{\chi}^2)^4}$ & $2 \left(  \frac{y}{2}  \right)^4 \frac{1}{m_{\chi}^2} \left| \psi_{\chi_1 \chi_1}^{\prime}(0) + \psi_{\chi_2 \chi_2}^{\prime}(0) \right|^2$ \\
\hline
\end{tabular}
\caption{Two-body decay squared matrix elements, phase-space integrated, final-state summed. For abbreviation, we define $g^{\prime} = Q_{\chi} g$. The ``$LSJC$'' indicates the quantum numbers of the orbital angular momentum, total spin, total angular momentum, and C-parity. Compared with the Tab.~\ref{TwoBodyDecay}, the parity is absent, because many states are no longer eigenstates of the parity. Although $m_{\chi_1} \neq m_{\chi_2}$, for simplicity, we still use $m_{\chi}$ as the mass of both $\chi_1$ and $\chi_2$ during our calculations. $E$ is the energy of either initiate particle, as we have calculated in the center of mass frame.}  \label{TwoBodyM2}
\end{table}

For those states which do not have a two-body decay channel, we need to calculate the three-body decay amplitudes. For generic values of $m_\chi, m_{\gamma^{'}}, m_{I}$ and $m_R$, rigorous analytical results of the transition probabilities are formidable to acquire because of the intricate phase space structures. We shall neglect the final particle masses by setting 
\[
m_{\gamma^{'}} = m_{I} = m_R = 0,
\]
to simplify the 3-body decay formulas. The analytical decay amplitudes and the decay rates, including the phase space integration of the final three particles are listed in Tab.~\ref{ThreeBodyM2}.  However one has to pay the penalty for this simplification. This is the appearance of the infrared divergences (IR) in the phase space integral. To see this, {let us introduce an explicit IR cutoff parameter $a$ in} the corner of the Dalitz-plot. The three momentums of the final state particles are denoted to be $k_1$, $k_2$, $k_3$., and $(k_1 + k_2 + k_3)^2 = 4 E^2$. Having ignored all the final state masses, the three Dalitz plot 
parameters then become
\begin{eqnarray}
k_1 \cdot k_2 &=& x_3 E^2, \nonumber \\
k_2 \cdot k_3 &=& x_1 E^2, \nonumber \\
k_1 \cdot k_3 &=& x_2 E^2.
\end{eqnarray}
Since $x_1+x_2+x_3 = 2$, there are only two independent parameters. The divergences usually exist in the $x_i=0$, $x_j=0$ $(i \neq j)$ region. If one divergence is located in the, e.g., $x_1=0$, $x_2=0$ region, we just modify the phase space integral $\int_0^2 d x_1 \int_0^{2-x_1} d x_2$ into $\int_a^2 d x_1 \int_a^{2-x_1} d x_2$ to avoid it, and leave all the $a$'s explicitly in Tab.~\ref{ThreeBodyM2}.

Notice that in our later numerical calculations, our $m_{\gamma^{\prime}}$ and $m_{R}$ are in the similar order of magnitude. It seems that the divergence we have encountered is merely an artifact of ignoring the final state masses, so a naive way to fix this problem would be to replace the IR cutoff by the lightest particle mass among the three final state particles. This is true when $m_{\gamma^{\prime},R} \gtrsim \mu_r \alpha^{\prime 2}$, and will introduce approximately $a = \frac{m_{\gamma^{\prime},R}}{\mu_r}$ as a natural {IR} cut-off. However, when one of the $m_{\gamma^{\prime},R} \lesssim \mu_r \alpha^{\prime 2}$, and when it is produced with the energy/momentum $\lesssim \mu_r \alpha^{\prime 2}$, {some propagators with the $\lesssim \mu_r \alpha^{\prime 2}$ energy/momentum scale shall inevitibly appear, and hence the prediction ability of our method here fails as the commonly used ``ladder approximation'' only considers the $| \chi_i \chi_j \rangle$ Fock's state and neglects many comparable $\mu_r \alpha^{\prime 2}$-scale energy/momentum terms. As is well-known in QCD, this IR divergence problem was first discovered in the context of $P$-wave charmonium decays long time ago Ref.~\cite{QCD1, QCD2, QCD3, QCD4}, and made a long standing problem until the color-octet mechanism was proposed by Bodwin, Braaten and Lepage (BBL) Ref.~\cite{NRQCDSolution, BBLNRQCD}. Similar with the QCD strategy, we need to regard $\mu_r \alpha^{\prime 2}$ or $\mu_r \alpha^{\prime}$ as a natural ``matching scale'', and when the energy/momentum of one of the decay product is below this scale, contributions from the higher Fock's state $| \chi_i \chi_j \gamma^{\prime}/I/R \rangle$ arise. Reducing to the phase space parameter, this is corresponding to the $a = \frac{|E_B|}{\mu_r} \sim \alpha^{\prime 2}$ or $a=\alpha^{\prime}$, where $E_B$ is the binding energy. Studying the precise manipulation below the IR-matching point is beyond the scope of this paper, however we know that the matching point scale will appear in the final results, and can be treated as an effective IR cut-off. In Tab.~\ref{ThreeBodyM2}, the results are logarithmly dependent on $a$ so they change little when $a$ varies from $\alpha^{\prime 2}$, $\alpha^{\prime}$ and $\frac{m_{\gamma^{\prime},R}}{\mu_r}$. With these considerations, we adopt {$a = \text{MAX}[\frac{|E_B|}{\mu_r}, \frac{m_R}{\mu_r}, \frac{m_{\gamma^{\prime}}}{\mu_r}]$} in this paper. }
{However actual calculations show that the contributions from the bound states with no 2-body final decay channels are negligible.}
  
\begin{table}
\begin{tabular}{|c|c|c|c|}
\hline
Channel & $^{2S+1}L_J , C$ & Kernel & Coefficient \\
\hline
$\gamma^{'} \gamma^{'} \gamma^{'}$ & $^1P_1 , -$ & $\frac{64 (136 - 7 \pi^2 - 8 \ln 4 + 48 \ln a)}{3 m_{\chi}^2}$ & $4 \left( \frac{g^{\prime}}{2} \right)^6 \frac{1}{m_{\chi}^2} \left| \psi_{\chi_1 \chi_2}^{\prime}(0) - \psi_{\chi_2 \chi_1}^{\prime}(0) \right|^2$ \\
\hline
\hline
$R \gamma^{'} \gamma^{'}$ & $^1S_0 , +$ & $\frac{128 [-132+48 a+\pi^2 +48(1+a) \ln \frac{2}{a} ]}{3 m_{\chi}^2}$ & $4 \left( \frac{y}{2} \frac{g^{\prime}}{2} \frac{g^{\prime}}{2} \right)^2 \left| \psi_{\chi_1 \chi_1}(0)-\psi_{\chi_2 \chi_2}(0)\right|^2$ \\
\hline
 & $^3P_0 , +$ & $\frac{1}{3} \frac{-128 \left[ 1072 - 432 a + 3 \pi^2 - 432 (1+a) \ln \frac{2}{a} \right]}{3 m_{\chi}^2}$ & $4 \left( \frac{y}{2} \frac{g^{\prime}}{2} \frac{g^{\prime}}{2} \right)^2 \frac{1}{m_{\chi}^2} \left| \psi_{\chi_1 \chi_1}^{\prime}(0)-\psi_{\chi_2 \chi_2}^{\prime} (0) \right|^2$ \\
\hline
\hline
$I \gamma^{'} \gamma^{'}$ & $^1S_0 , -$ & $\frac{32 (-132+\pi^2+48 \ln \frac{2}{a}) }{3 m_{\chi}^2} + \frac{512 (1+\ln \frac{2}{a} ) a}{m_{\chi}^2}$ & $4 \left( \frac{y}{2} \frac{g^{\prime}}{2} \frac{g^{\prime}}{2} \right)^2 \left| \psi_{\chi_1 \chi_2}(0)+\psi_{\chi_2 \chi_1}(0) \right|^2$ \\
\hline
 & $^3P_0 , -$ & $\begin{array}{l} \frac{1}{3} \left[ \frac{-32 (1072 + 3 \pi^2 - 432 \ln \frac{2}{a} )}{3 m_{\chi}^2} \right. \\ \left. + \frac{4608 (1+\ln \frac{2}{a}) a}{m_{\chi}^2} \right] \end{array}$ & $4 \left( \frac{y}{2} \frac{g^{\prime}}{2} \frac{g^{\prime}}{2} \right)^2 \frac{1}{m_{\chi}^2} \left| \psi_{\chi_1 \chi_2}^{\prime} (0)+\psi_{\chi_2 \chi_1}^{\prime} (0)\right|^2$ \\
\hline
\hline
$I R \gamma^{'}$ & $^1S_0 , +$ & $\frac{16 (-228 + 23 \pi^2)}{3 m_{\chi}^2}$ & $4 \left( \frac{y}{2} \frac{y}{2} \frac{g^{\prime}}{2} \right)^2 \left| \psi_{\chi_1 \chi_1}(0)+\psi_{\chi_2 \chi_2}(0) \right|^2$ \\
\hline
 & $^3P_1 , +$ & $ \begin{array}{l} \frac{1}{3} \left[ \frac{16 (56+67 \pi^2 - 96 \ln 2 + 256 \ln a)}{3 m_{\chi}^2} \right. \\ \left.- \frac{128 ( 1+24 \ln 2 - 72 \ln a) a}{3 m_{\chi}^2} \right] \end{array}$ & $4 \left( \frac{y}{2} \frac{y}{2} \frac{g^{\prime}}{2} \right)^2 \frac{1}{m_{\chi}^2} \left| \psi_{\chi_1 \chi_1}^{\prime}(0)+\psi_{\chi_2 \chi_2}^{\prime}(0) \right|^2$ \\
\hline
\hline
$RRR$ & $^3P_0 , +$ & $\begin{array}{l} \frac{2}{3 m_{\chi}^2}  \left[ 9 \lambda_{\Phi} (32 + \lambda_{\Phi})  \right. \\+ 8 \pi^2 (-8 + 9 \lambda_{\Phi})  \\ + 128 (-28 + 75 \ln 2) \\ \left. - 9600 \ln a \right] \end{array}$ & $4 \left( \frac{y}{2} \right)^6 \frac{1}{m_{\chi}^2} \left| \psi_{\chi_1 \chi_1}^{\prime}(0)-\psi_{\chi_2 \chi_2}^{\prime}(0) \right|^2$ \\
\hline
\hline
$IRR$ & $^3P_0 , -$ & $\begin{array}{l} \frac{2}{27 m_{\chi}^2}  \left[ 72 \pi^2 (8+\lambda_{\Phi}) \right. \\ + 9 \lambda_{\Phi} (32+\lambda_{\Phi})   \\  + 128 (-320 + 231 \ln 2) \\ \left.- 29568 \ln a \right] \end{array}$ & $4 \left( \frac{y}{2} \right)^6 \frac{1}{m_{\chi}^2} \left| \psi_{\chi_1 \chi_2}^{\prime}(0)+\psi_{\chi_2 \chi_1}^{\prime}(0) \right|^2$ \\
\hline
\hline
$IIR$ & $^3P_0 , +$ & $\begin{array}{l} \frac{2}{27 m_{\chi}^2}  \left[ 72 \pi^2 (8+\lambda_{\Phi}) \right. \\ + 9 \lambda_{\Phi} (32+\lambda_{\Phi})  \\ + 128 (-320 + 231 \ln 2) \\ \left.- 29568 \ln a \right] \end{array}$ & $4 \left( \frac{y}{2} \right)^6 \frac{1}{m_{\chi}^2} \left| \psi_{\chi_1 \chi_1}^{\prime}(0)-\psi_{\chi_2 \chi_2}^{\prime}(0) \right|^2$ \\
\hline
\hline
$III$ & $^3P_0 , -$ & $\begin{array}{l} \frac{2}{27 m_{\chi}^2}  \left[ 72 \pi^2 (8+\lambda_{\Phi}) \right. \\ + 9 \lambda_{\Phi} (32+\lambda_{\Phi})  \\ + 128 (-320 + 231 \ln 2) \\ \left.- 29568 \ln a \right] \end{array}$ & $4 \left( \frac{y}{2} \right)^6 \frac{1}{m_{\chi}^2} \left| \psi_{\chi_1 \chi_2}^{\prime}(0)+\psi_{\chi_2 \chi_1}^{\prime}(0) \right|^2$ \\
\hline
\end{tabular}
\caption{Three-body decay squared matrix elements, phase-space integrated, final-state summed. Here, for abbreviation, we define $g^{\prime} = Q_{\chi} g$. The ``$LSJC$'' indicates the quantum numbers of the orbital angular momentum, total spin, total angular momentum, and C-parity. We still use $m_{\chi}$ to replace the $m_{\chi_{1,2}}$ as in the Tab.~\ref{TwoBodyM2}, and since the three-body modes are not counted in the annihilation processes, we only show the results in the $E \rightarrow m_{\chi}$ limit.} \label{ThreeBodyM2}
\end{table}

With these informations on the relevant amplitudes, we can finally calculate the decay widths
of DM bound states
\begin{eqnarray}
\Gamma = \frac{f_s}{128 \pi^2 m_{\chi}} \int_{\text{phase}} d(\text{phase}) \frac{|\mathcal{M}_{\text{Tab}}|^2}{2 \mu_r},	\label{DecayWidth}
\end{eqnarray}
where $\int_{\text{phase}} d(\text{phase}) |\mathcal{M}_{\text{Tab}}|^2$ is obtained by multiplying the expressions listed in the ``Kernel'' and ``Coefficient'' columns in Table~\ref{TwoBodyM2} for each two-body decay channel, and in Table~\ref{ThreeBodyM2} multiplied by an additional $\frac{m_{\chi}^2}{4 \pi}$ factor for each three-body decay channel. $f_s$ is the factor for the final state identical particles. $f_s = \frac{1}{2!}$ for two  identical particles in the final state, and $f_s = \frac{1}{3!}$ for three identical particles in the final state.

\subsection{Boltzmann Equations}

Now let us discuss  the Boltzmann equations for our model. In principle, we need to write down the Boltzmann equations of all the components, including $\chi_1$, $\chi_2$, $\gamma^{\prime}$, $R$ and all of the bound states as well as their temperatures. 
This will result in a set of complicated coupled equations, which is beyond our current computing resources. Therefore, we shall simplify the situation with the following considerations:
\begin{itemize}
\item When $T_{\gamma} \gg m_{\gamma^{\prime}}$, $m_R$, where $T_{\gamma}$ is the photon temperature of the plasma, although $\gamma^{\prime}$ or $R$ might have already frozen out from the thermal plasma quite early, being nearly massless particles, they cool down approximately synchronously with the plasma due to their red-shifts. Therefore, the kinetic equilibrium keeps $T_{\chi} = T_{\gamma^{\prime}}=T_R \approx T_{\gamma}$, where $T_{\chi}$ is the temperature of $\chi_1$, $\chi_2$, and $T_{\gamma^{\prime}}$, $T_R$ are the temperatures of the $\gamma^{\prime}$ and $R$.
\item $\chi_1 \chi_1 \rightarrow \chi_2 \chi_2$ processes by interchanging the $\gamma^{\prime}$ particles are rapid enough to keep a number density equilibrium between $\chi_1$ and $\chi_2$. 
\item When $T_{\gamma} \lesssim m_{\gamma^{\prime}}$, $m_R$, the light components become non-relativistic and their number densities drop dramatically. The kinetic equilibrium between $m_{\gamma^{\prime}}$ and $\chi_{1,2}$ is broken. The red-shift of $\chi_1$ and $\chi_2$ will imply $T_{\chi} \propto T_{\gamma}^2$.
\end{itemize}
Also based upon the steps listed in Ref.~\cite{EllisFengLuo}, we define
\begin{eqnarray}
F_{X} = \frac{\langle \Gamma \rangle_{X}}{\langle \Gamma \rangle_{X} + \langle \Gamma \rangle_{X \text{,dis}}},
\end{eqnarray}
where $\langle \Gamma \rangle_{X}$ and $\langle \Gamma \rangle_{X \text{,dis}}$ are the thermal averaged decay width and the dissociation rate of the bound state, respectively.  Their definitions are
\begin{eqnarray}
\langle \Gamma \rangle_{X} &=& \frac{K_1(\frac{m_X}{T_{\chi}})}{K_2(\frac{m_X}{T_{\chi}})} \Gamma_X, \nonumber \\
\langle \Gamma \rangle_{X \text{,dis}} &=& \frac{1}{2} \sum_{i,j} \langle \sigma v \rangle_{T_{\chi}, \chi_i \chi_j \rightarrow X+\gamma^{\prime}/R} \frac{n_{\chi_i}^{\text{eq}}(T_{\chi}) n_{\chi_j}^{\text{eq}}(T_{\chi})}{n_X^{\text{eq}}(T_{\chi})}, \label{DissociationFormula}
\end{eqnarray}
where $K_1$ and $K_2$ are the Bessel functions, and $\Gamma_X$ is the decay width calculated from (\ref{DecayWidth}).  $i$,$j=1,2$ indicate the $\chi_1$, $\chi_2$ components, and $n_{A}^{\text{eq}}(T)$ is the thermal-equivalent number density of component $A$ at the temperature $T$. The $\langle \sigma v \rangle_{T, \chi_i \chi_j \rightarrow X + \gamma^{\prime}/R}$ is the thermal-averaged bound state formation cross section: 
\begin{eqnarray}
\langle \sigma v \rangle_{\chi_i \chi_j \rightarrow X + \gamma^{\prime}/R} & = & 
\frac{1}{n_{\chi_i} n_{\chi_j}} \frac{4 T}{32 \pi^4} \int ds^{\prime} s^{\prime \frac{3}{2}} K_1 \left( \frac{\sqrt{s^{\prime}}}{T} \right) \lambda(1, \frac{m_{\chi_1}^2}{s^{\prime}}, \frac{m_{\chi_2}^2}{s^{\prime}} )  
\nonumber 
\\
& \times & \sigma_{\chi_i \chi_j \rightarrow X + \gamma^{\prime}/R} (s^{\prime}),
\end{eqnarray}
where $n_{\chi_i}$ is the number density of component $\chi_i$. $\sigma_{\chi_i \chi_j \rightarrow X + \gamma^{\prime}/R}$ is calculated following the steps listed in Section \ref{BSFProcesses}, and $\lambda(1, a, b) = (1-a-b)^2-4 a b$. We should note that in Ref.~\cite{EllisFengLuo}, there is also an additional term $\langle \sigma v \rangle_{\tilde{g} \tilde{R} \rightarrow \tilde{g} g} n_{\tilde{g}}$, which should be something like, e.g.,  $\langle \sigma v \rangle_{\chi_i  X \rightarrow \chi_j \gamma^{\prime}} n_{\chi_i}$ in our paper. We ignore this term because this term will be suppressed by $n_{\chi}$ compared with the $\langle \Gamma \rangle_{X, \text{dis}}$, when the temperature drops below the $\chi_{i,j}$ mass.

For a pair of DM particle $\chi_i \chi_j$, their effective cross section times velocity is defined as
\begin{eqnarray}
\langle \sigma v \rangle_{\chi_i \chi_j, \text{eff}} = \langle \sigma v \rangle_{\chi_i \chi_j, \text{anni}} + \sum_X F_X \langle \sigma v \rangle_{\chi_i \chi_j \rightarrow X + \gamma^{\prime}/R},
\end{eqnarray}
where $\langle \sigma v \rangle_{\chi_i \chi_j, \text{anni}}$ indicates the annihilation rate including the Sommerfeld effect. In this paper, we only consider the 2-body final state when calculating the $\langle \sigma v \rangle_{\chi_i \chi_j, \text{anni}}$. Although Ref.~\cite{SommerfeldComplete1, SommerfeldComplete2} provided a complete unitarity-preserved calculation of the Sommerfeld effect in the large boost situation, we only perform a rough numerical scan on the parameter space while missing the resonances in this paper, so the methods in Ref.~\cite{NimaSommerfeld} are sufficient. Therefore, the phase space integrated squared amplitudes are again from the Tab.~\ref{TwoBodyM2}, and the $\psi_{\chi_i \chi_j}^{(\prime)}(0)$ there plays the role of the ``boost factor''. For the multi-component dark matter models, Ref.~\cite{WrongskianMethod1, WrongskianMethod} performed a Wrongskian-based method to acquire these zero-point wave functions (or derivatives). {However, in this paper, we shall use the definitions of these wave function at the origin straightforwardly.} For the acquired scattering wave functions $\chi_{lk}$ given by (\ref{ScatteringForm}) satisfying the asymptotic conditions (\ref{ScatteringFormAsymptotic1}) or (\ref{ScatteringFormAsymptotic2}), 
\begin{eqnarray}
\psi_i (0) & = & \lim\limits_{x \rightarrow0} \frac{\chi_{ilk} ( x )}{x} ~~~~~~~~~~(\textrm{for} ~l=0) ,  \label{psi01}
\\
\psi_i^{\prime} (0) & = & i \mu_r v \lim\limits_{x \rightarrow0} \frac{3 \chi_{ilk} ( x )}{k_i x^2} ~~(\textrm{for} ~l=1) .  \label{psi02}
\end{eqnarray} 
Here $i$ is the index of the wave function components. {We have testified the accuracy of straightforwardly applying the definitions in (\ref{psi01}-\ref{psi02}) of the Coulomb potential and the no-interaction limit by comparing the numerical results with the analytical solution, and found they are precisely compatible. Since the Yukawa potential is somehow the intermediate state between Coulomb interaction and no interaction limit, and for our two component situation, the potential does not deviate from the Yukawa potential very much, so we expect a sufficient precision for this simple ``method''.} Once we obtain the squared amplitudes, we then calculate the cross section in the center of momentum frame with
\begin{eqnarray}
\sigma_{\chi_i \chi_j \text{,anni}} v = \frac{f_s}{4 E_{\chi_1} E_{\chi_2}} \frac{2 |\vec{p}|}{32 \pi E_{\text{cm}}} \int_{\Omega} d\Omega_{\text{cm}} |\mathcal{M}_{\chi_i \chi_j,\text{Tab}} |^2,
\end{eqnarray}
where $E_{\chi_i}$ is the energy of component $\chi_i$, $|\vec{p}|$ is the magnitude of the 3-momentum of either particle in the center of momentum frame, and $E_{\text{cm}}$ is the center of mass energy of the initial particle pair.  The angular integration $\int_{\Omega} d\Omega_{\text{cm}} |\mathcal{M}_{\chi_i \chi_j\text{Tab}} |^2$ {is calculated by summing over} all the annihilation channels listed in Tab.~\ref{TwoBodyM2} for the initial particles $\chi_i \chi_j$. To calculate the $\mathcal{M}_{\text{Tab}}$, we need the values of the wave functions  at the origin $\psi_{\chi_i \chi_j}(0) $ or their first derivatives $\psi^{\prime}_{\chi_i \chi_j}(0)$ of the scattering state. These can be extracted from the radial  wave functions $\chi_{1lk}(x)$ and $\chi_{2lk}(x)$ according to (\ref{psi01}-\ref{psi02}), and the coefficients $A_{li}$, $B_{li}$ should also be multiplied for the corresponding initial states. Thus, sommerfeld effects have also been automatically included through all these processes.

Besides the amplitudes listed in the Tab.~\ref{TwoBodyM2}, we should also note that there is an interference term between $S$ and $D$ waves which contribute to the $IR$ final state. This is absent in the Tab.~\ref{TwoBodyM2}, because it is not relevant to the decay width calculations. The ``Kernel'' part of the squared amplitude is calculated to be
\begin{eqnarray}
& & \frac{2 \pi v^2}{9 (E^2+m_{\chi}^2)^4} (-4 E^8 T+ E^8 T^2 - 96 E^7 m_{\chi} + 24 E^7 m_{\chi} T - 128 E^6 m_{\chi}^2 + 20 E^6 T m_{\chi}^2 \nonumber \\
&+& 4 E^6 m_{\chi}^2 T^2  - 224 E^5 m_{\chi}^3 + 80 E^5 m_{\chi}^3 T - 128 E^4 m_{\chi}^4 + 52 E^4 m_{\chi}^4 T+ 6 E^4 m_{\chi}^4 T^2 + 56 E^3 m_{\chi}^5 T \nonumber \\
&+& 28 E^2 m_{\chi}^6 T + 4 E^2 m_{\chi}^6 T^2 + m_{\chi}^8 T^2 ),
\end{eqnarray}
where $E$ is again the energy of each particle in the center of momentum frame, and $T = \frac{2 Q_{\chi} g}{y}$. In the ``Coefficient'' part, there should include the Sommerfeld boost factor ``$\psi(0) \psi^{\prime \prime}(0)$''. However, the interference term is sub-dominant, so we do not consider this term and set the boost factor to be 1, so the total ``Coefficient'' becomes $2 \left(\frac{y}{2} \right)^4$.

Then the total effective annihilation cross section is given by
\begin{eqnarray}
\langle \sigma v \rangle_{T_{\chi}, \text{eff}} = \sum_{i,j} \frac{n_{\chi_i}^{\text{eq}}(T_{\chi}) n_{\chi_j}^{\text{eq}}(T_{\chi})}{n^{\text{eq }}(T_{\chi})^2} \langle \sigma v \rangle_{\chi_i \chi_j, \text{eff}}, \label{EffectiveSigmaV}
\end{eqnarray}
where $n^{\text{eq}} = \sum_i n_{\chi_i}^{\text{eq}}$. The Boltzmann equation is then written in the familiar form
\begin{eqnarray}
\frac{dY}{dx_f} = - \frac{z s}{H(m_{\chi})} \left( 1- \frac{x_f}{3 g_s^*} \frac{d g_s^*}{dx_f} \right) \langle \sigma v \rangle_{T_{\chi}, \text{eff}} (Y^2 - Y_{eq}^2), \label{Boltzmann}
\end{eqnarray}
where $H(T)$ is the Hubble constant at temperature $T$, $m_{\chi}$ is the mass of the dark matter, $x_f = \frac{m_{\chi}}{T_{\gamma}}$, and $g_s^*$ is the effective degree of freedom entering the entropy density. $Y$ is the effective dark matter particle number per co-moving volume, or $\frac{n}{s}$, where $n$ is the particle's number density, and the $s$ is the plasma's entropy density. To solve  (\ref{Boltzmann}), we also need to know the relationship between $T_{\chi}$ and $T_{\gamma}$. This requires another equation. However, in our paper, we simplify this by setting the kinetic decoupling constant $T_{\text{kd}} = \text{min} \lbrace m_{\gamma^{\prime}}, m_R \rbrace $, and
\begin{eqnarray}
T_{\chi} = \left\lbrace \begin{array}{cc}
T_{\gamma} & \text{ , when $T_{\gamma} > T_{\text{kd}}$} \\
\frac{T_{\gamma}^2}{T_{\text{kd}}} & \text{ , when $T_{\gamma} \leq T_{\text{kd}}$}
\end{array}  \right. .
\end{eqnarray}

\section{Numerical Results and Discussions}
\subsection{Numerical Results}
Using the internal functions embedded in the micrOMEGAs\cite{micrOMEGAs1, micrOMEGAs2}, we solve the differential equation in (\ref{Boltzmann}). Among the parameters for the numerical analysis, we choose two different values of DM masses: $m_\chi =3$ TeV and $m_\chi =10$ TeV. Then we vary $\alpha_g = \frac{(Q_{\chi} g)^2}{4 \pi}$ and $\frac{2 c_1}{c_2} = \frac{2 y^2}{(Q_{\chi} g)^2}$ to show the results for 
$\Omega h^2$ in Figs. \ref{Omegah2_1500} and \ref{Omegah2_10000} for $m_\chi =3$ TeV and $m_\chi =10$ TeV, 
respectively.  

In these figures, we have adopted
\[
\xi_1=\frac{1}{1.03} \times \left[ \frac{(2 c_1 + c_2)^2 \alpha^{\prime}}{2} \right]^{-1} , ~~~\rm{and} ~~~\xi_2=1.5 \xi_1 .
\]
The reason for this choice are the following.  First let us notice that $\delta \gamma^2 \ll 1$ and $\xi_{1,2} \ll 1$ in our concerned parameter space to be discussed, so that the binding energies of the ground states can be approximated by the hydrogen atom's formula. For the radial equation (\ref{RadialEquation}), this is estimated to be $E_{\text{gnd.~stat.}} \simeq -\frac{(2 c_1 + c_2)^2 \alpha^{\prime 2} \mu_r}{2}$. Reduce this by the inverted ``Bohr radius'' $\frac{-E_{\text{gnd.~stat.}}}{\kappa} = \frac{\gamma^2}{\alpha^{\prime}} \approx \frac{(2 c_1 + c_2)^2 \alpha^{\prime}}{2}$, and recall that $\xi_{1,2} = \frac{\kappa}{m_{R,\gamma^{\prime}}}$. It is easy to see that if $\xi_1 \lesssim \left[ \frac{(2 c_1 + c_2)^2 \alpha^{\prime}}{2} \right]^{-1}$, the emission of $R$ is kinematically prohibited in the low temperature situation to escape the CMB constraints. When $\xi_{2} >  \left[ \frac{(2 c_1 + c_2)^2 \alpha^{\prime}}{2} \right]^{-1}$, significant $\gamma^{\prime}$ emission occurs and will induce the second-stage annihilation processes to be discussed below. However, we will keep $\frac{\xi_{2}}{\xi_{1}}$  small enough to evade the unitarity constraints on $\lambda_{\Phi}$. Finally, $\xi_1=\frac{1}{1.03} \times \left[ \frac{(2 c_1 + c_2)^2 \alpha^{\prime}}{2} \right]^{-1}$ and $\xi_2=1.5 \xi_1$ are enough for both these two purposes. Details of the unitarity bounds mentioned before will be addressed  in section \ref{Constraints}.

\begin{table}
\begin{tabular}{c|c}
\begin{minipage}[t]{0.45\textwidth}
\includegraphics[width=0.99\textwidth]{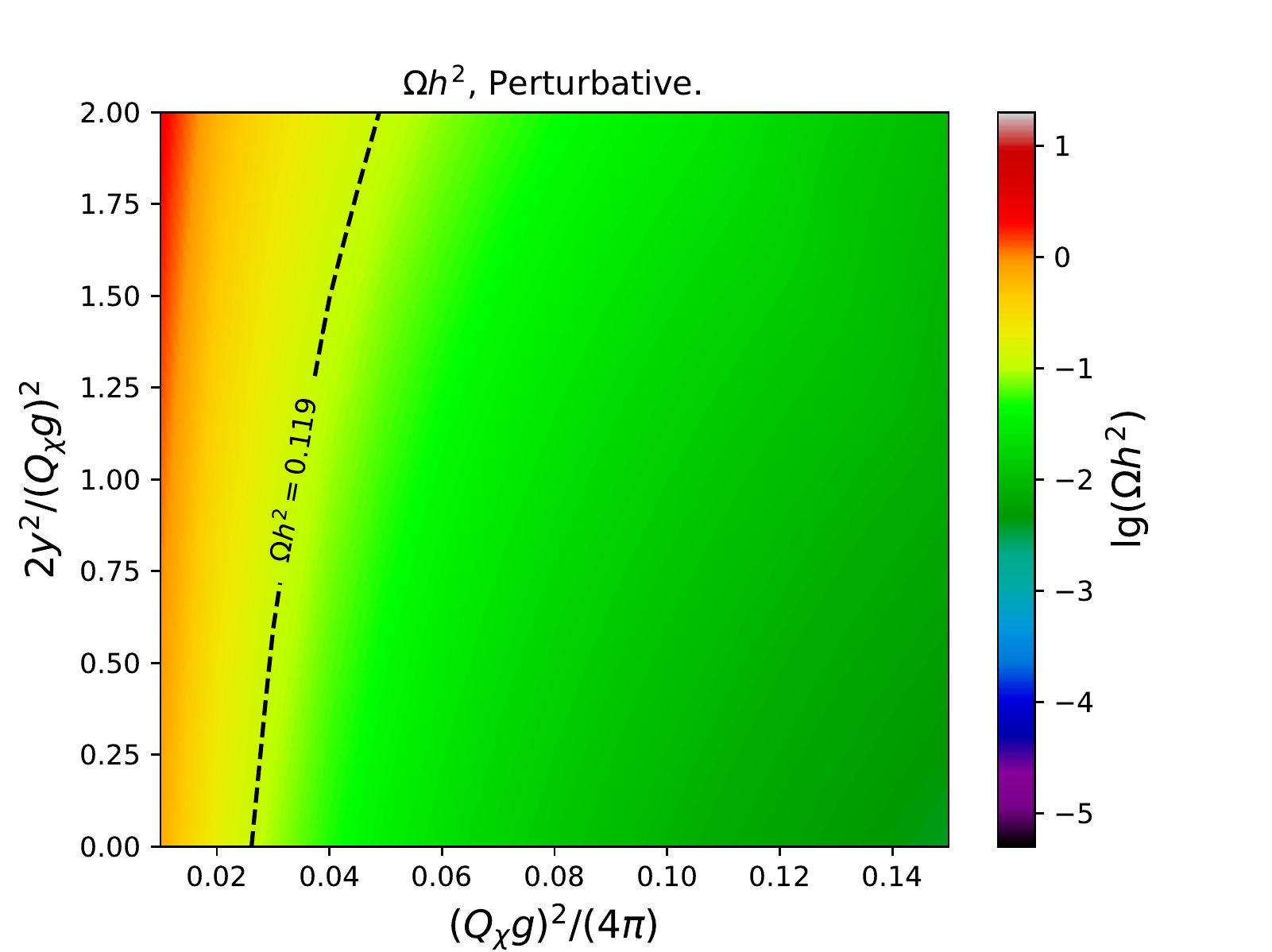}
\includegraphics[width=0.99\textwidth]{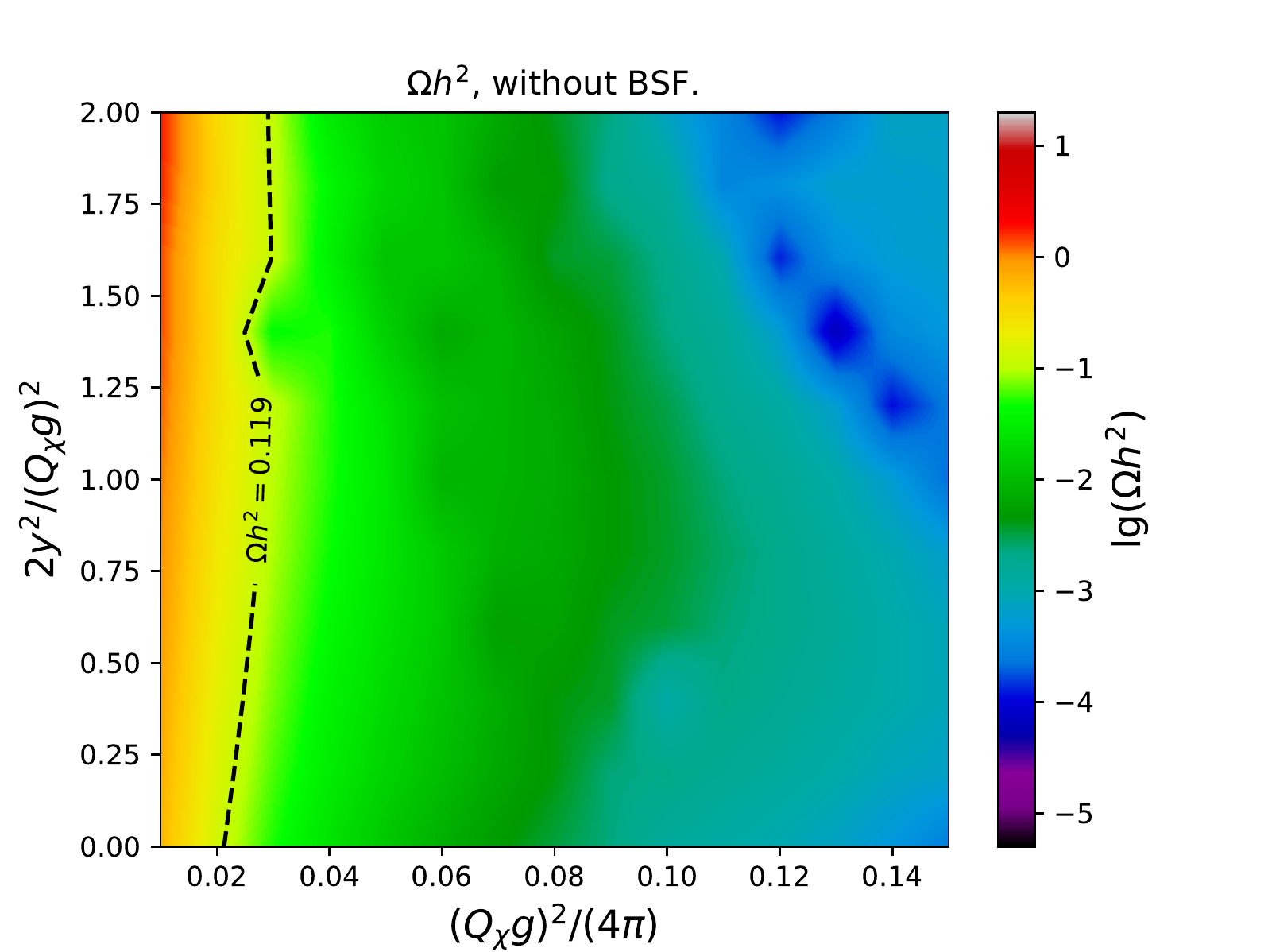}
\includegraphics[width=0.99\textwidth]{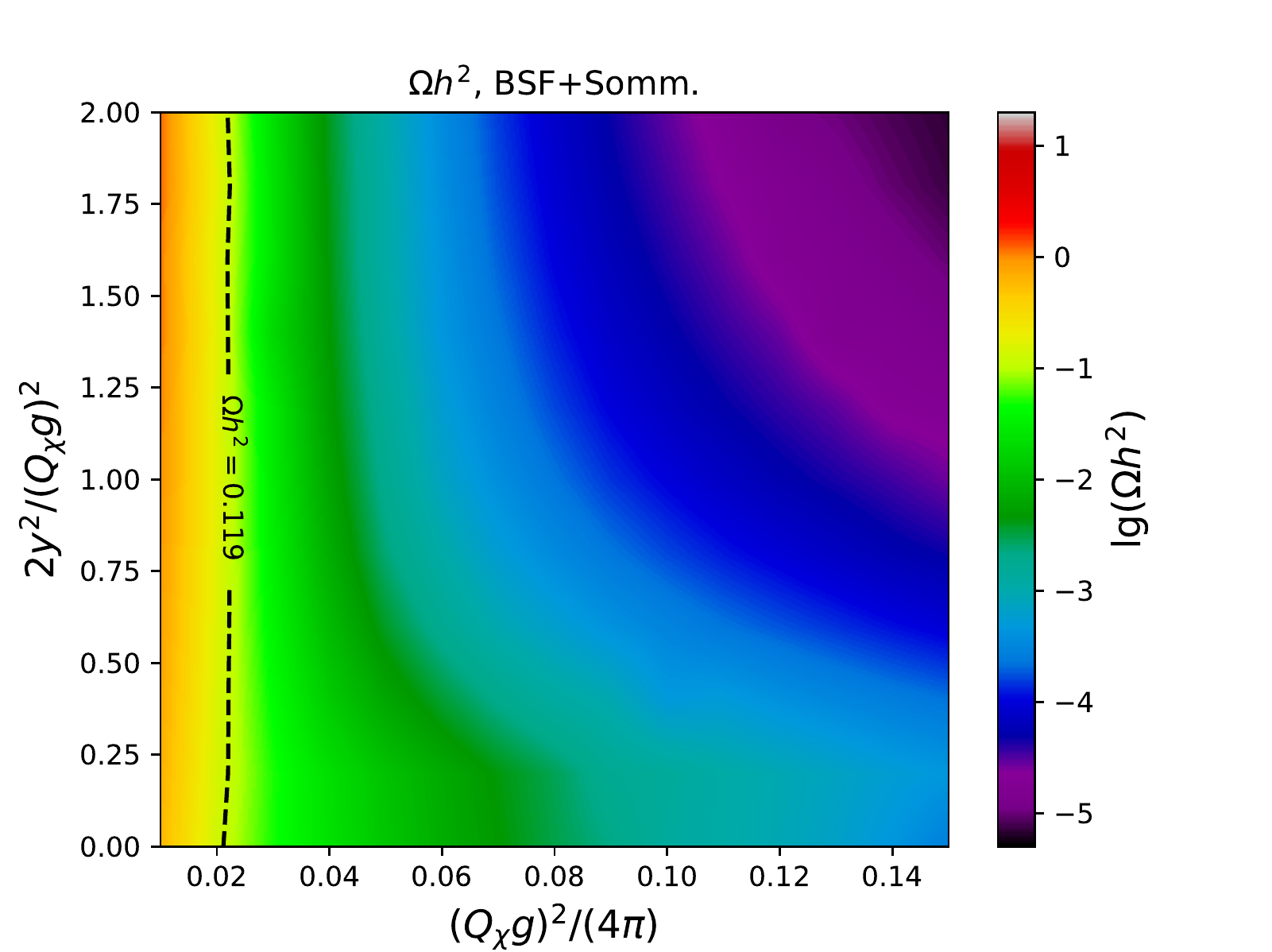}
\captionof{figure}{Dark matter relic density $\Omega h^2$ in the $m_{\chi}=3$ TeV, $\xi_1=\frac{1}{1.03} \times \left[ \frac{(2 c_1 + c_2)^2 \alpha^{\prime}}{2} \right]^{-1}$ and $\xi_2=1.5 \xi_1$ situation. $c_2 = \frac{(Q_{\chi} g)^2}{4 \pi \alpha^{\prime}}$ and $\frac{c_1}{c_2} = \frac{y^2}{(Q_{\chi} g)^2}$ vary. The pannels are the tree-level perturbative results, the Sommerfeld results, and the bound state formation effects included results from up to down, respectively.} \label{Omegah2_1500}
\end{minipage} &
\begin{minipage}[t]{0.45\textwidth}
\includegraphics[width=0.99\textwidth]{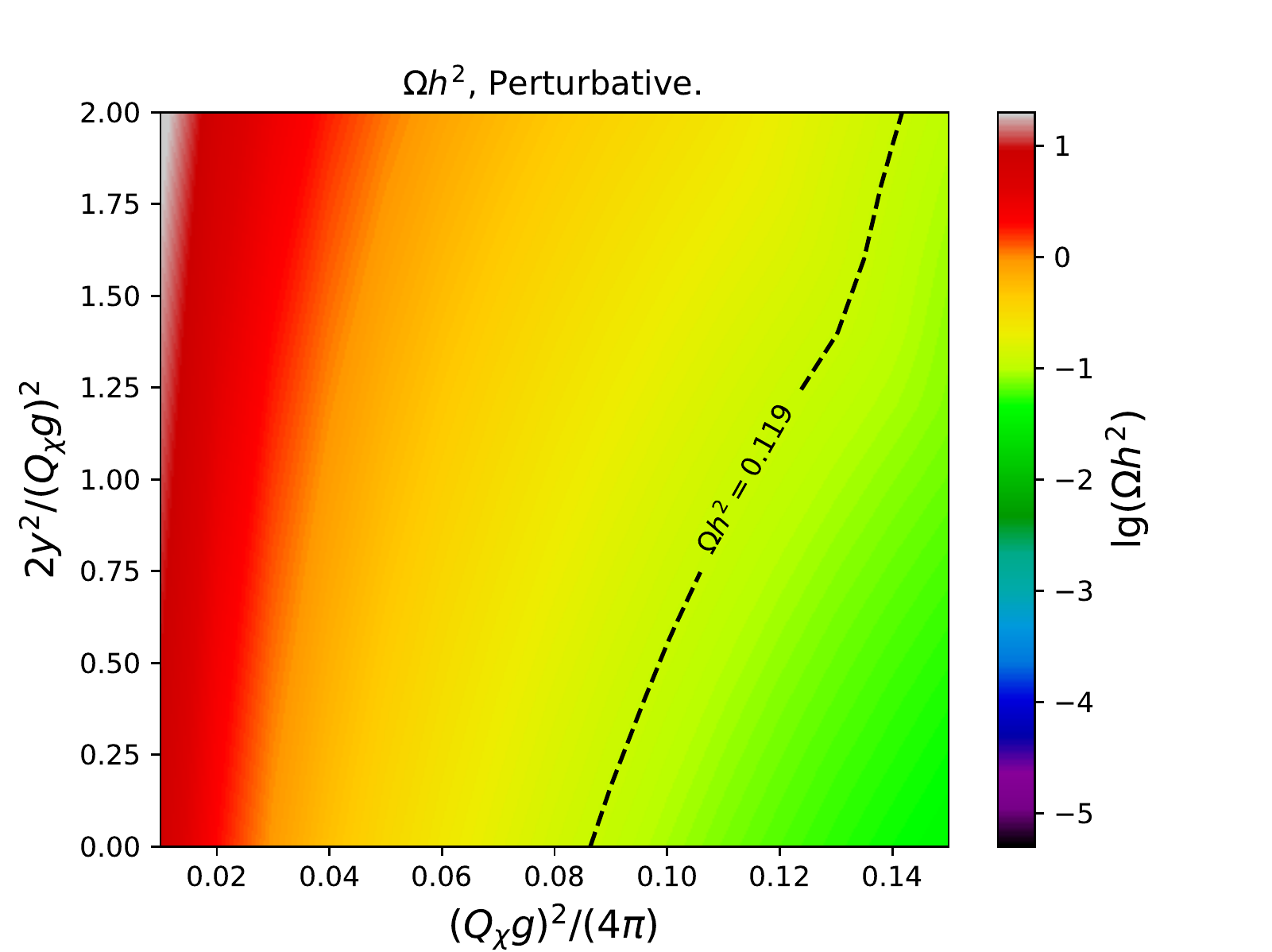}
\includegraphics[width=0.99\textwidth]{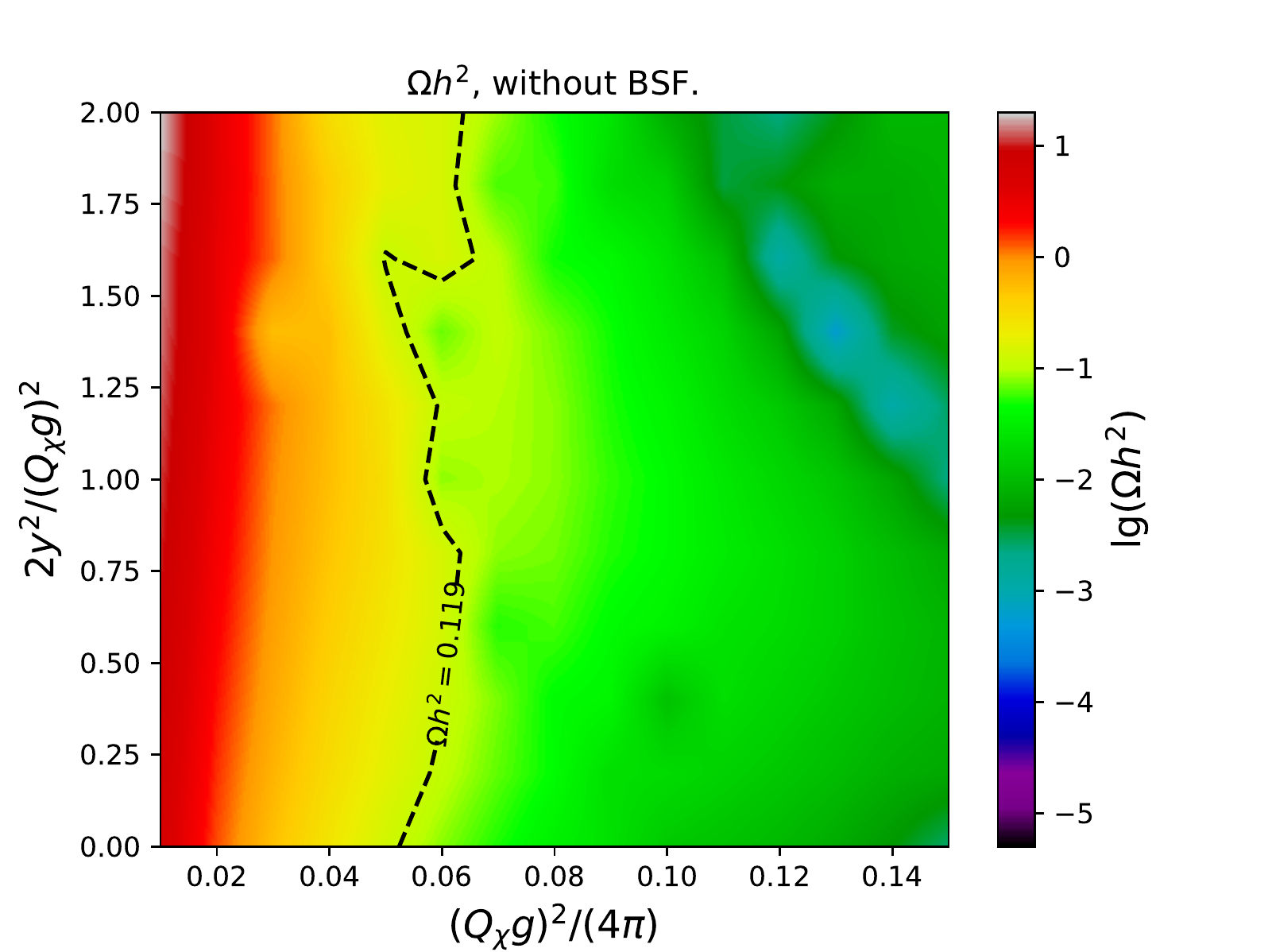}
\includegraphics[width=0.99\textwidth]{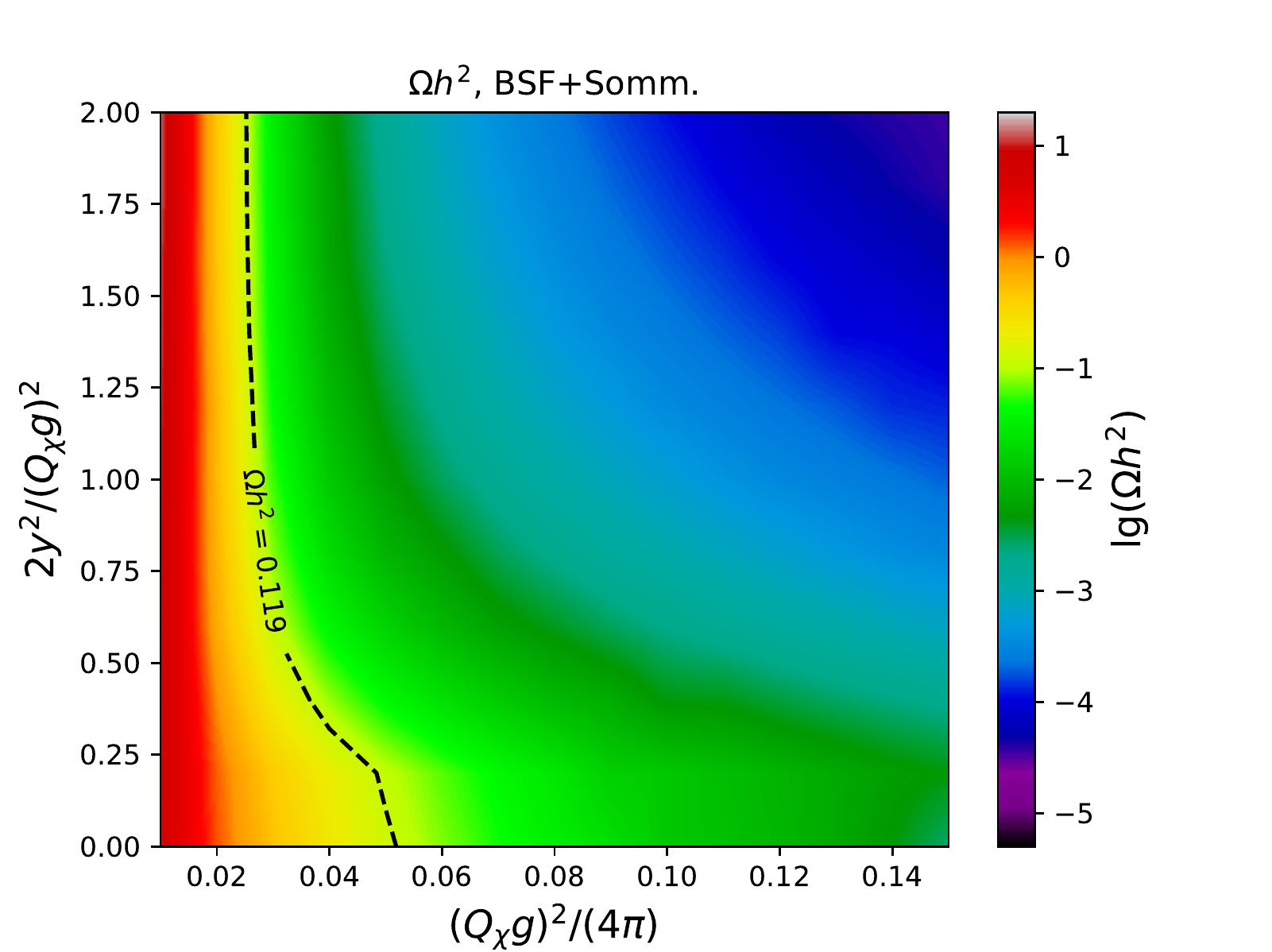}
\captionof{figure}{$m_\chi =10$ TeV. Other parameters and the meaning of the three pannels are the same as Fig.~\ref{Omegah2_1500}.} \label{Omegah2_10000}
\end{minipage}
\end{tabular}
\end{table}

We can easily see that the Sommerfeld effect can  modify the relic density significantly, while the bound state formation effects further alter the results. To see how this can happen, we plot the thermally averaged cross section times velocity and the evolution of the dark matter relic density in Figs.~\ref{SigmaVContribution} and \ref{OmegaEvolution}, respectively.

\begin{table}
\begin{tabular}{c|c}
\begin{minipage}[t]{0.45\textwidth}
\includegraphics[width=0.99\textwidth]{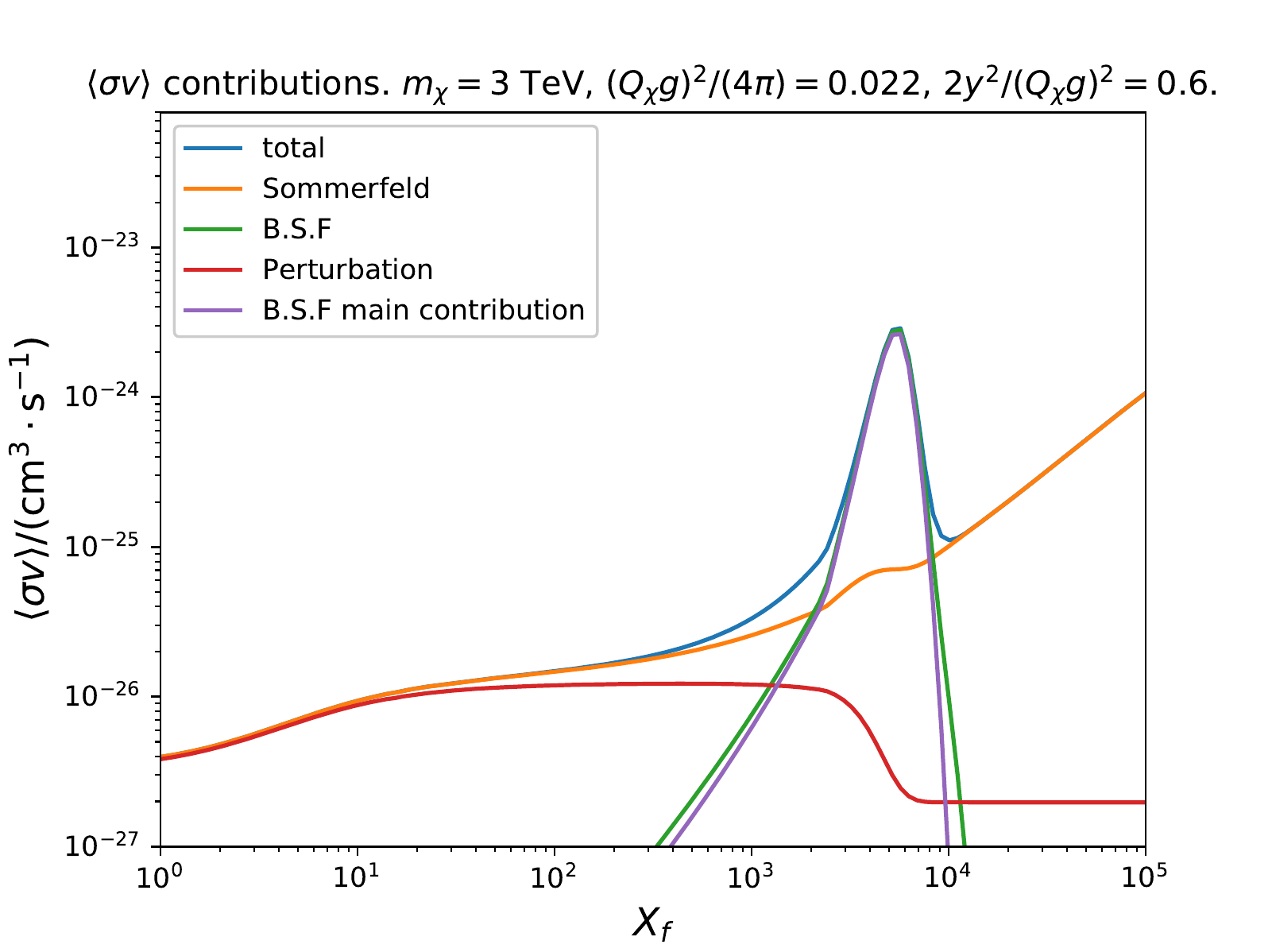}
\includegraphics[width=0.99\textwidth]{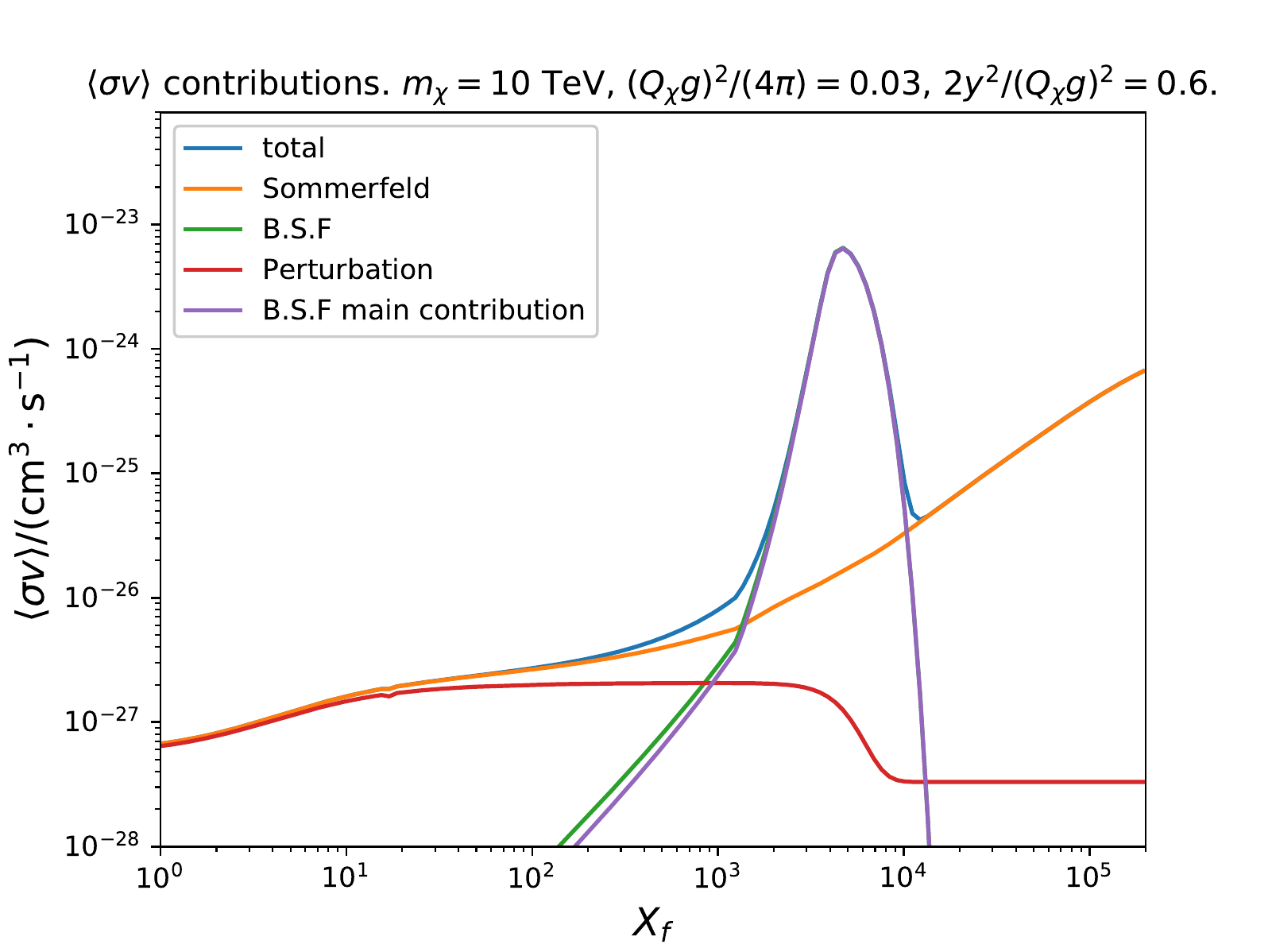}
\includegraphics[width=0.99\textwidth]{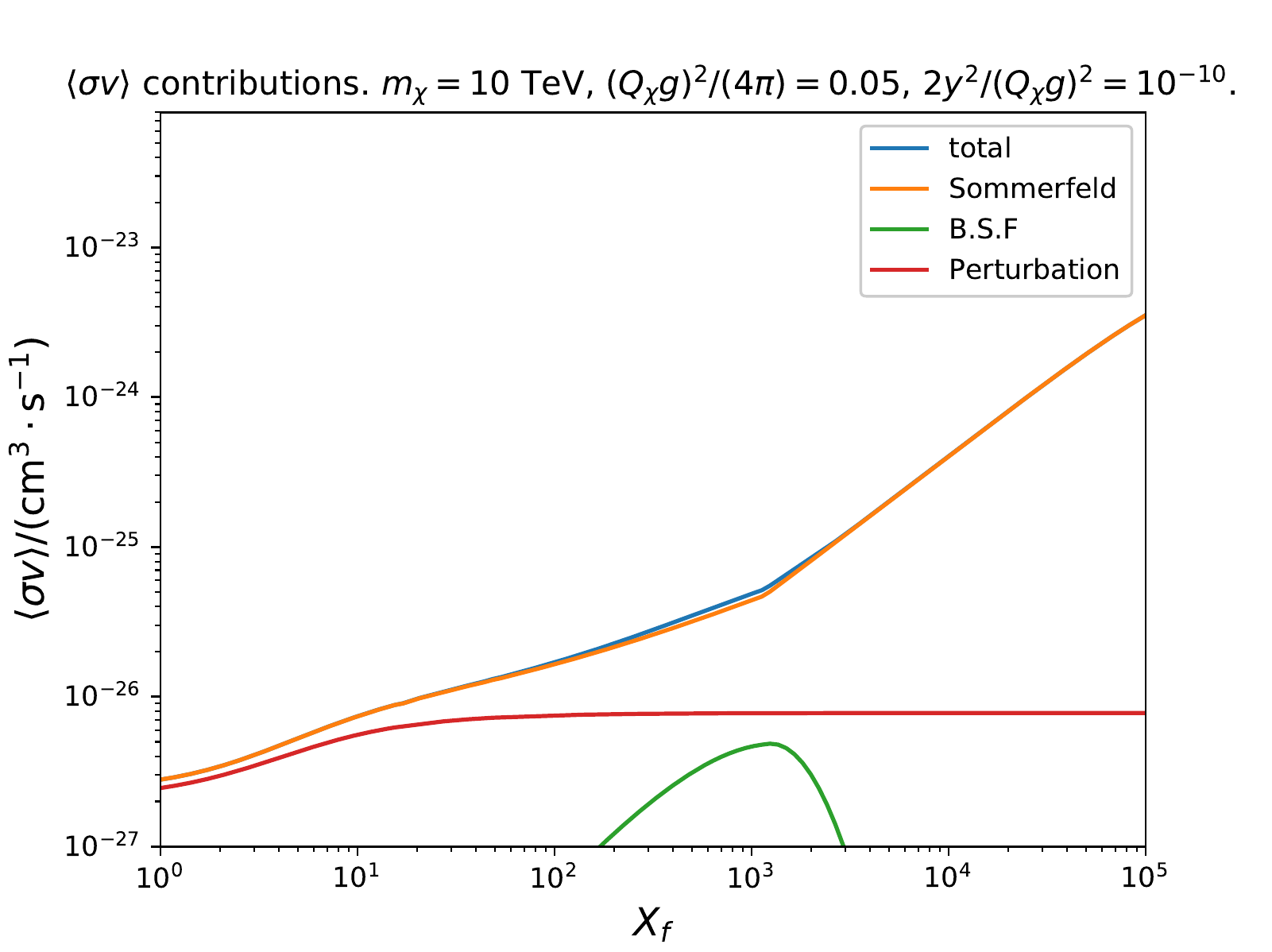}
\captionof{figure}{Contributions to the $\langle \sigma v \rangle$ as the temperature evolves. Other parameters not mentioned in the titles are the same as Fig.~\ref{Omegah2_1500}. 	``B.S.F main contribution'' indicates the $\chi_1 \chi_2 \rightarrow 1^1 S_0^{s +} + \gamma^{\prime}$ channel, which is the main contribution to the bound state formation channels. Parameters are chosen for $\Omega h^2$ is close to $0.11$.} \label{SigmaVContribution}
\end{minipage}
&
\begin{minipage}[t]{0.45\textwidth}
\includegraphics[width=0.99\textwidth]{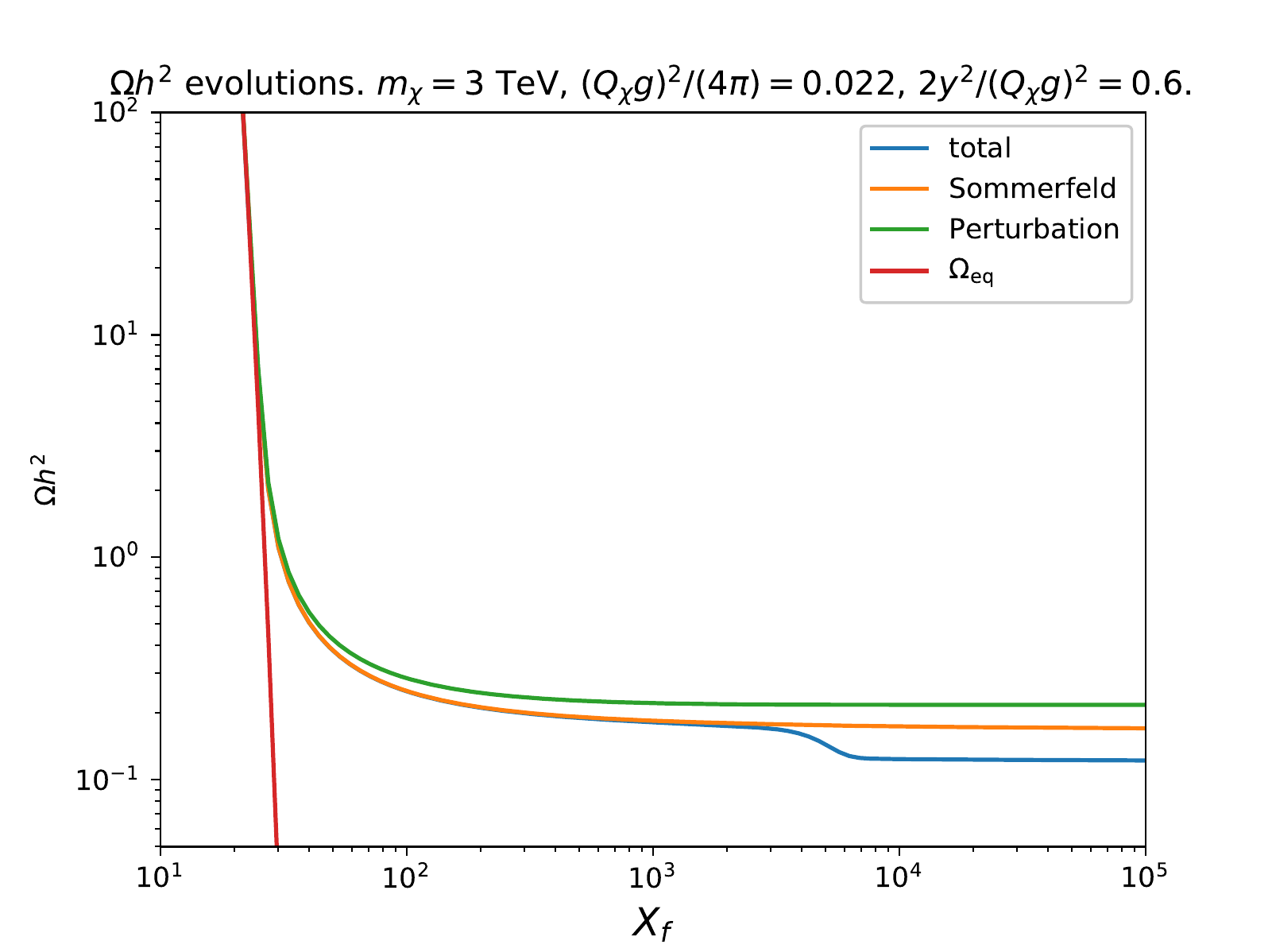}
\includegraphics[width=0.99\textwidth]{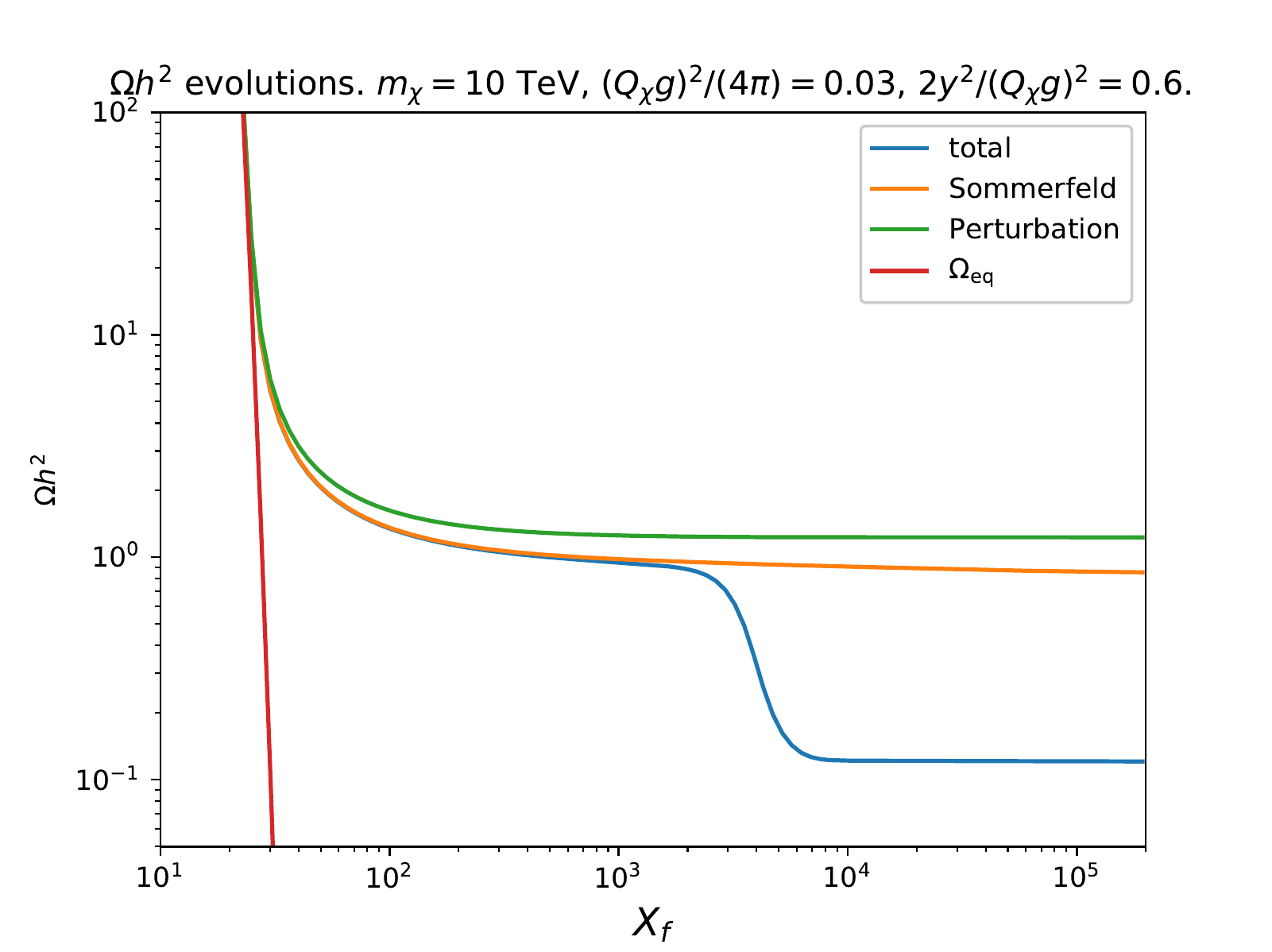}
\includegraphics[width=0.99\textwidth]{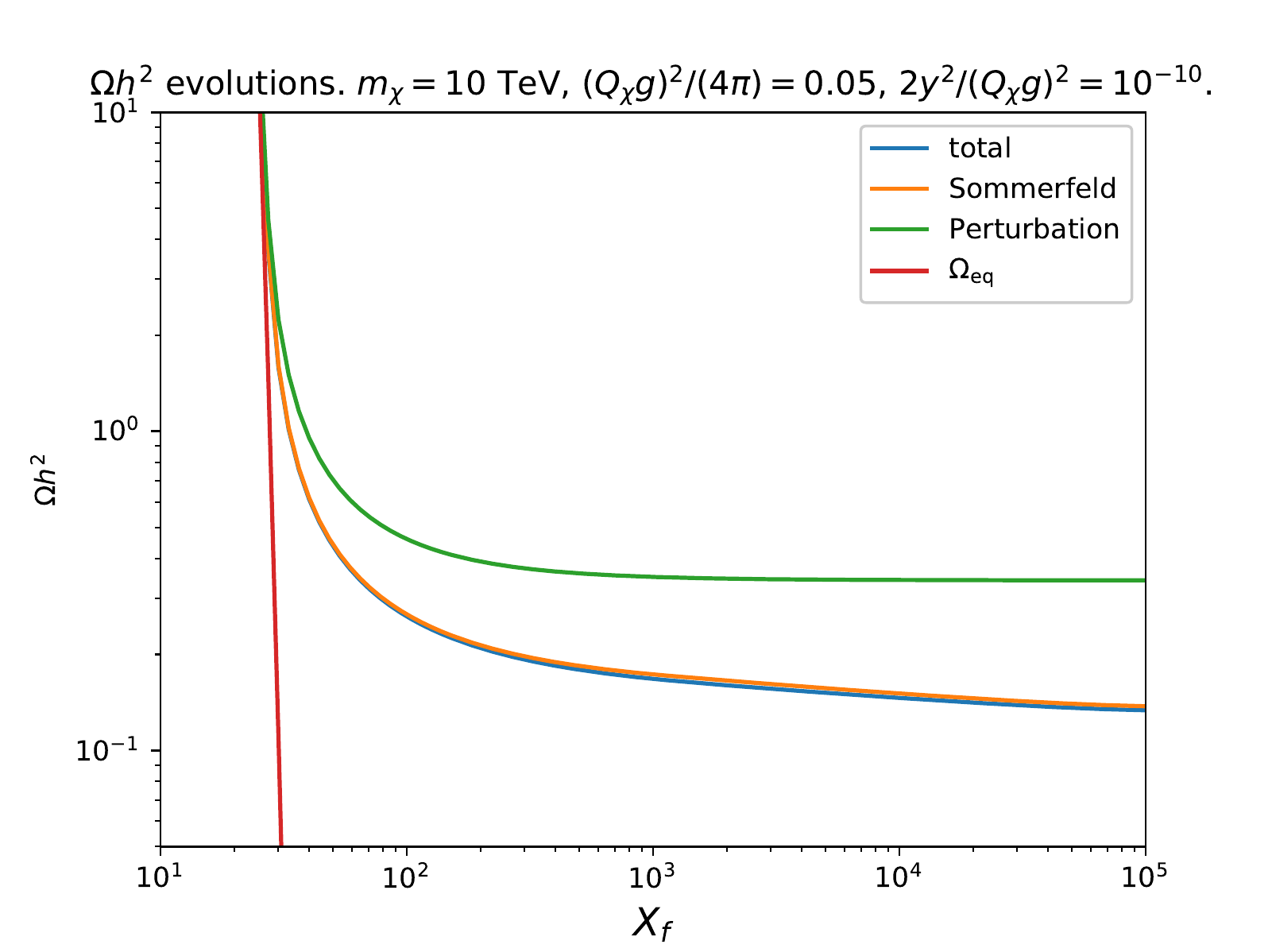}
\captionof{figure}{Evolutions of the $\Omega h^2$ as the temperature evolves. The parameters are the same as the corresponding panels in Fig.~\ref{SigmaVContribution}.} \label{OmegaEvolution}
\end{minipage}
\end{tabular}
\end{table}

In Figs.~\ref{SigmaVContribution} and \ref{OmegaEvolution}, we show and compare the contributions of the perturbative calculations, the calculations including  Sommerfeld effects, and including also the bound state formation effects. {In the top and middle panels, we adopt the $m_{\chi}=3 \rm{ TeV}$ and $m_{\chi}=10 \text{ TeV}$ respectively. $\frac{(Q_{\chi} g)^2}{4 \pi}$ and $\frac{2 y^2}{(Q_{\chi} g)^2}$ are shown in the figures. For the $m_{\chi}=3 \text{ TeV}$ case,  $\xi_1 = 34.5$, $\xi_2 = 51.7$, corresponding to $m_{R} = 0.957 \text{ GeV}$ and $m_{\gamma^{\prime}} = 0.638 \text{ GeV}$. For the $m_{\chi}=10 \text{ TeV}$ case, $\xi_1 = 25.3$, $\xi_2 = 37.9$, corresponding to $m_{R}=5.93 \text{ GeV}$ and $m_{\gamma^{\prime}} = 3.96 \text{ GeV}$. Both of these two benchmark points can be safe within the constraints, especially for the dark Higgs boson decay bound in the BBN constraint to be described in Sec.~\ref{Constraints}.} We can also learn that when $y$ takes some moderate value, the main contributions from the dark matter bound state formations mainly become effective in a later time. In the usual re-annihilation processes induced by Sommerfeld enhancement in the literature, e.g., in the Ref.~\cite{reannihilation1, reannihilation2, reannihilation3, reannihilation4, reannihilation5}, the re-annihilation can only happen after the ``kinematical decoupling'' of the dark matter particles, when the kinematical temperature of the dark matter particles accelerate to cool down due to the rapid kinematical energy reduction of the non-relativistic particle compared with radiations. Detailed calculations show that the annihilation cross section times velocity $\langle \sigma v \rangle_{\rm{eff}}$ of the dark matter should be at least proportional to $\frac{1}{v^t}$ where $t > 1$ for a successful re-annihilation process. This can only happen in some resonance area of the Sommerfeld enhancements. Compared with the usual re-annihilation process induced by Sommerfeld enhancement, in our model, the main contribution to enhance the effective total annihilation cross section times velocity is the bound state formation process though the $\chi_1 \chi_2 \rightarrow 1^1 S_0^{s +} + \gamma^{\prime}$ channel. This process becomes significant because of the large longitudinal $\gamma^{\prime}$ emissions, which is equivalent to the Goldstone emissions, and described by the (\ref{GoldStoneEmission}). This becomes large due to its non-zero 0th order expansions. The dark matter bound state formation cross section $\langle \sigma v \rangle_{\chi_1 \chi_2 \rightarrow 1^1 S_0^{s +} + \gamma^{\prime}} \sim \frac{1}{v^t}$, where $t \gg 1$ when $x_F \lesssim 10^{4}$, inducing a significant second-stage annihilation. Then, numerical calculation shows that when $v \lesssim 0.01$, $\langle \sigma v \rangle_{\chi_1 \chi_2 \rightarrow 1^1 S_0^{s +} + \gamma^{\prime}}$ rapidly saturate before the breaking the unitarity bound. Finally, after $T \ll \delta m$, we will have $\frac{n_{\chi_2}^{\text{eq}}}{n_{\chi_1}^{\text{eq}}} \sim e^{-\frac{\delta m}{T}} \ll 1$, so that the contributions from the $\chi_1 \chi_2$ channels will be suppressed according to the (\ref{EffectiveSigmaV}). Finally, the re-annihilation processes cease when {$x_F \gtrsim 10^5$}.

On the lowest pannels of the Figs.~\ref{SigmaVContribution} and \ref{OmegaEvolution}, we plot a benchmark point for the $y \ll 1$ situation. In this situation, the global symmetry on $\chi_{1,2}$ as well as the Ward identity $M_{\mu} P^{\mu}_{\phi}=0$ in the symmetric phase recovers, therefore the longitudinal contributions in the (\ref{M0Equation}) disappears. Compared with the longitudinal contributions, the emissions of the transverse $\gamma^{\prime}$ described by the (\ref{Jpsd}-\ref{Jmds}) will be suppressed by the $v^2$, so the re-annihilation process vanishes. However, a sufficiently small relic abundance of the dark matter requires a larger $\frac{(Q_{\chi} g^{\prime})^2}{4 \pi}$, inducing a much larger bound state formation rate in the freeze-out epoch. At the parameter point we selected in the lowest panels of Figs.~\ref{SigmaVContribution} and \ref{OmegaEvolution}, the bound state formation corrects the relic density result by several percent. If we assume that the $\chi_1$ particles only contribute to a fraction of the now observed dark matter density, the corrections on the relic density result can be much larger.

\subsection{Constraints from the Experiments} \label{Constraints}

In this paper, we have ignored the $\lambda_{\Phi H} |\Phi|^2 H^{\dagger} H$ and $\epsilon F^{\mu \nu} B_{\mu \nu}$ terms to simplify the relic density calculations. As have been mentioned, these are the only interactions that communicate between the dark and visible sectors. Practically, $\lambda_{\Phi H}$ receives loop corrections and is unable to be non-zero in all energy scales. A strict $\epsilon=0$ also induces a dark photon relic which is not the main topic to be described in this paper. For the non-zero $\lambda_{\Phi H}$ and $\epsilon$, the dark matter direct detection bounds constrain both of these two interactions. Higgs exotic decay data mainly confine $\lambda_{\Phi H}$, and $\epsilon$ is limited through various collider experiments. One should also be careful to avoid the big bang nucleosynthesis (BBN) to be disturbed by the long-life dark Higgs and dark photon boson decay. Although the details are beyond the discussions of this paper, we will briefly go through all these constraints to show the possibility to evade all of them.

According to Ref.~\cite{ADMWise}, the spin-independent (SI) cross section for a fermionic dark matter particle scattering with a nucleon is given by
\begin{eqnarray}
\sigma_{\text{SI}} \simeq \frac{y^2 f^2 m_N^4 \mu_{\phi h}^2}{\pi m_h^4 m_{\phi}^4},
\end{eqnarray}
where $m_h$ is the SM Higgs boson mass, $m_N$ is the mass of a nucleon, and $f \simeq 0.35$. $\mu_{\phi h}$ is the mixing parameter between a SM Higgs boson and a dark Higgs boson. In our model, it is given by
\begin{eqnarray}
\mu_{\phi h} = \lambda_{\Phi H} v_{\Phi}.
\end{eqnarray}
If, e.g., we would like $\sigma_{\text{SI}} \lesssim 10^{-45} \text{cm}^2$, this requires $y^2 \mu^2_{\phi h} \lesssim 2 \times 10^{-10} \text{GeV}^2$ if $m_{\phi} \sim 0.3 \text{GeV}$. Then, $\lambda_{\Phi H} v_{\Phi} \lesssim 1.5 \times 10^{-5} \text{ GeV}$. If we would like the dark photon mass to be $m_{\gamma^{\prime}} \sim 0.5 \text{ GeV}$, this implies $v_{\Phi} \sim 0.5 \text{ GeV}$, therefore $\lambda_{\Phi H} \lesssim 10^{-5} \sim 10^{-6}$.

Such a constraint is much more stringent than the Higgs exotic decay results. The sum of $h \rightarrow R R$ and $h \rightarrow \gamma^{\prime} \gamma^{\prime}$ partial width is
\begin{eqnarray}
\Gamma_{h \rightarrow R R} + \Gamma_{h \rightarrow \gamma^{\prime} \gamma^{\prime}} &=& \frac{(\lambda_{\Phi H} v_{\rm{EW}})^2}{32 \pi m_h} + \frac{2 Q_{\Phi}^4 g^4 v_{\Phi}^2}{16 \pi m_h} \left[ 2 + \frac{(m_h^2-2 m_{\gamma^{\prime}}^2)^2}{4 m_{\gamma^{\prime}}^4} \right] \sin^2 \theta_{R h}, \nonumber \\
&\simeq& \frac{(\lambda_{\Phi H} v_{\rm{EW}})^2}{16 \pi m_h}, \label{hExoticDecay}
\end{eqnarray}
where $v_{\rm{EW}}$ is the electroweak vev. In the above we have applied the $m_{\gamma^{\prime}} = |Q_{\Phi}| g v_{\Phi}$, and the mixing angle between the SM Higgs and dark Higgs boson $\sin \theta_{R h} \simeq \frac{\lambda_{\Phi H} v_{\Phi} v_{\text{EW}}}{m_h^2}$. Higher order $\frac{v_{\Phi}^2}{m_h^2}$ terms had also been neglected. It is easy to realize that the main contributions to the (\ref{hExoticDecay}) are $h \rightarrow R R$ and $h \rightarrow \gamma^{\prime}_{\text{L}} \gamma^{\prime}_{\text{L}}$, and can be analysed by $h \rightarrow R R/I I$, or be combined into $h \rightarrow \Phi \Phi^*$ channels with the Goldstone equivalence theorem. We just estimate the bounds $\Gamma_{h \rightarrow R R} +\Gamma_{h \rightarrow \gamma^{\prime} \gamma^{\prime} }  \lesssim \Gamma_{h_{\text{SM}}} = 4.07 \text{ MeV}$ (For the standard model Higgs widths, one can see the reviews in Ref.~\cite{PDG, CernHiggs1, CernHiggs2, CernHiggs3}, which requires $\lambda_{\Phi H} \lesssim 2 \times 10^{-2}$. Therefore, the direct detection bounds on $\lambda_{\Phi H}$ are far beyond the sensitivity of the collider searches for Higgs exotic decay channels.

The coupling $\lambda_{\Phi}$ is mainly constrained by the unitarity bound, or somehow equivalently,  perturbative unitarity. A rigorous discussion would involve the complete calculations of the S-wave amplitudes  ($a_0$'s) for the channels, 
$R R \leftrightarrow R R$, $R R \leftrightarrow \gamma^{\prime}_L \gamma^{\prime}_L$, $\gamma^{\prime}_L \gamma^{\prime}_L \leftrightarrow \gamma^{\prime}_L \gamma^{\prime}_L$, $R \gamma^{\prime}_L \leftrightarrow R \gamma^{\prime}_L$. The unitarity bounds are extracted from $|a_0|^2 < 1$\cite{Unitarity1, Unitarity2, Unitarity3}, or $\rm{Re}(a_0) < \frac{1}{2}$\cite{TightUnitarity1, TightUnitarity2}. In this paper, we derive this bound in the limit of large center of mass frame energy. In such a case, the $\gamma^{\prime}_L$ can be directly replaced by its corresponding Goldstone boson $I$ due to the Goldstone equivalence theorem. Furthermore, the dark gauge symmetry can also be regarded as have been recovered in this limit, so we can directly calculate the amplitudes for 
the following processes, 
\[
\frac{1}{\sqrt{2}} \Phi \Phi \rightarrow \frac{1}{\sqrt{2}}\Phi \Phi , ~~
\frac{1}{\sqrt{2}}\Phi^* \Phi^* \rightarrow \frac{1}{\sqrt{2}}\Phi^* \Phi^* , ~~
\Phi^* \Phi \rightarrow \Phi^* \Phi , 
\] 
for simplicity. Then we can obtain the unitarity bounds on $\lambda_\Phi$: $\lambda_{\Phi} < 4 \pi$ from $|a_0|^2 < 1$, or $\lambda_{\Phi} < 2 \pi$ from ${\rm Re}(a_0) < \frac{1}{2}$.

From (\ref{vevAcquired}) and (\ref{BosonMasses}),  we can easily derive that
\begin{eqnarray}
\lambda_{\Phi} = \left( \frac{\xi_2}{\xi_1} \right)^2 \frac{Q_{\Phi}^2 g^{\prime 2}}{2} = \left( \frac{\xi_2}{\xi_1} \right)^2 2 \pi \alpha^{\prime} \frac{Q_{\Phi}^2}{Q_{\chi}^2}.
\end{eqnarray}
Here, $\alpha^{\prime} = \frac{Q_{\chi}^2 g^{\prime 2}}{4 \pi}$ as we have suggested in subsection \ref{U1Break2Z2}. Therefore, the adopted $\xi_2 = 1.5 \xi_1$ values in Figs.~\ref{Omegah2_1500}  and \ref{Omegah2_10000} can induce $\lambda_{\Phi} = 18 \pi \alpha^{\prime} < 4 \pi$, which in turn results in  $\alpha^{\prime} < 0.22$ for the $|a_0|^2<1$ bound, and $\lambda_{\Phi} = 18 \pi \alpha^{\prime} < 2 \pi \rightarrow \alpha^{\prime} < 0.11$ for the ${\rm Re}(a_0)<\frac{1}{2}$ bound. However, this bound changes a lot as the $\frac{\xi_2}{\xi_1}$ varies in the range $1.2$-$2$. Our calculations show that at least the relic density results in the lowest panels in Fig.~\ref{Omegah2_1500}- \ref{Omegah2_10000} do not change a lot within this range when $m_R$ is set to be slightly larger than the threshold energy of $\chi_1 \chi_1 \rightarrow X + R$. Therefore, in order not to mislead the readers as if it is a universal bound on $\frac{(Q_{\chi} g)^2}{4 \pi}$ for all of the $\frac{\xi_2}{\xi_1}$ values, we do not explicitly show the unitarity bounds on Fig.~\ref{Omegah2_1500}-\ref{Omegah2_10000}.

The BBN constrains the lifetime of the $R$ particle. \footnote{One can also consult the axion(-like) particle results in Ref.~\cite{DarkPhotonConfinement2} for a careful constraints on light scalar particles. However, this is beyond our current scope.} Roughly speaking, if $\tau_{R} \lesssim 1 \text{ sec}$, the cosmological $R$ particles disappear before the BBN epoch. The non-hadronic partial width of the $R$ decay width is calculated to be \cite{ADMWise}
\begin{eqnarray}
\Gamma_{R \rightarrow 2(e/\mu/\gamma)} &=& \left\lbrace \frac{G_F m_e^2 m_{R}}{4 \sqrt{2} \pi} \left( 1-\frac{4 m_e^2}{m_{R}^2} \right)^{\frac{3}{2}} \Theta(m_{\phi} - 2 m_{e}) +  \frac{ G_F m_{\mu}^2 m_{R}}{4 \sqrt{2} \pi} \left( 1-\frac{4 m_{\mu}^2}{m_{R}^2} \right)^{\frac{3}{2}} \Theta(m_{R} - 2 m_{e}) \right. \nonumber \\
 &+& \left. \frac{G_F \alpha^2 m_{R}^3}{128 \sqrt{2} \pi^3} \left[ A_1 + 3 A_{1/2}  \right]^2  \right\rbrace \frac{\mu_{\phi h}^2 v^2}{m_h^4}.
\end{eqnarray}
The previous upper bound we had acquired $y^2 \mu_{\phi h}^2 \lesssim 2 \times 10^{-10} \text{GeV}^2$ will result in $\tau_{\phi} \gtrsim 0.02 \text{ sec}$, for $y \lesssim 1$ and $m_{\phi} = 0.3 \text{ GeV}$ which is above the dimuon threshold. This is enough for the dark Higgs bosons to decay away before the beginning of the BBN.

Compared with the dark Higgs boson $R$ with the width suppressed by the Yukawa couplings of the electrons and muons, the couplings between the dark photon $\gamma^{\prime}$ and the charged leptons are universal. Currently, for the $m_{\phi} > 2 m_{\mu} \approx 0.2 \text{ GeV}$, the lifetime of the dark photon is expressed as\cite{DarkPhotonConfinement1}
\begin{eqnarray}
\tau_{\gamma^{\prime}} \simeq \frac{3}{\epsilon^2 \alpha} = 6 \times 10^5 \text{ yr} \times \frac{10 \text{ MeV}}{m_{\gamma^{\prime}}} \times \frac{10^{-35}}{\epsilon^2 \alpha}.
\end{eqnarray}
A rather tight constraint on $\epsilon < 10^{-10}$ implies a safe $\tau_{\gamma^{\prime}} \gtrsim 0.1 \text{ sec}$, which can still be much smaller than the beginning period of the BBN. The results in Ref.~\cite{DarkPhotonConfinement1, DarkPhotonConfinement2} show that the $\epsilon$ can be larger than $10^{-7}$ or $10^{-5}$, further reducing the $\tau_{\gamma^{\prime}}$ to avoid the dark photon decay after the beginning of BBN.

The non-zero $\epsilon$ can also give rise to the direct detection signals. Since the photon couplings with the $\chi_{1,2}$ induced from the $\gamma^{\prime}$-$\gamma$ mixing always flip the $\chi_{1}$-$\chi_{2}$, the dark matter scattering with the nucleon becomes inelastic. We rely on the dark photon results in the Ref.~\cite{InelasticPhoton} to show that most of the parameter space discussed in this paper is safe with the indirect detection bound, although Ref.~\cite{InelasticExp1, InelasticExp2, InelasticExp3} show more recent general experimental data. Since it is difficult to follow the detailed numerical processes in Ref.~\cite{InelasticPhoton}, we only compare the results qualitatively. In the Fig.~\ref{Omegah2_1500}-\ref{Omegah2_10000}, the correct dark matter's relic density requires $\frac{(Q_{\chi} g)^2}{4 \pi} \gtrsim 0.02$, so $m_{\gamma^{\prime}} = \frac{\mu_r \alpha^{\prime}}{\xi_2} \gtrsim 0.3 \rm{ GeV}$. For most region of our interested parameter space, say, when $\frac{2 y^2}{(Q_{\chi} g)^2} > 0.01$, $m_{\chi_2}-m_{\chi_1} = 2 y v_{\Phi}$ is calculated to be well above 1 MeV. From Fig.~5 of Ref.~\cite{InelasticPhoton}, this is much larger than the analysed $E_R \sim 1$-$1000$ keV, so the sensitivity of the detector on such a dark matter is significantly suppressed. From the tendency of two panels in the Fig.~8 when the $\delta m$ varies, we can also expect to release the dark matter direct detection bound when $\delta m \gg 1$ MeV, exposing scarcely constraint compared with other non-DM constraints.

The lowest panels in Fig.~\ref{SigmaVContribution}-\ref{OmegaEvolution} correspond to $y \ll 1$ when $m_{\chi_2} - m_{\chi_2} \ll 1$ MeV. Such a small mass difference is expected to be excluded according to the Ref.~\cite{InelasticPhoton}. However, these radical parameters are calculated only for a comparison of the contributions from the dark matter bound state formation processes. A larger $y \sim 0.01$ can still leave the main results intact, while amplify $m_{\chi_2} - m_{\chi_2} \propto y v_{\Phi}$ to suppress the direct detection signals.

Ref.~\cite{InelasticPhoton, DarkPhotonLoop1, DarkPhotonLoop2} introduced the loop-level dark photon scattering cross sections. A rigorous study should include the interference between the tree-level and loop-level amplitudes, which is beyond our scope. The loop induced scattering cross section contributions are much less sensitive on the $m_{\chi_2}-m_{\chi_1}$ values, and might be the main contribution when $m_{\chi_2}-m_{\chi_1} \gg 1$ MeV. Here, we adopt the loop-level formula in Ref.~\cite{DarkPhotonLoop1, DarkPhotonLoop2}
\begin{eqnarray}
\sigma_{\chi p \rightarrow \chi p} = \frac{a_X^2}{2 \pi} \frac{m_{\chi}^2 m_p^2}{(m_{\chi} + m_p)^2},
\end{eqnarray}
where
\begin{eqnarray}
a_X = \frac{\epsilon e Q_{\chi} g}{m_{\chi}^2}.
\end{eqnarray}
$\epsilon \lesssim 10^{-3}$ and $m_{\chi} \gtrsim 100 \text{ GeV}$ are sufficient to reduce $\sigma_{\chi p \rightarrow \chi p}  \lesssim 10^{-46} \text{cm}^2$. Obviously, our choices of the parameters can easily evade the direct detection bounds. 

Finally, the strongly interacting dark matter (SIDM) models are robustly constrained by the CMB spectrum \cite{CMBBound}. Since our model includes quite light mediators $R$ and $\gamma^{\prime}$, and our dark matter-mediator interactions are expected to be significant enough to induce the sufficient dark matter bound state formation rates, our model can be regarded as a SIDM. The enhancement of the dark matter annihilation cross section in a SIDM in the very late epoch may disturb the CMB observations,  if the mediator particles decay into the SM particles faster than the Hubble rate.
Then at the recombination epoch,
\begin{eqnarray}
\frac{\langle \sigma v \rangle_{\text{rec}}}{N_{\chi}} \lesssim 4 \times 10^{-25} \text{cm}^3 \text{s}^{-1} \left( \frac{f_\text{eff}}{0.1} \right)^{-1} \left( \frac{m_{\chi}}{100 \text{ GeV}} \right), \label{CMBBound}
\end{eqnarray}
where $\langle \sigma v \rangle_{\text{rec}}$ is the annihilation rate at recombination epoch, $N_{\chi}=1(2)$ for Majorana (Dirac) dark matter. Generally, $f_{\text{eff}} \gtrsim 0.1$ for the SM particles in the final states other than neutrinos. Generally, such a constraint is very difficult to escape. Ref.~\cite{Selection} provides a model that is similar with ours. In their model, there is no dark photon, so the $\chi_1 \chi_1$ annihilation is controlled by the p-wave channels to the $RR$ or $II$. In our model, the mass difference between $\chi_1$ and $\chi_2$ is sufficient for the $\chi_2$ to nearly disappear in the recombination epoch, therefore only the s-wave $\chi_1 \chi_1 \rightarrow \gamma^{\prime} \gamma^{\prime}$ might affect the CMB \cite{CMBMatsuiSuggest}. The bound state formation processes $\chi_1 \chi_1 \rightarrow X+\gamma^{\prime}/R$ are also set to be prohibited by the insufficient threshold energy in the parameter space we had studied in this paper. This is Sommerfeld enhanced and might contribute to the $\langle \sigma v \rangle_{\text{rec}}$. Fortunately, from (\ref{Potential12}) we can learn that the Sommerfeld boost factor is mainly controlled by the Yukawa coupling terms. Firstly, we are interested in the $y < Q_{\chi} g$ region, and a sufficiently small but moderate $y$ is enough for us to acquire a significant bound state formation effect while small enough  Sommerfeld boost factor. Secondly, $m_{R} = \sqrt{2} \mu$ is independent of all the other parameters, and can be adjusted to reduce the saturated Sommerfeld factor. All of these can help us easily evade the bound in (\ref{CMBBound}).

Finally, we discuss about the cluster constraints on dark matter self-interactions. A complete calculation involves the full solution to the Schr\"{o}dinger equation. Practically, it is done by a calculation up to $l \lesssim 20$ in the partial wave expansions. In the parameter space adopted in this paper, the kinetic energy for a dark matter particle inside a halo is far below the threshold to produce  $\chi_2$. In this situation, only the $R$-mediated force takes place, and for the correct dark matter relic density, the Yukawa coupling strength should be weaker than the usual values in the literature. From Ref.~\cite{Cluster0}, we can see that it is quite easy to evade such bounds,
\[
\frac{\sigma_\text{T}}{m_{\chi}} \lesssim 1 \text{cm}^2 \text{g}^{-1}
\] 
in the usual case, let alone our smaller Yukawa coupling situations.

\subsection{Comparison with Other Similar Situations in the Literature }

Before closing, let us compare our results based on the model Lagrangian (2.1) with those in other approaches in the literature. We shall emphasize that the model (2.1) is based on the spontaneously breaking of a local dark gauge symmetry, respecting the unitarity of the amplitudes involving massive dark photons through dark Higgs contributions. Some earlier studies ignored the contributions from the longitudinal dark photon.

In the literature, most studies of excited fermionic DM are based on the model without dark Higgs, based on the following model Lagrangian :
\begin{eqnarray}
\mathcal{L}_{\rm w/o ~dark ~Higgs} = -\frac{1}{4} F^{\prime \mu \nu} F^{\prime}_{\mu \nu} 
+ \frac{1}{2} m_{A'}^2 A_\mu^{'} A^{'\mu} 
+ \overline{\chi} D\!\!\!\!/ \chi - m_{\chi} \overline{\chi} \chi 
+ (\delta  \  \overline{\chi^C} \chi + \text{h.c.})  . \label{BasicLag_simp}
\end{eqnarray}

In this case dark photon is assumed  to get its mass by St\"{u}ckelberg mechanism, and the mass splitting ($\delta$) between $\chi_1$ and $\chi_2$  is simply generated by the soft $U(1)$-breaking dim-3 operator put in by hand, and the dark $U(1)$ symmetry is only approximate and explicitly broken in (\ref{BasicLag_simp}). On the other hand, in our case the original local dark  $U(1)$ gauge symmetry is broken down to its $Z_2$ subgroup 
by dark Higgs mechanism. And  the above Lagrangian (\ref{BasicLag_simp}) can be obtained from (2.1) by integrating out the physical dark Higgs boson after the dark $U(1)$  symmetry breaking.  Our model is based on spontaneous breaking of local dark gauge symmetry, and thereby keep all the nice features of gauge theories such as unitarity and renormalizability. 
 
Ref.~\cite{SoftBreaking1, SoftBreaking2} made attempts on solutions for some line-shaped photon excesses observed from near the galactic center. Both these papers are based upon one single model, in which the dark Higgs sector is eliminated compared with our model. There the mass difference of the nearly-degenerate dark matter particles is introduced with a soft breaking term, which breaks the dark $U(1)$ symmetry explicitly. Interestingly, Ref.~\cite{SoftBreaking2} utilized the annihilation process $\chi \chi \rightarrow \gamma^{'} \gamma^{'}$ followed by $\gamma^{'} \rightarrow \text{SM}$ to explain the continuous gamma-ray spectrum from the galactic center, but detailed calculations were not proceeded. We performed such calculations in Appendix \ref{UnitarityDetails} and found that the contributions from the longitudinal dark photon can become catastrophic. This is similar to the case of the unitarity violation of Higgsless case in the SM in the longitudinal $W$ scattering. The simplest way to cure this problem is to introduce a dark Higgs boson $\Phi$, and generate the dark photon mass.  And the mass difference of the dark matter particles originates from the $\Phi \overline{\chi^C} \chi + \text{h.c.}$ term as in our model, (2.1). In this situation, in addition to the purely dark photon annihilation channels, $R \gamma^{\prime}$, $R R$ final states also arise. From Tab.~\ref{TwoBodyM2}, we can learn that the $R \gamma^{\prime}_L$, in which $\gamma^{\prime}_L$ indicates the longitudinal polarization of a dark photon, or equivalently $R I$ channel contribute to an additional s-wave annihilation term, which can be significant compared with the $\gamma^{\prime}_T \gamma^{\prime}_T$ process, in which $\gamma^{\prime}_T$ indicates a transverse polarization of a dark photon.

Ref.~\cite{SoftBreaking3} provides another approach to the excited fermion DM scenario. 
There the dark matter mass difference originate from the vacuum expectation value of the dark Higgs boson, however the dark matter and dark Higgs interaction term is non-renormalizable:
\begin{eqnarray}
{\cal L} \supset - \frac{y}{\Lambda} (\phi^\dagger \phi^\dagger \overline{\chi^C} \chi + H.c. ) \label{NonRenormalizable}
\end{eqnarray}
which induces the mass splitting $\delta = y v_\phi^2 / \Lambda$ after $\phi$ develops a nonzero VEV $v_\phi$. \footnote{$\phi$ in (5.11) is not the same as $\Phi$ in Eq. (2.1), since their $U(1)$ charge assignmets are different from ours.}  
There the renormalizable dim-4 operator (the last term in Eq. (2.1)) was ignored without any reasons.
In fact, the coupling described by (\ref{NonRenormalizable}) requires $m_{\gamma^{\prime}} \ll \Lambda$ for the effective ``Yukawa coupling constant'' $\frac{2 y v_{\phi}}{\Lambda}<1$, as well as a cross section well below the unitary bound. The amplitude is calculated to be $\propto \frac{1}{\Lambda^2} \propto \frac{g^2 \delta m}{m_{\gamma^{\prime}}^2}$. This is somehow similar to the results without a Higgs cancellation in Appendix \ref{UnitarityDetails}. In fact, the non-renormalizable model introduced by Ref.~\cite{SoftBreaking3} still has a diagram similar to the third one with a s-channel Higgs boson listed in Fig.~\ref{Annihilations}, however the $h$-$\gamma^{\prime}$-$\gamma^{\prime}$ coupling constant is only one half compared with our model. Therefore, a precise cancellation does not occur, so the $\frac{g^2 \delta m}{m_{\gamma^{\prime}}^2}$ term remains, and the resulting amplitude is not unitary at high 
energy limit, unlike our case.

In the Ref.~\cite{SoftBreaking1, SoftBreaking2, SoftBreaking3, DMBS32}, compared with our paper, the longitudinal contribution to the off-diagonal element of the potential matrix which is proportional to $c_1$ in our (\ref{RadialPotentialS}) is absent. This is a good approximation if $\delta m \ll m_{\gamma^{\prime}}$. However, in Ref.~\cite{SoftBreaking1, SoftBreaking2}, the longitudinal contribution can be absorbed by redefining the $\alpha_D$, with all the phenomenologies unchanged.

Finally, we need to mention that although Ref.~\cite{Petraki1, Petraki2, DMBS28} had classified nearly all the emission product cases during the dark matter bound state formation, the longitudinal dark photon emission is not included there. In their models involving a massive dark vector boson, the Ward-Takahashi identity was utilized, e.g., in the Eqn.~(3.3) of Ref.~\cite{Petraki2}. Therefore, only transverse polarization dark photon emission was concerned. In our paper, we have applied a general version of the Ward-Takahashi identity given by (\ref{WDIdentity}). Therefore, the longitudinal polarization of a dark photon is considered through an equivalent Goldstone boson emission term. This term can be neglected in the Ref.~\cite{DMBS32}, once the mass difference of the dark matter is much smaller than the dark photon mass.

\section{Conclusions  and Future Prospect}

In this paper, we considered $Z_2$ fermion dark matter model with a pair of nearly-degenerate 
fermions defined by Lagrangian (2.1). The Yukawa potential induced by the real scalar $R$ 
between the same component particle pairs $\chi_1 \chi_1$ and $\chi_2 \chi_2$ are attractive, 
while it is repulsive between different component pair $\chi_1 \chi_2$. 
The longitudinal vector boson, or equivalently, Goldstone boson, also contributs to an 
additional off-diagonal term in the potential matrix.

In our model, besides the emission of the scalar $R$ and the usual transverse vector boson 
$\gamma^{\prime}$ to form a dark matter bound state, emission of a longitudinal dark photon 
$\gamma^{\prime}$, or somehow equivalently a Goldstone boson $I$ also arises because of our 
dark charge assignment of the dark matter particle. The mass difference between the two components plays a crucial role in this process. Unlike Ref.~\cite{Petraki1, Petraki2}, the zeroth ``mono-pole'' contributions to the (\ref{Is}-\ref{Idsm}) in our paper are non-zero, because we either need to replace the direct inner product by something inserted with a $\sigma_{1,3}$, or need to compute the inner products of the wave functions acquired under different potentials. Finally, we find that the contribution from the longitudinal vector boson emission is extremely important. This leads to a re-annihilation process, reducing the relic abundance of the dark matter significantly. 
Because this reannihilation process ceased before the BBN, the following cosmological parameters remain undisturbed \footnote{We choose benchmark parameters such that $\Delta m \gg T_{\rm BBN}$ and $\Gamma_{\chi_2} \gg H_{\rm BBN}$. Therefore we can safely ignore 
$\chi_1 + \chi_2 \leftrightarrow X + \gamma^{'}$. And $\chi_1 + \chi_1 \rightarrow X + R$ is set to be kinematically forbidden.}.

Constrained by our current computing resources and programming techniques, we are only able to calculate a simple approximation (\ref{Boltzmann}) rather than a complete version of Boltzmann equation, like those appeared in Ref.~\cite{reannihilation5}. In the future, we are planning to make more complete calculations by considering various effects which were not included in this paper. For example, the deviation of the velocity distributions of each component from the Maxwell-Boltzmann distributions should also be taken into account. A careful scanning of the parameter space considering all the experimental constraints and the properties including the SIDM scenarios will also be considered.

\appendix
\section{A Dark Matter Annihilation Problem in the Higgsless Soft-Breaking Model} \label{UnitarityDetails}

If we get rid of the Higgs sector, and introduce the soft mass terms for the dark photon and the mass differences between $\chi_1$ and $\chi_2$, we can still calculate the Higgsless diagrams of the dark matter pair $\chi_1 \chi_1 \rightarrow \gamma^{\prime}_{L} \gamma^{\prime}_{L}$. The polarization of the longitudinal $\gamma^{\prime}$ can be written as
\begin{eqnarray}
\epsilon_L(k) = \left( \frac{|\vec{k}|}{m_{\gamma^\prime}},~\frac{E}{m_{\gamma^{\prime}}} \frac{\vec{k}}{|\vec{k}|}\right) \xrightarrow{E \gg m_{\gamma}^{\prime}} \frac{k}{m_{\gamma^{\prime}}} + O\left( \frac{m_{\gamma^{\prime}}}{E} \right),
\end{eqnarray}
where $k=(E,~\vec{k})$ is the 4-momentum of the dark photon. For simplicity, we replace $\epsilon_L(k)$ with $\frac{k}{m_{\gamma^{\prime}}}$, and omit the sub-dominant $O\left( \frac{m_{\gamma^{\prime}}}{E} \right)$ terms.
\begin{figure}
\includegraphics[width=0.32\textwidth]{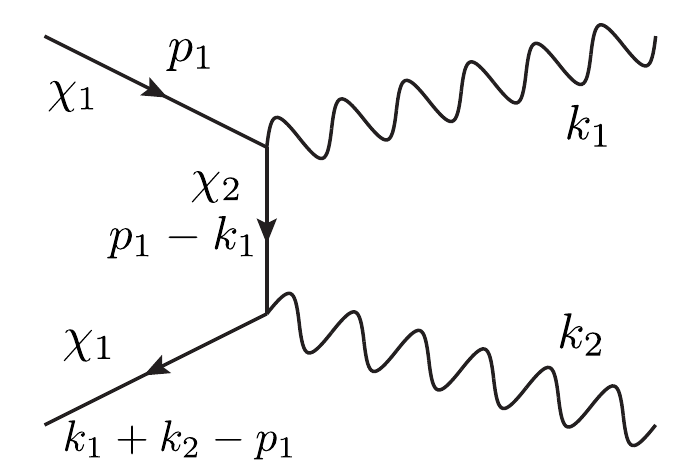}
\includegraphics[width=0.32\textwidth]{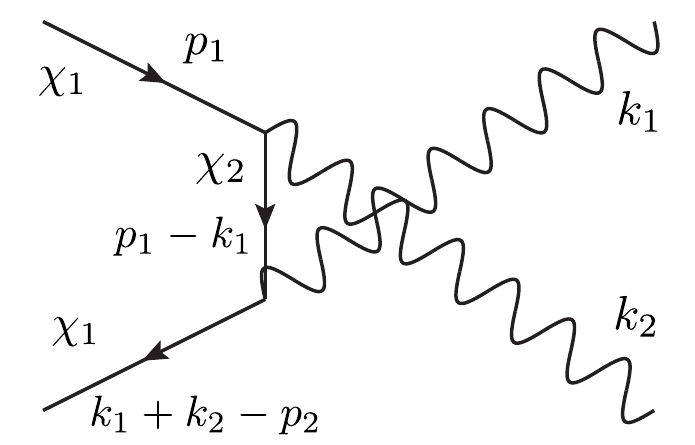}
\includegraphics[width=0.32\textwidth]{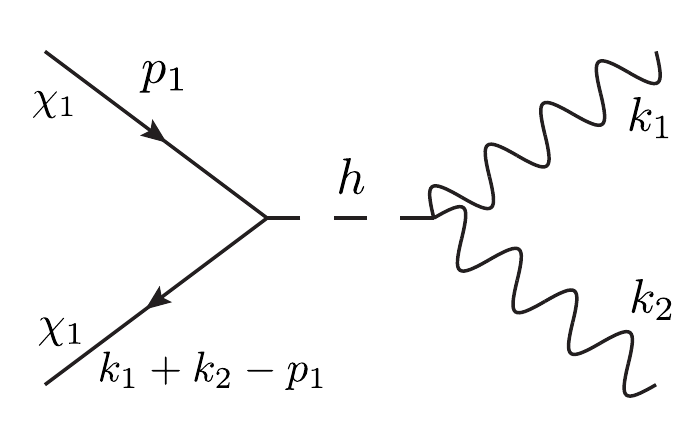}
\caption{$\chi_1 \chi_1 \rightarrow \gamma^{\prime} \gamma^{\prime}$ diagrams. Compared with the Higgsless soft-breaking model, the third diagram arises in our model} \label{Annihilations}
\end{figure}

We then compute the first and second diagrams shown in (\ref{Annihilations}). The $\chi_1 \chi_1 \rightarrow \gamma^{\prime}_L \gamma^{\prime}_L$ amplitude is given by
\begin{eqnarray}
\mathcal{M}_{\chi_1 \chi_1 \rightarrow \gamma^{\prime}_L \gamma^{\prime}_L} &=& \frac{Q_{\chi}^2 g^2}{m_{\gamma^{\prime}}^2} \left[ \overline{v}(k_1+k_2-p_1) k\!\!\!/_2 \frac{i}{p\!\!\!/_1-k\!\!\!/_1-m_{\chi_2}} k\!\!\!/_1 u(p_1) \right. \nonumber \\
& & +\left. \overline{v}(k_1+k_2-p_1) k\!\!\!/_1 \frac{i}{p\!\!\!/_1-k\!\!\!/_2-m_{\chi_2}} k\!\!\!/_2 u(p_1) \right].
\end{eqnarray}
Now we can replace the slashed $k_1$ with $p_1 - (p_1 - k_1)$ in the first term, and with 
$ (p_1-k_2)-(p_1-k_1-k_2)$ in the second term.  Applying the equation of motion and cancelling the redundant terms, we acquire
\begin{eqnarray}
\mathcal{M}_{\chi_1 \chi_1 \rightarrow \gamma^{\prime}_L \gamma^{\prime}_L} &=& \frac{Q_{\chi}^2 g^2}{m_{\gamma^{\prime}}^2} (m_{\chi_1}-m_{\chi_2}) \left[ \overline{v}(k_1+k_2-p_1) k\!\!\!/_2 \frac{i}{p\!\!\!/_1-k\!\!\!/_1-m_{\chi_2}} u(p_1) \right. \nonumber \\
& & - \left. \overline{v}(k_1+k_2-p_1) \frac{i}{p\!\!\!/_1-k\!\!\!/_2-m_{\chi_2}} k\!\!\!/_2 u(p_1) \right].
\end{eqnarray}
Then we continue to replace the slashed $k_2$ with $p_1-k_1-(p_1-k_1-k_2)$ in the first term, and with $p_1-(p_1-k_2)$ in the second term. Again, applying the equation of motion, we acquire
\begin{eqnarray}
\mathcal{M}_{\chi_1 \chi_1 \rightarrow \gamma^{\prime}_L \gamma^{\prime}_L} &=& \frac{Q_{\chi}^2 g^2}{m_{\gamma^{\prime}}^2} (m_{\chi_1}-m_{\chi_2})^2 \left[ - \overline{v}(k_1+k_2-p_1) \frac{i}{p\!\!\!/_1-k\!\!\!/_1-m_{\chi_2}} u(p_1) \right. \nonumber \\
& & - \left. \overline{v}(k_1+k_2-p_1) \frac{i}{p\!\!\!/_1-k\!\!\!/_2-m_{\chi_2}} u(p_1) \right] + \frac{Q_{\chi}^2 g^2}{m_{\gamma^{\prime}}^2} (m_{\chi_1}-m_{\chi_2}) \nonumber \\
& & 2 i \overline{v}(k_1+k_2-p_1)  u(p_1).
\end{eqnarray}
The first term is proportional to $\frac{(m_{\chi_1}-m_{\chi_2})^2}{m_{\gamma^{\prime}}^2}$, and the second one is proportional to $\frac{m_{\chi_1}-m_{\chi_2}}{m_{\gamma^{\prime}}^2}$. In the moderate parameter space $m_{\chi_1}-m_{\chi_2} \sim m_{\gamma^{\prime}}$, the second term will induce a $\propto \frac{1}{m_{\gamma^{\prime}}^2}$ cross section, and would violate immediately the unitarity bound if $m_{\gamma^{\prime}} \ll m_{\chi_{1,2}}$.

The introduction of a dark Higgs boson can cancel this dangerous term. A direct calculation of the third diagram in Fig.~\ref{Annihilations} gives rise to
\begin{eqnarray}
\mathcal{M}_{\chi_1 \chi_1 \rightarrow h\rightarrow \gamma^{\prime}_L \gamma^{\prime}_L} = \frac{y^{\prime} A}{m_{\gamma^{\prime}}^2} \frac{i k_1 \cdot k_2}{(k_1+k_2)^2-m_h^2} \overline{v}(k_1+k_2-p_1) u(p_1),
\end{eqnarray}
where $y^{\prime}$ and $A$ are the $\chi \chi h$ and $h \gamma^{\prime} \gamma^{\prime}$ couplings, 
respectively. In the large momentum limit, we have 
$k_1 \cdot k_2 \approx \frac{(k_1 + k_2)^2 - m_h^2}{2}$.   Therefore, if
\begin{eqnarray}
y^{\prime} A = -4 Q_{\chi}^2 g^2 (m_{\chi_1} - m_{\chi_2}),
\end{eqnarray}
the unitarity violating term is precisely cancelled. Notice that in our model, $m_{\chi_1} - m_{\chi_2} = - 2 y v_{\Phi}$. If $y^{\prime} = y$, $A = 8 Q_{\chi}^2 g^2 v_{\Phi}$, the unitarity breaking term is automatically cancelled in our model.   This shows that the $Z_2$ DM model with dark Higgs mechanism behaves better than the models withourt dark Higgs.
\section{Modified Ward Identity in the Broken Phase}  \label{WTProof}

\begin{figure}
\centering
\includegraphics[width=0.8\textwidth]{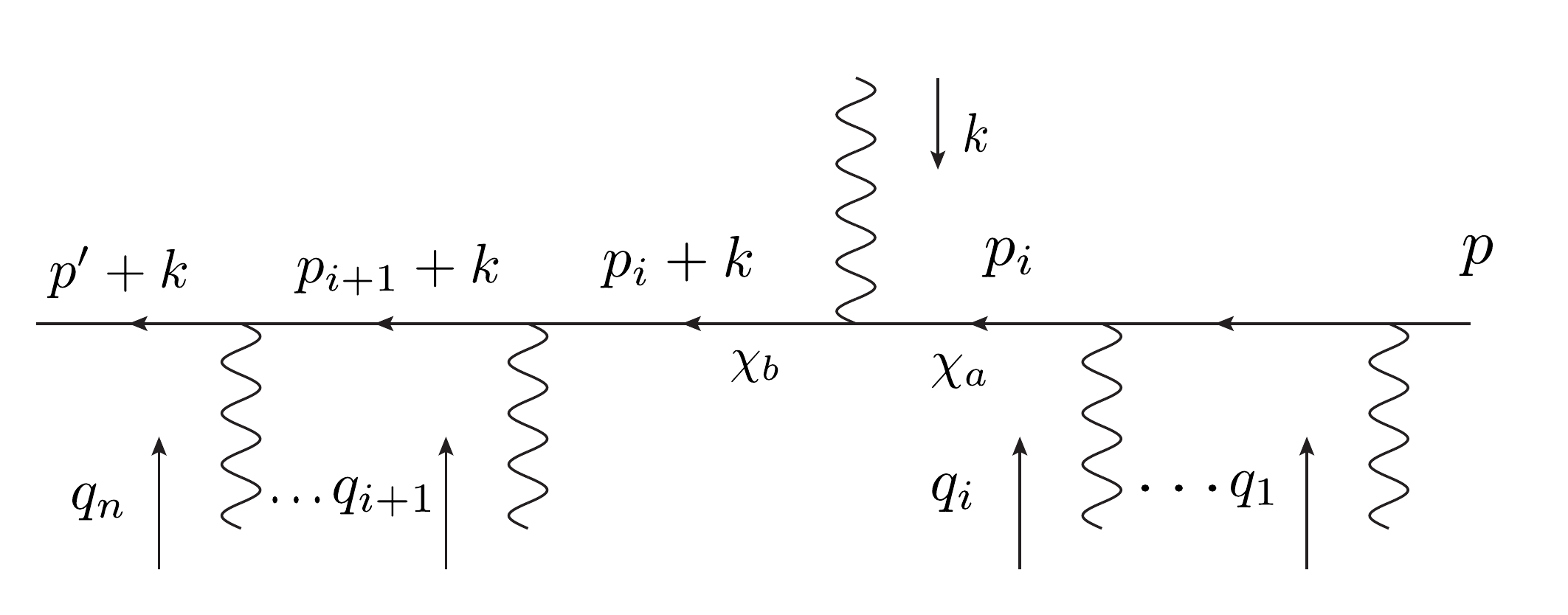}
\caption{The same as one of the key Figure appeared in Ref.~\cite{Peskin}. 
Here the polarization vector of the photon with the symbol of its momentum ``$k$'' will replaced by this momentum.} \label{WIdentity}
\end{figure}

In (\ref{WDIdentity}), we modified the Ward identity by adding a non-conserving term on the right-handed side. A path-integral proof can be found in Ref.~\cite{WTIdentityBrokenPhase} for a general $R_{\xi}$ gauge. Inpsired by the Section 7.4 of Ref.~\cite{Peskin}, we only provide a diagrammatic proof in the unitary gauge depending on the model of this paper for brevity rather than in a lengthy general $R_{\xi}$ gauge situation. Let us consider an external photon, with its polarization vector $\epsilon_{\mu}(k)$ replaced by the momentum $k_{\mu}$. We have to decompose the amplitude into several parts. We will see that some of the parts remain, which can be understood to be the Goldstone boson equivalent replacements. Other parts will be expected to be cancelled by the corresponding parts from some other diagrams. If we enumerate every type of the diagram connections,  except the ``Goldstone equivalent terms'', and if we find out that each ``cancelled'' part in one diagram's amplitude has a corresponding ``countered'' term in another diagram, we complete the proof. We need to note that in this appendix, we only enumerate the several types of the diagram connections which are essential to the ``ladder-approximation''. It looks as if the external photon to be inserted is ``sliding'' along the ``charged skeleton'' of the diagrams.

The first connection type is provided in Fig.~\ref{WIdentity}, characterized by the external line neighboured with two vector bosons. Notice that
\begin{eqnarray}
i Q_{\chi} g k_{\mu} \gamma^{\mu} = i Q_{\chi} g \left[ (p\!\!\!/_i + k\!\!\!/ - m_{\chi_b}) - (p\!\!\!/_i - m_{\chi a}) + (m_{\chi_b} - m_{\chi_a}) \right].
\end{eqnarray}
Therefore, part of the amplitude as shown in the (7.65) in Ref.~\cite{Peskin} should be modified to
\begin{eqnarray}
& & \frac{i}{p\!\!\!/_i+k\!\!\!/-m_{\chi_b}} (i Q_{\chi} g k\!\!\!/) \frac{i}{p\!\!\!/_i -m_{\chi_a}} = \mathcal{M}_{\rm{cancel},1}^{\gamma \gamma} + \mathcal{M}_{\rm{cancel},2}^{\gamma \gamma} + \mathcal{M}_{\rm{GS}}^{\gamma \gamma},
\end{eqnarray}
where
\begin{eqnarray}
& \mathcal{M}_{\rm{cancel},1}^{\gamma \gamma} &= -Q_{\chi} g \frac{i}{p\!\!\!/_i -m_{\chi_{a}}}  \nonumber \\ 
& \mathcal{M}_{\rm{cancel},2}^{\gamma \gamma} &= Q_{\chi} g  \frac{i}{p\!\!\!/_i+k\!\!\!/-m_{\chi_b}}  \nonumber \\
& \mathcal{M}_{\rm{GS}}^{\gamma \gamma}  &= \frac{i}{p\!\!\!/_i+k\!\!\!/-m_{\chi_b}} i Q_{\chi} g (m_{\chi_b} - m_{\chi_a}) \frac{i}{p\!\!\!/_i -m_{\chi_{a}}}. \label{ModifiedWithoutDarkHiggs}
\end{eqnarray}

\begin{figure}
\includegraphics[width=0.32\textwidth]{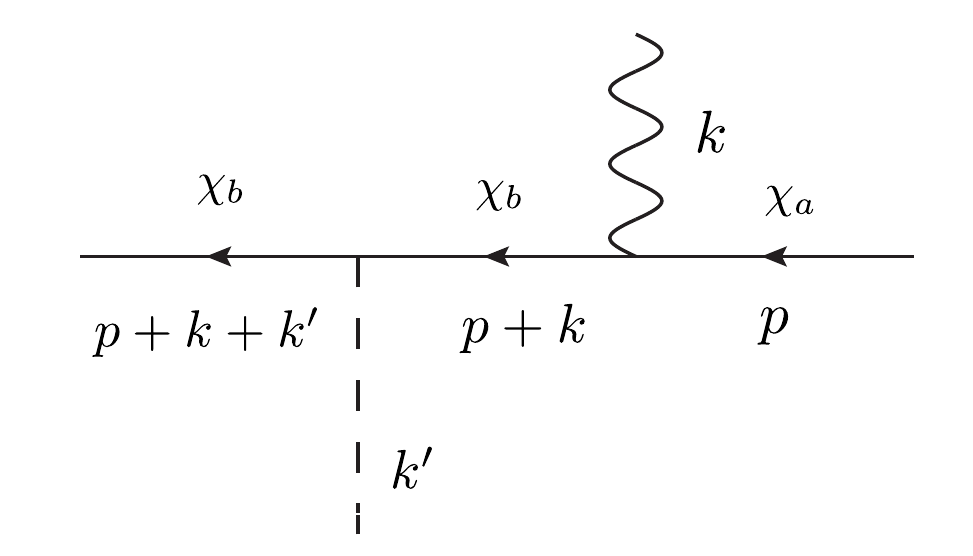}
\includegraphics[width=0.32\textwidth]{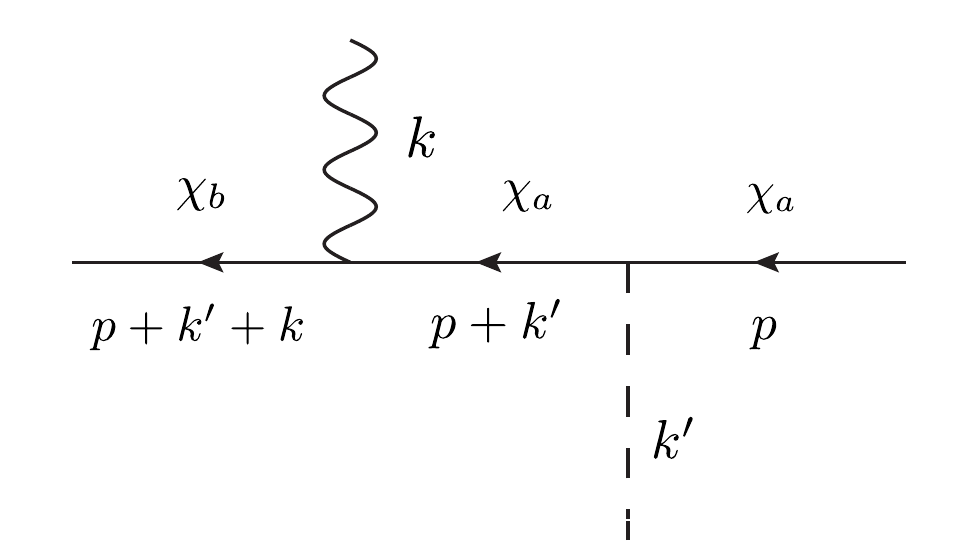}
\includegraphics[width=0.32\textwidth]{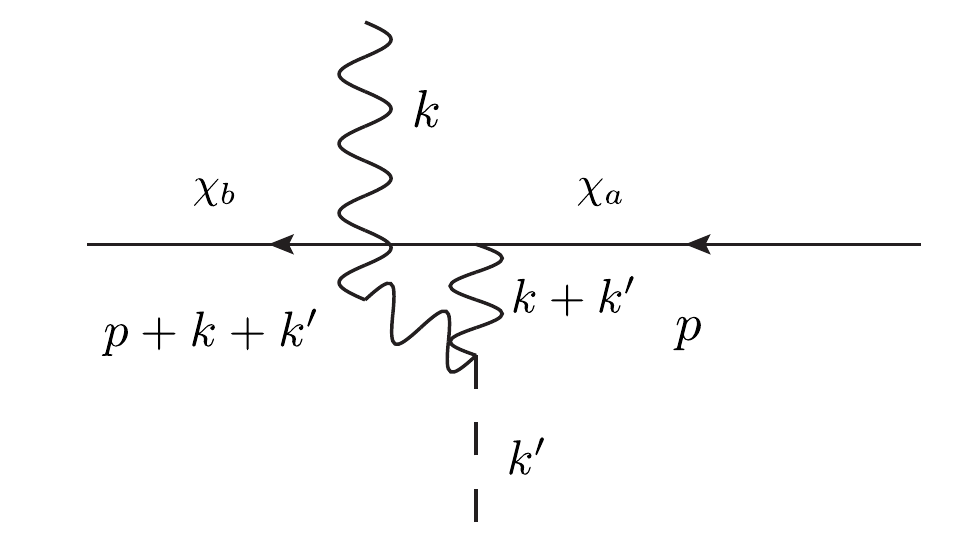}
\caption{Dark photon outer leg ``skipping'' over a dark Higgs vertex.} \label{SkipHiggs}
\end{figure}

\begin{figure}
\includegraphics[width=0.48\textwidth]{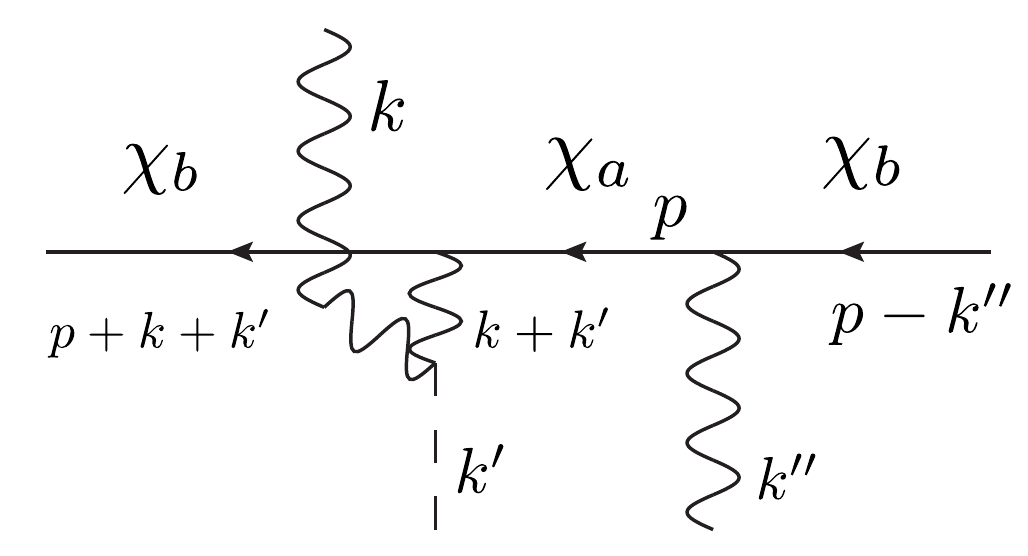}
\includegraphics[width=0.48\textwidth]{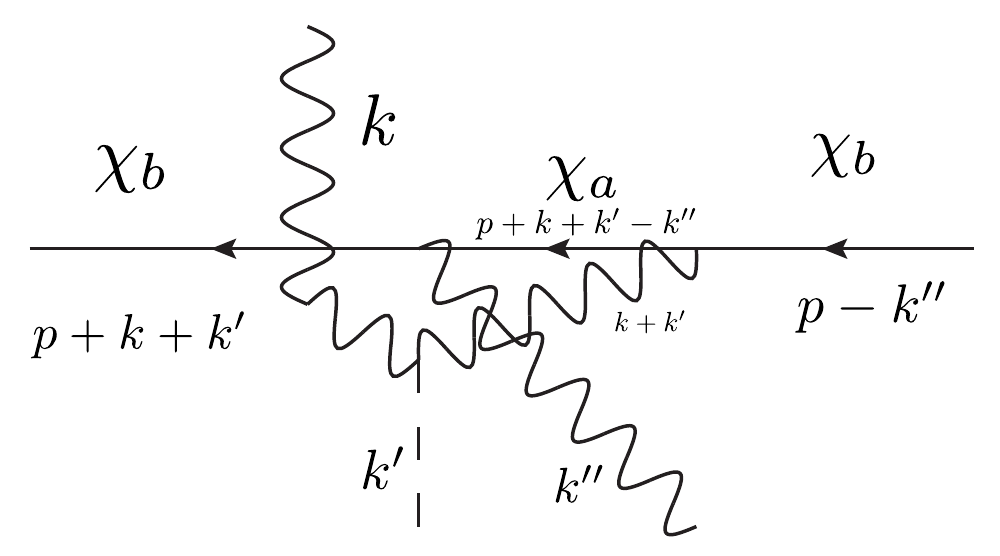}
\caption{Treat the $k+k^{\prime}$ line as a new ``external'' line recursively when it has a vector neighbour.} \label{Recursive}
\end{figure}

\begin{figure}
\includegraphics[width=0.32\textwidth]{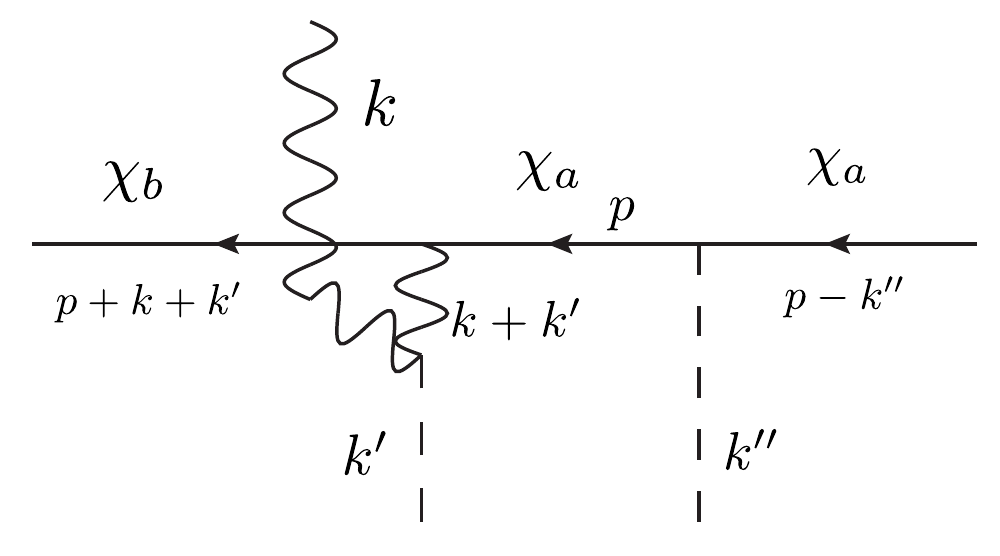}
\includegraphics[width=0.32\textwidth]{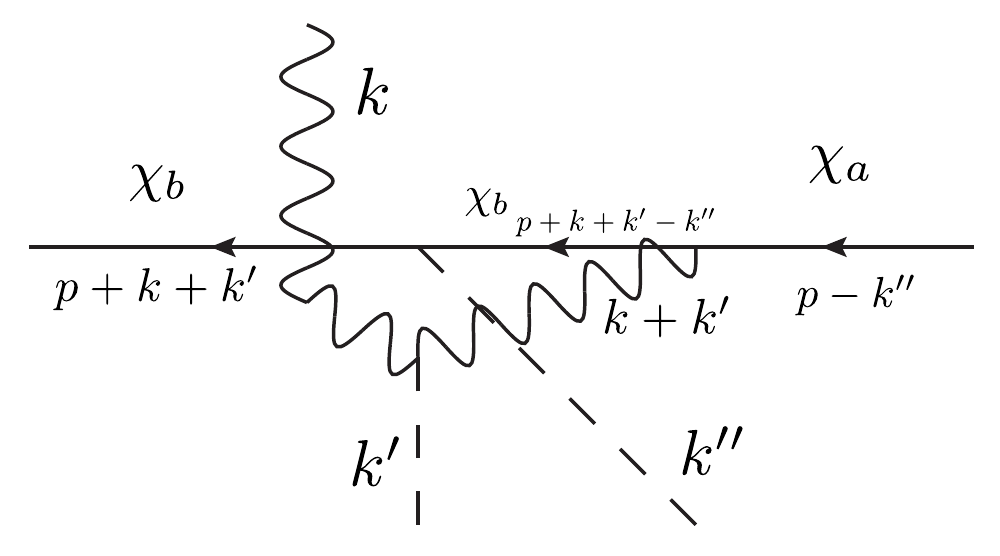}
\includegraphics[width=0.32\textwidth]{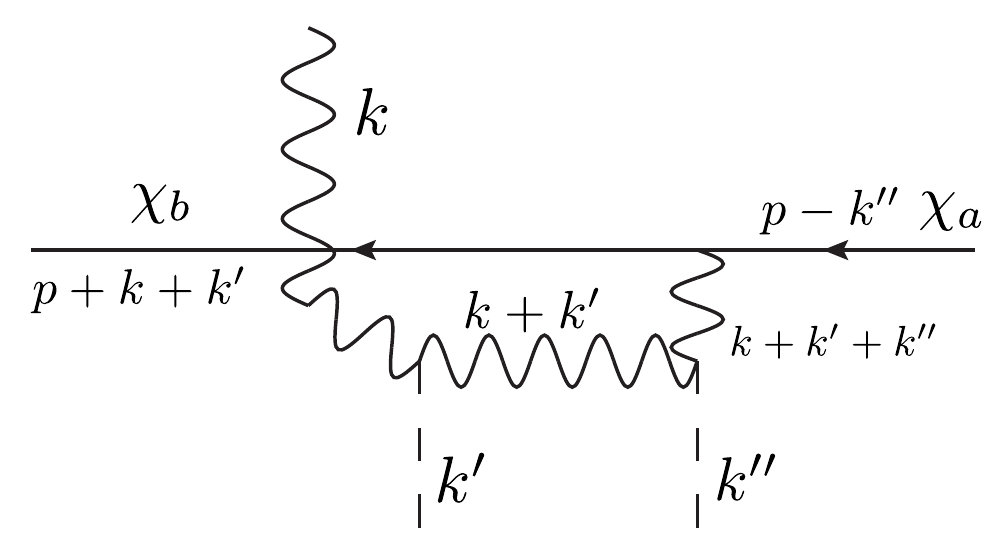}
\caption{An example to treat the $k+k^{\prime}$ line as a new ``external'' line recursively when it has a scalar neighbour.} \label{Recursive1}
\end{figure}

Just similar to the Section 7.4 of Ref.~\cite{Peskin}, one immediately find out that $\mathcal{M}_{\rm{cancel},1}^{\gamma \gamma}$ is cancelled by the corresponding term when the external vector boson ``slide'' to the $p_{i+1}+k$ fermionic line, and $\mathcal{M}_{\rm{cancel},1}^{\gamma \gamma}$ is cancelled when the vector boson ``slide'' to the $p_{i-1}$ fermionic line. $\mathcal{M}_{\rm{GS}}^{\gamma \gamma}$ has the structure of the Goldstone emission term which are not to be cancelled. However, unlike the simplest situation of the QED, a dark photon external leg might ``skip over'' a Higgs vertex. See Fig.~\ref{SkipHiggs} for the diagrams. Since the $\chi_1 \chi_1 R$ and $\chi_2 \chi_2 R$ couplings differ by an extra minus sign, the corresponding terms in the first two diagrams in Fig.~\ref{SkipHiggs} can no longer cancel with each other. We then need to take into account the third diagram in Fig.~\ref{SkipHiggs}. Parts of the amplitude by the third diagram give the result
\begin{eqnarray}
\frac{i}{p\!\!\!/+k\!\!\!/+k\!\!\!/^{\prime}-m_{\chi_b}} \left(4 Q_{\chi}^3 g^3 v_{\Phi} \right)\left[k\!\!\!/ - \frac{k \cdot (k+k^{\prime}) (k\!\!\!/ + k\!\!\!/^{\prime})}{m_{\gamma^{\prime}}^2} \right] \frac{i}{p\!\!\!/-m_{\chi_a}} \frac{(-i)}{(k+k^{\prime})^2-m_{\gamma^{\prime}}}. \label{SkipHiggsTerm}
\end{eqnarray}
The middle factor can be reduced to
\begin{eqnarray}
2 Q_{\chi}^2 g^2 m_{\gamma^{\prime}} \left[ (k\!\!\!/-k\!\!\!/^{\prime}) - \frac{(k+k^{\prime}) \cdot (k-k^{\prime})}{m_{\gamma^{\prime}}^2} (k\!\!\!/+k\!\!\!/^{\prime}) + \frac{m_{\gamma^{\prime}}^2-(k+k^{\prime})^2}{m_{\gamma^{\prime}}^2} (k\!\!\!/+k\!\!\!/^{\prime}) \right]. \label{DecompositionUnitary}
\end{eqnarray}
Therefore, we decomposite (\ref{SkipHiggs}) at first into two parts,
\begin{eqnarray}
\mathcal{M}^{R\gamma 3}_{\rm{GS}} + \mathcal{M}^{R \gamma 3}_{\rm{cancel}},
\end{eqnarray}
where
\begin{eqnarray}
\mathcal{M}^{R\gamma 3}_{\rm{GS}} = \frac{2 i Q_{\chi}^2 g^2 m_{\gamma^{\prime}} }{p\!\!\!/+k\!\!\!/+k\!\!\!/^{\prime}-m_{\chi_b}} \left[ (k\!\!\!/-k\!\!\!/^{\prime}) - \frac{(k+k^{\prime}) \cdot (k-k^{\prime})}{m_{\gamma^{\prime}}^2} (k\!\!\!/+k\!\!\!/^{\prime}) \right] \frac{i}{p\!\!\!/-m_{\chi_a}} \frac{(-i)}{(k+k^{\prime})^2-m_{\gamma^{\prime}}}, \nonumber
\end{eqnarray}
which is actually the same structure for a Goldstone boson replacing the $k$-momentum dark photon, and then times a $m_{\gamma^{\prime}}$. $\mathcal{M}^{R \gamma}_{\rm{cancel}}$ is given by
\begin{eqnarray}
\mathcal{M}^{R \gamma 3}_{\rm{cancel}} = \frac{i}{p\!\!\!/+k\!\!\!/+k\!\!\!/^{\prime}-m_{\chi_b}} \left(2 Q_{\chi}^2 g^2 \right) \frac{k\!\!\!/+k\!\!\!/^{\prime}}{m_{\gamma^{\prime}}} \frac{i}{p\!\!\!/-m_{\chi_a}} (i). \nonumber
\end{eqnarray}
It is then decomposed into three parts, 
\begin{eqnarray}
\mathcal{M}^{R \gamma 3}_{\rm{cancel}} = \mathcal{M}^{R \gamma 3}_{\rm{cancel},1} + \mathcal{M}^{R \gamma 3}_{\rm{cancel},2} + \mathcal{M}^{R \gamma 3}_{\rm{cancel},3}, \nonumber
\end{eqnarray}
where
\begin{eqnarray}
\mathcal{M}^{R \gamma 3}_{\rm{cancel},1} &=& - \left(2 Q_{\chi}^2 g^2 \right) \frac{1}{m_{\gamma^{\prime}}} \frac{i}{p\!\!\!/-m_{\chi_a}}, \nonumber \\
\mathcal{M}^{R \gamma 3}_{\rm{cancel},2} &=& \left(2 Q_{\chi}^2 g^2 \right) \frac{1}{m_{\gamma^{\prime}}} \frac{i}{p\!\!\!/+k\!\!\!/+k\!\!\!/^{\prime}-m_{\chi_b}},  \nonumber \\
\mathcal{M}^{R \gamma 3}_{\rm{cancel},3} &=& \frac{i}{p\!\!\!/+k\!\!\!/+k\!\!\!/^{\prime}-m_{\chi_b}} \left(2 Q_{\chi}^2 g^2 \right) \frac{m_{\chi_a}-m_{\chi_b}}{m_{\gamma^{\prime}}} \frac{i}{p\!\!\!/-m_{\chi_a}} (i), \nonumber \\
&=& \frac{i}{p\!\!\!/+k\!\!\!/+k\!\!\!/^{\prime}-m_{\chi_b}} \left(2 Q_{\chi} g y  \cdot \rm{sgn}(m_{\chi_a}-m_{\chi_b}) \right) \frac{i}{p\!\!\!/-m_{\chi_a}} (i). \label{SkipHiggsTermFinal}
\end{eqnarray}
Notice that $\frac{2 Q_{\chi}^2 g^2 (m_{\chi_a} - m_{\chi_b})}{m_{\gamma^{\prime}}} = 2 Q_{\chi} g y \cdot \rm{sgn}(m_{\chi_a}-m_{\chi_b})$ in the third term $\mathcal{M}^{R \gamma 3}_{\rm{cancel},3}$. To cancel the $\mathcal{M}^{R \gamma 3}_{\rm{cancel},3}$, we calculate the first diagram in Fig.~\ref{SkipHiggs}.
\begin{eqnarray}
\mathcal{M}^{R \gamma 1} = \mathcal{M}^{R \gamma 1}_{\rm{GS}} + \mathcal{M}^{R \gamma 1}_{\rm{cancel},1} + \mathcal{M}^{R \gamma 1}_{\rm{cancel},2},
\end{eqnarray}
where we do not show the $\mathcal{M}^{R \gamma 1}_{\rm{cancel},1}$ for brevity, and 
\begin{eqnarray}
\mathcal{M}^{R \gamma 1}_{\rm{cancel},1} &=& -i Q_{\chi} g y \cdot \rm{sgn}(m_{\chi_b}-m_{\chi_a}) \frac{i}{p\!\!\!/_i+k\!\!\!/-m_{\chi_b}} \frac{i}{p\!\!\!/-m_{\chi_a}}, \nonumber \\
\mathcal{M}^{R \gamma 1}_{\rm{cancel},2} &=& i Q_{\chi} g y \cdot \rm{sgn}(m_{\chi_b}-m_{\chi_a}) \frac{i}{p\!\!\!/_i + k\!\!\!/ - m_{\chi_b}} \frac{i}{p\!\!\!/ + k\!\!\!/ -m_{\chi_a}}. \label{RGamma1}
\end{eqnarray}
A similar calculation of the second diagram in Fig.~\ref{SkipHiggs} will also contribute to term exactly equivalent to the $\mathcal{M}^{R \gamma 1}_{\rm{cancel},1}$, so their summation cancels the $\mathcal{M}^{R \gamma 3}_{\rm{cancel},3}$ in the (\ref{SkipHiggsTermFinal}). $\mathcal{M}^{R \gamma 1}_{\rm{cancel},2}$ in the (\ref{RGamma1}) will be cancelled with other diagrams recursively by the same discussions above if the next right neighbour of the external line with the momentum $k$ is again a dark photon or a dark Higgs boson. Therefore, in (\ref{SkipHiggsTermFinal}), only $\mathcal{M}^{R \gamma 3}_{\rm{cancel},1}$ and $\mathcal{M}^{R \gamma 3}_{\rm{cancel},2}$ remain. These terms will never be cancelled if we fix all the other skeletons of diagrams. However,  $\mathcal{M}^{R \gamma 3}_{\rm{cancel},1}$ and $\mathcal{M}^{R \gamma 3}_{\rm{cancel},2}$ looks very similar to the $\mathcal{M}^{\gamma \gamma}_{\rm{cancel},1}$ and $\mathcal{M}^{\gamma \gamma}_{\rm{cancel},2}$ in the (\ref{ModifiedWithoutDarkHiggs}), and they only differ by a factor of $\frac{2 Q_{\chi} g}{m_{\gamma^{\prime}}}$. This prompts us that in this case, the $k+k^{\prime}$ internal vector boson line can be treated as a new ``external line'', carrying exactly the same $k+k^{\prime}$ momentum, however missing the corresponding ``Goldstone equivalent term''. We can then reuse all the above discussions recursively to cancel these terms by ``sliding'' the new $k+k^{\prime}$ ``external line'' along the fermionic skeleton. For example, if the right neighbour of the $k+k^{\prime}$ line is a vector boson line, as in the first diagram of the Fig.~\ref{Recursive}, the second diagram in the Fig.~\ref{Recursive} will then contribute to a term through a similar decomposition like (\ref{ModifiedWithoutDarkHiggs}), which cancels the $\mathcal{M}^{R \gamma 3}_{\rm{cancel},2}$ in the (\ref{SkipHiggsTermFinal}). If the right neighbour of the $k+k^{\prime}$ line is a dark Higgs line, as in the first diagram in the Fig.~\ref{Recursive1}, the rest diagrams in the Fig.~\ref{Recursive1} will contribute to terms through the similar decompositions like the whole (\ref{SkipHiggsTermFinal}) to cancel $\mathcal{M}^{R \gamma 3}_{\rm{cancel},2}$. Of course, this will leave another term similar to $\mathcal{M}^{R \gamma 3}_{\rm{cancel},2}$ which is not cancelled, then repeat the recursive processes to cancel this\dots

Finally, after ``sliding'' the external line with the momentum $k$ to the ends of the fermionic lines, the diagrammatic result becomes
\begin{eqnarray}
\frac{k_{\mu}}{m_{\gamma^{\prime}}} \mathcal{M}^{\mu} = \overline{\mathcal{M}}_{\text{GS}} + \mathcal{M}_{\text{begin}} - \mathcal{M}_{\text{end}}, \label{CompleteWTIdentity}
\end{eqnarray}
in which $\overline{\mathcal{M}}_{\text{GS}}$ are the corresponding diagrams that replace all of the dark photon $k$-leg with a Goldstone boson. In this paper, as we have mentioned, we have ignored the contributions that dark boson radiation from the interchanging mediators due to the extra vertices multiplied when calculating the wave function overlap, so only emissions from the fermions are calculated. Therefore, (\ref{CompleteWTIdentity}) finally becomes the $\overline{\mathcal{M}}_{\text{GS}}$ term in the (\ref{WDIdentity}). $\mathcal{M}_{\text{begin/end}}$ are diagrams without the dark photon insertion, and the mass input of the fermionic line had been shifted by increasing/reducing $k$. Note that when we are calculating the diagrams in Fig.~\ref{EmittingGammaPrime}, we need to amputate the initial/final bound state poles to extract the S-matrix by adopting the coefficients of the double-poles, but $\mathcal{M}_{\text{begin/end}}$ only indicate one bound state propagator, which contain only single-poles. Therefore, $\mathcal{M}_{\text{begin/end}}$ does not contribute to the transition amplitude in (\ref{WDIdentity}).

One might consider a proof in the general $R_{\xi}$ gauge. At first we show the Feynmann gauge propagator,
\begin{eqnarray}
\frac{-i}{k^2 - m_{\gamma^{\prime}}^2 } \left[ g^{\mu \nu} - \frac{k^{\mu} k^{\nu}}{k^2 - \xi m_{\gamma^{\prime}}^2} (1-\xi) \right]. 
\label{PropagatorDecomposition}
\end{eqnarray}
Rather than give a lengthy and complete proof, we only note that the $\xi$-dependent part of this propagator can also be decomposed similarly as in the (\ref{DecompositionUnitary}), with all the $m_{\gamma^{\prime}}^2$ replaced by $k^2 - m_{\gamma^{\prime}}^2$, and a $1-\xi$ factor should also be supplemented in the second and third term. Goldstone and ghost propagators are also required. Then, with the similar logic, proof in the general $R_{\xi}$ gauge can be derived.

Before closing this appendix section, we need to note that in order to derive the complete modified Ward identity, not only the ``straight'' ladder diagrams are considered. Fig.~\ref{Recursive} prompts us that all the diagrams, including all of the possible ``crossed ladders'' and all the ``vertex correction'' diagrams, should be taken into account. However, the previous processes of the proof tells us that all the terms which are expected to disappear due to the cancellation among different diagrams contain less fermionic propagator(s) then the originally ladder diagrams, so they indicate the ``off-shell corrections'' of the internal fermionic lines. (And in fact, in the general $R_{\xi}$ gauge, these terms are all $\xi$-dependent.) Although in the large $\xi$-limit (unitary gauge), these terms cannot be simply omitted (as will be discussed in the next appendix section), if we apply the ``on-shell approximation''  to keep only the nearly ``pinched pole'' contributions of all the fermionic lines, the remained ``straight-ladder'' contributions still satisfy the modified Ward identity. All the remained terms in the other ``crossed-ladder'', ``vertex correction'', etc., diagrams all contribute less ``pinched poles'' then the ``straight ladder'', thus they are safely omitted (Similar to the discussions in section II-A, Ref.~\cite{BoundStateIntegrate}). 

\section{Gauge Dependence of the Bound State} \label{GaugeIndependence}

\begin{figure}
\includegraphics[width=0.48\textwidth]{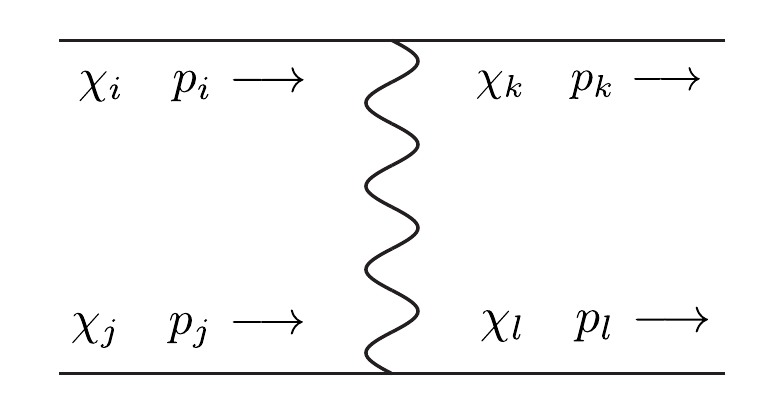}
\includegraphics[width=0.48\textwidth]{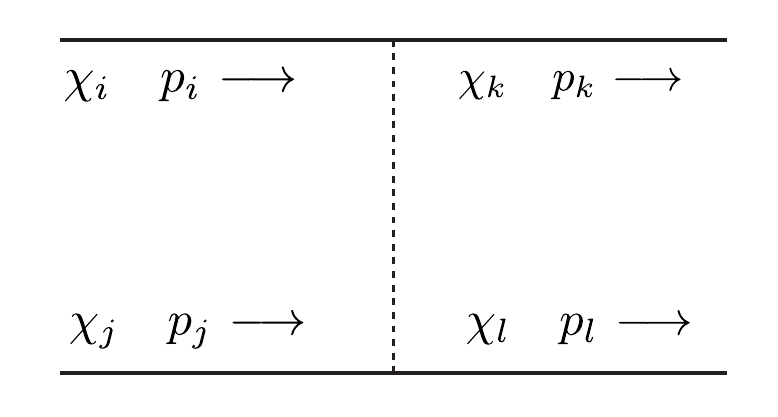}
\caption{2-2 diagram interchanging a dark photon (left) and a Goldstone boson (right).}
\label{22InLadder}
\end{figure}
We have calculated the (\ref{PotentialLongitudinal}) with the 2-2 diagrams shown in 
Fig.~\ref{22InLadder} by assuming the on-shell external legs (on-shell approximation\cite{Petraki1}). 
The $\xi$-dependent terms in both diagrams disappear by the on-shell conditions. 
The most convenient way to describe the potential is working in the Feynman gauge. 
The left-panel in Fig.~\ref{22InLadder} exchanging the dark photon leads to (\ref{PotentialTransverse}), 
while the right-panel exchanging the Goldstone boson contributes to the (\ref{PotentialLongitudinal}). In the unitary gauge, equivalently, $\frac{k^{\mu} k^{\nu}}{m_{\gamma^{\prime}}^2}$ term in the dark photon propagator contributes to the (\ref{PotentialLongitudinal}).

However, the diagrams in Fig.~\ref{22InLadder} are only a fraction inside the ladder diagrams, and usually the fermionic external legs in them are slightly off-shell. Therefore, the $\xi$-dependent $k^{\mu} k^{\nu}$ terms in the dark photon propagator cannot be simply contracted with the fermionic lines. The ``off-shellness'' of a fermionic component is typically $\sim \mu \alpha^{\prime 2}$ in a bound state. If $\xi \ll \frac{1}{\alpha^{\prime 2}}$, and notice that $k \sim \mu \alpha^{\prime}$, the $\frac{k^{\mu} k^{\nu}}{k^2-m^2_{\gamma \prime}} (1-\xi) \ll 1$, this term can be safely neglected, and the final result is (nearly) $\xi$-independent. However, if $\xi \gg \frac{1}{\alpha^{\prime 2}}$, e.g., in the unitary gauge, since $\mu \alpha^{\prime 2}$ is  much larger than the typical dark photon mass $m_{\gamma^{\prime}}$, we will have a large deviation $\propto \frac{\mu^2 \alpha^{\prime 4}}{m_{\gamma^{\prime}}^2}$ compared with the on-shell approximation, which cannot be simply neglected. Although we can confine ourself in the  $\xi \ll \frac{1}{\alpha^{\prime 2}}$ area to apply the ``on-shell approximation'' to directly write down the (\ref{PotentialTransverse}-\ref{PotentialLongitudinal}) both for on-shell and slightly off-shell fermionic legs, we still want to see whether the large $\xi$, especially the unitary gauge will ruin all of our previous discussions. Since Feynmann gauge $\xi=1$ is the most ``natural'' gauge to generate the potentials in (\ref{PotentialTransverse}-\ref{PotentialLongitudinal}), we will compare other $\xi$ with the Feynmann gauge to illustrate the gauge independence of the physical results in the following text of this appendix.

Showing a complete proof on this issue is far beyond the main topic of this paper, so we only give the most important instructions. For a general $R_\xi$ gauge, we again apply the standard trick in (\ref{ModifiedWithoutDarkHiggs}) to decompose the $k^{\mu} k^{\nu}$ terms in all of the exchanging dark photon propagators (\ref{PropagatorDecomposition}). Notice that only the $m_{\chi_a}-m_{\chi_b}$ term preserves all the fermionic propagators, and combined with the (\ref{PropagatorDecomposition}), part of its contributions can be attributed to the Feynmann gauge Goldstone, while the other part is cancelled with the general $R_{\xi}$ gauge Goldstone propagators. The $(p\!\!\!/_i+k\!\!\!/-m_{\chi_b})$ and $(p\!\!\!/_i-m_{\chi_a})$ terms in (\ref{ModifiedWithoutDarkHiggs}) will kill one of the fermionic propagators when it is internal, and will be finally cancelled with other terms if we sum over all the possible mediator connections between the bound state component lines. Finally, we acquire a complete series of Feynmann-gauge diagrams with all the gauge dependent terms only remaining in the external legs. These terms are proportional to the $p\!\!\!/_i - m_i$ for each external fermionic leg with the momentum $p_i$ and mass $m_i$.

For the scattering state, all the external legs are on-shell, so all the gauge dependent $p\!\!\!/_i - m_i$ terms disappear. Therefore, the scattering state calculations are eventually $\xi$-independent. On the other hand, for the bound state, the external leg $p\!\!\!/_i - m_i$ remains. With the symbols of (2.16) in Ref.~\cite{BoundStateIntegrate}, we define $L_{\xi}$ as the diagram summation over all possible mediator connections, and $L_{\xi=1}$ will be the Feynmann gauge results. After decomposing and cancelling all the dark photon and dark Goldstone propagator corresponding terms in $L_{\xi}$, we acquire such kind of format:
\begin{eqnarray}
L_{\xi} = L_{\xi=1} + i S_L(\xi) L_{\xi=1} - i L_{\xi=1} S_R(\xi) - S_L(\xi) L_{\xi=1} S_R(\xi) + \text{ finite terms},
\end{eqnarray}
where $S_{L,R}(\xi)$ contains the $p\!\!\!/_i-m_i$ terms for the external legs. We did not write down the loop integrals explicitly between $S_{L,R}(\xi)$ and $L_{\xi=1}$ for brevity. Notice that $S_L$ and $S_R$ should have the same dependence on the corresponding external momentums. Therefore, it is easy to see that the general $L_{\xi}$ and $L_{\xi=1}$ share the same pole if we define
\begin{eqnarray}
\Psi_{P, \xi}(p) = \Psi_{P, \xi=1}(p) + S_L(\xi, P, q) \Psi_{P, \xi=1}(q), \nonumber \\
\Psi^{*}_{P, \xi}(p) = \Psi_{P, \xi=1}^*(p) + \Psi^*_{P, \xi=1}(q) S_R(\xi, P, q), \label{WaveFunctionTransform}
\end{eqnarray}
where the $q$ should be integrated out in a loop integration, and $\Psi_{P, \xi=1}$ is the wave function of some bound state extracted from $L_{\xi=1}$,
\begin{eqnarray}
L_{\xi=1} \approx \frac{\Psi_{P, \xi=1} \Psi_{P, \xi=1}}{P^0-\sqrt{\vec{P}^2+M^2}} + \dots
\end{eqnarray}
Then $L_{\xi}$ can be written in the form of
\begin{eqnarray}
L_{\xi} \approx \frac{\Psi_{P, \xi} \Psi_{P, \xi}^*}{P^0-\sqrt{\vec{P}^2+M^2}} + \dots, \label{BoundStatePropagator}
\end{eqnarray}
which has exactly the same pole structures with the $L_{\xi=1}$, and $\Psi_{P, \xi}$ is the corresponding wave function. This means that the bound state energies are exactly gauge (or $\xi$-) independent. However, the wave functions need to be transformed according to (\ref{WaveFunctionTransform}). We should also note that for large $\xi$, Bethe-Salpeter wave function $\Psi_{P, \xi}$ can not even be reduced to the equal-time schr\"{o}dinger equation, so we need to solve the Bethe-Salpeter equation in this situation. This is because when $\xi \ll \frac{1}{\alpha^{\prime 2}}$, large awkward time-dependent gauge transformations are introduced, disrupting the time dependence of the wave functions.

We then point out that the $\xi$-dependence on the wave functions do not affect the physical S matrix calculations. For an example, if we want to calculate the state transition characterized by the diagrams in Fig.~\ref{EmittingGammaPrime}, we acquire
\begin{eqnarray}
L_{\xi, \text{out}} K_{\xi} L_{\xi, \text{in}}, \label{StateTransition}
\end{eqnarray}
where $K_{\xi}$ is the perturbative kernel to emit the dark photon. Note that after summing over all possible mediator connections to attribute all of the $\xi$-dependent terms to the external legs, and then adopt the poles of the initial and final states, the internal mediators in (\ref{StateTransition}) finally become Feynmann gauge propagators, and the final result is of the format
\begin{eqnarray}
\frac{\Psi_{P_i, \xi} \Psi_{P_i, \xi=1}^*}{P_i^0-\sqrt{\vec{P_i}^2+M^2}} K_{\xi=1} \frac{\Psi_{P_o, \xi=1}^* \Psi_{P_o, \xi}}{P_o^0-\sqrt{\vec{P_o}^2+M^2}}.
\end{eqnarray}
According to the principles of the LSZ reduction formula, the $\frac{\Psi_{P_{i,o}, \xi}}{P_{i,o}^0-\sqrt{\vec{P_{i,o}}^2+M^2}}$ can be regarded as the bound state propagator as well as the ``renormalzation factor'' compared with the (\ref{BoundStatePropagator}), and it should be amputated, leaving us only the $ \Psi_{P_i, \xi=1}^* K_{\xi=1} \Psi_{P_o, \xi=1}$. Then what if we calculate the $ \Psi_{P_i, \xi}^* K_{\xi} \Psi_{P_o, \xi}$? Note that when we transform all the wave functions according to (\ref{WaveFunctionTransform}), the $S_{L,R}$ will also exert on the interaction kernel $K$ by adding lots of off-shell terms proportional to $p\!\!\!/_i-m_i$. This changes $K_{\xi=1}$ to $K_{\xi}$. Therefore, $ \Psi_{P_i, \xi=1}^* K_{\xi=1} \Psi_{P_o, \xi=1} = \Psi_{P_i, \xi}^* K_{\xi} \Psi_{P_o, \xi}$,  which is exactly the gauge (or $\xi$-) independent physical S-matrix. Finally, we can see that the off-shell corresponding terms $S_{L,R}$ does not contribute to this matrix element because their contributions to the $\Psi_{P_i, \xi}$,  $K_{\xi}$,  $\Psi_{P_o, \xi}$ respectively cancel with each other.

Furthermore, although we need to calculate all the possible mediator connections between the bound state component fermions, contributions other than ladder diagrams are actually ignored in our paper for the same reason demonstrated in section II-A from Ref.~\cite{BoundStateIntegrate}. We can easily verify that most of the $\xi$-dependent off-shell contributions in the ladder diagrams are actually also absorbed by these non-ladder diagrams. The remained terms are attributed to the external legs and does not disturb all of the physical results as we have discussed. Therefore, we finally recover the ``on-shell'' approximation principle in the most general $R_\xi$ gauge even for large $\xi \gg \frac{1}{\alpha^{\prime 2}}$: when calculating the diagrams in Fig.~\ref{22InLadder}, we shall safely assume all the external legs to be on shell. Because all the off-shell contributions are $\xi$-dependent and will be exactly cancelled and absorbed into the non-ladder diagrams and the physical state definitions.

\section{Reasons for Neglecting the Emission From the Mediators} \label{Reason}

\begin{figure}
\centering
\includegraphics[width=0.46\textwidth]{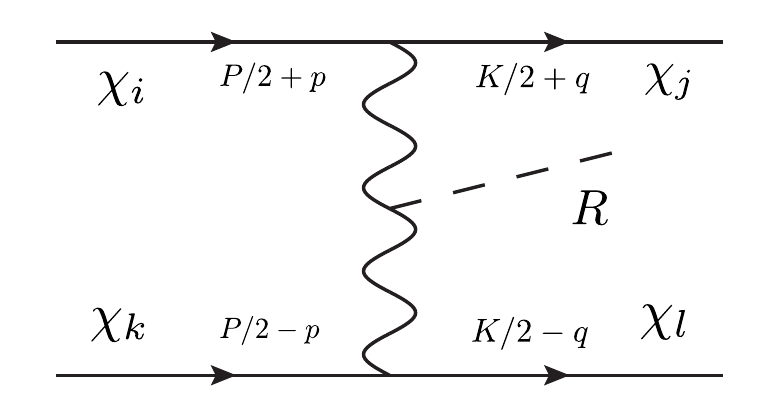}
\includegraphics[width=0.46\textwidth]{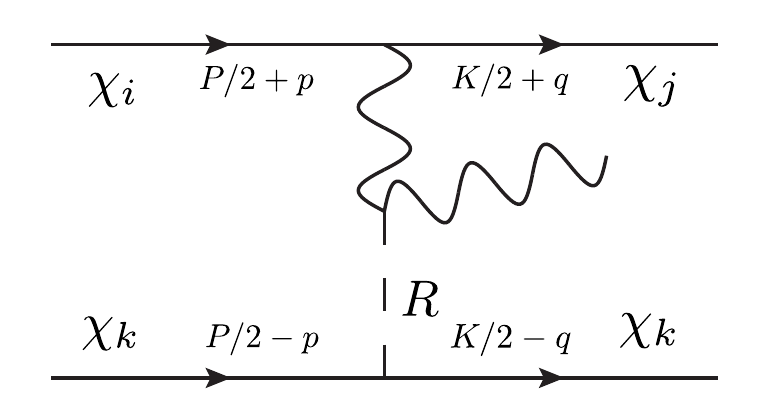}
\caption{Original diagrams of the emission from the mediators. \label{OriginalEmission}}
\includegraphics[width=0.46\textwidth]{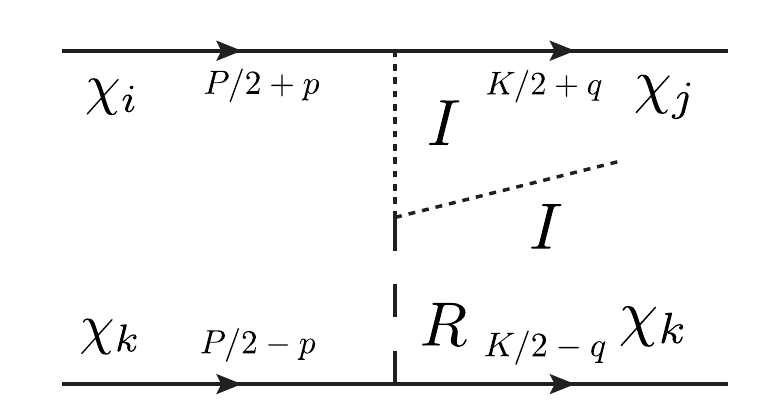}
\includegraphics[width=0.46\textwidth]{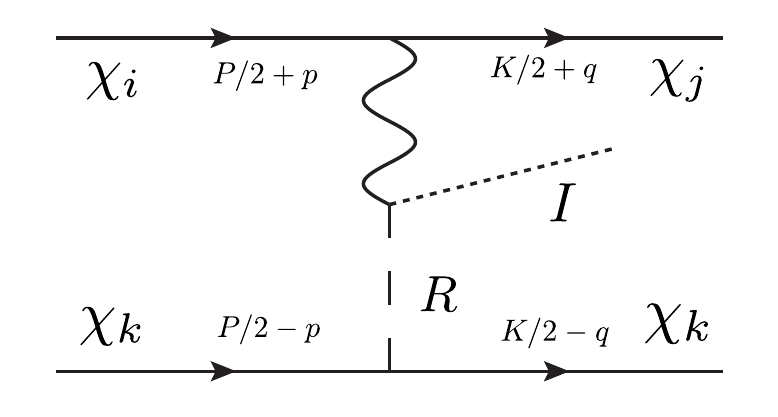}
\includegraphics[width=0.46\textwidth]{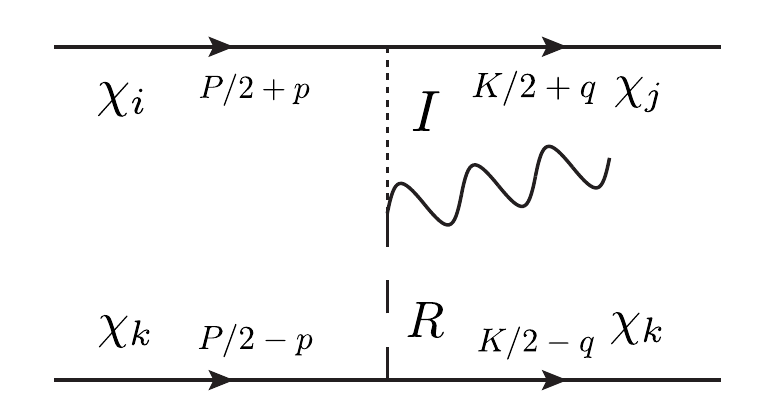}
\caption{Derived diagrams of the emission from the mediators. \label{DerivedEmission}}
\end{figure}

Ref.~\cite{DMBS28} had calculated the emission from one of the ``ladder'' mediators. Naively, such terms contain more vertices and can be considered as higher-order contributions. However, the momenta exchanging among these diagrams $p-q$ are of the order $\mu \alpha^{\prime}$, which appears in the propagator as $\frac{1}{|\vec{p}-{\vec{q}}|^n}$, canceling the extra coupling constants in the numerator. Ref.~\cite{DMBS28} calculated and showed that such a contribution is nearly equivalent to the diagrams where vector bosons emit from the bounded components.

In our situation, we have only two diagrams where a dark Higgs boson or a dark photon emit from a mediator, which are listed in Fig.~\ref{OriginalEmission}. All the out leg fermions are nearly on-shell, as the process described in Ref.~\cite{Petraki1}. The unitary gauge propagator for a dark photon $\frac{-i}{k^2-m_{\gamma^{\prime}}^2+i \epsilon}\left( g_{\mu \nu} - \frac{k_{\mu} k_{\nu}}{m_{\gamma^{\prime}}^2} \right)$ can be decomposed into two parts, $\frac{-i g_{\mu \nu}}{k^2-m_{\gamma^{\prime}}^2 + i \epsilon}$ and $\frac{i}{k^2-m_{\gamma^{\prime}}^2 + i \epsilon} \left( \frac{k_{\mu} k_{\nu}}{m_{\gamma^{\prime}}^2} \right)$, roughly symbolize the transverse and longitudinal (or Goldstone) contributions. Contracting out the $k_{\mu}$ and $g_{\mu \nu}$ generate the original diagrams in Fig.~\ref{OriginalEmission}, and extra three diagrams listed in Fig.~\ref{DerivedEmission}, in which all of the dark photons are $\frac{-i g_{\mu \nu}}{q^2 - m_{\gamma^{\prime}}^2}$ ``transverse'' propagators.

In each diagram of Fig.~\ref{OriginalEmission} and the first diagram in Fig.~\ref{DerivedEmission}, one of the vertices include a $v_{\Phi}$. {For the second diagram in Fig.~\ref{DerivedEmission}, note that since $K-P \sim \mu_r \alpha^{\prime 2}$, therefore only $p-q$ with their spatial components $\vec{p}-\vec{q} \sim \mu_r \alpha^{\prime}$ are picked up in the $R$ and $\gamma^{\prime}$ propagators. The metrix $g^{\mu \nu}$ in the $\gamma^{\prime}$ propagator contracts this $p-q$ with the $\chi_i \chi_j \gamma^{\prime}$ vertex and gives a $m_{\chi_i}-m_{\chi_j}$. Ignoring the mediator masses, this diagram becomes
\begin{eqnarray}
\propto y Q_{\Phi} Q_{\chi} g^2 (m_{\chi_i}-m_{\chi_j}) \frac{1}{(\vec{p}-\vec{q})^4}. \label{MediatorEmission}
\end{eqnarray}
Comparing this with (2.23a), (2.24) in Ref.~\cite{DMBS28}, we need to replace their ``$(T_1^b)_{i^{\prime} i} (T_2^c)_{j^{\prime} j} \times 8 g_s^{\rm{BSF}} g_s^2 \frac{\vec{q}-\vec{p}}{(\vec{q}-\vec{p})^4}$'' by (\ref{MediatorEmission}). Hence the Eqn.~(B.1) in Ref.~\cite{DMBS28}, will be replaced by 
\begin{eqnarray}
y Q_{\Phi} g (2 \mu_r) (m_{\chi_i}-m_{\chi_j}) \int \frac{d^3 \vec{p}}{(2 \pi)^3} \frac{d^3 \vec{q}}{(2 \pi)^3} \frac{1}{(\vec{p}-\vec{q})^4} \psi_{\text{s/d},nlm}^*(\vec{p}) (\sigma^1) \phi_{\text{d/s},\vec{k}} (\vec{q}).
\end{eqnarray}
the $\frac{1}{(\vec{p}-\vec{q})^4}$ term will induce an integration
\begin{eqnarray}
\int \frac{d^3 \vec{q}}{(2 \pi)^3} \frac{e^{-i \vec{q} \cdot \vec{r}}}{(\vec{q}^2)^2}.\label{InferredIntegral}
\end{eqnarray}
This gives an infrared divergence. By analysing the precise momenta flows (which means considering both the masses of the mediator's propagators and the sub-leading $O(\mu_r \alpha^{\prime 2})$ momenta ignored by usual bound state ladder approximation analysis), we can learn that the physical cut-off on this divergence can be very complicated and depends on $\frac{1}{2} \mu_r v^2$, $m_{\chi_1}-m_{\chi_2}$, and $\mu_r \alpha^{\prime 2}$, $m_{R}$ and $m_{\gamma^{\prime}}$. We define the effective infrared cut-off scale $\Lambda_{\text{IR}} \sim \text{max}\left\lbrace \frac{1}{2} \mu v^2,~m_{\chi_1}-m_{\chi_2},~\mu \alpha^{\prime 2},~m_{R},~m_{\gamma^{\prime}} \right\rbrace$. With this cut off, (\ref{InferredIntegral}) will then become $\sim \frac{i}{2 \pi^2} \frac{\pi e^{-\Lambda_{\text{IR} } r}}{4 \Lambda_{\text{IR}}}$. Since usually $\Lambda_{\text{IR}} \ll \kappa = \mu \alpha^{\prime}$, the numerator  $e^{-\Lambda_{\text{IR} } r} \sim 1$ in the usual spatial area where the bound state wave function dominates, therefore (\ref{MediatorEmission}) finally becomes
\begin{eqnarray}
\sim y Q_{\Phi} g (2 \mu_r) (m_{\chi_i}-m_{\chi_j}) \frac{1}{8 \pi \Lambda_{\text{IR}}} \int d^3 \vec{r} \psi_{\text{s/d},nlm}(\vec{r}) \sigma^1 \phi_{\text{d/s}, \vec{k}}(\vec{r}).	\label{MediatorEmissionFinal}
\end{eqnarray}
The ``$(2^5 \pi \alpha_s^{\text{BSF}} M^2/\mu^{1/2})$'' factor in (2.25) of Ref.~\cite{DMBS28} will also be replaced by $4 \sqrt{2} \sqrt{\mu_r} Q_{\chi}$.  Connecting this with \ref{MediatorEmissionFinal} and compare with the  (\ref{GoldStoneEmission}), we can finally see a suppression factor of $\frac{y Q_{\Phi} g (|m_{\chi_1}-m_{\chi_2}|) }{4 \pi \Lambda_{\text{IR}}} = \frac{y^2 m_{\gamma^{\prime}}}{2 \pi \Lambda_{\text{IR}}}$. Since $\Lambda_{\text{IR}}$ should be roughly the same order of magnitude with $m_{\gamma^{\prime}}$, and in this paper we are interested in the $y < Q_{\chi} g$ parameter space, and considering the further factor of $\frac{1}{2 \pi}$, we can assert that the final result of this mediator emission channel is roughly $\frac{\Lambda_{\text{IR}}}{\mu_r}$ suppressed compared with the (\ref{GoldStoneEmission}). }

Only the last diagram in Fig.~\ref{DerivedEmission} might be problematic. Here the transverse dark photon was emitted from the mediator. The momenta dependence of this diagram is exactly similar to the (2.23a) in Ref.~\cite{DMBS28}, with all the other group factors disappeared. And the factor of 8 that equation also disappear in our model due to the different Lorentz structures of the vertices. This means that our result will be suppressed by at least a factor of $\frac{1}{8}$. Compared with the (B.4d) in Ref.~\cite{DMBS28}, we also have a $\frac{y^2}{Q_{\Phi}^2 g^2}<1$ factor, which is usually smaller than 1 to evade the CMB recombination bound described in (\ref{CMBBound}), giving an extra suppression. Finally, a practical calculation had shown that the transverse dark photon emitting process is far from the main contribution compared with other bound state formation channels. Therefore, we also neglect this diagram.

One might concern whether the extended Ward identity is satisfied if we drop out these diagrams, since from the section \ref{WTProof}, we have learned that the extended Ward identity establishes only when we sum over all the possible diagrams in the same order. However, according to Eqn.~(2.12) in Ref.~\cite{BoundStateIntegrate}, we actually acquiesce to drop out all the off-shell contributions when we are calculating the schr\"{o}dinger equations to resum the ladder diagrams. Analysis in Appendix \ref{GaugeIndependence} also reveals that we can safely apply the ``on-shell'' approximation to calculate the perturbative kernel in a general R-$\xi$ gauge. It is easy to verify that if all the external fermions to be on-shell, all the $\xi$-dependent terms then disappear separately for a $\chi_i \chi_j \gamma^{\prime}$ kernel and the emission from mediator described in Fig.~\ref{OriginalEmission}. Therefore, Ward identity can still be satisfied even if we neglect some of the diagrams.

\section{Quantum Numbers of the DM Bound States} \label{QuantumNumbers}

In this paper, before the spontaneously symmetry breaking, the Lagrangian we have adopted in (\ref{BasicLag}) keeps both the parity and C-parity conserved, if we define the intrinsic parity of the dark Higgs boson $\Phi$ to be odd, so $P \Phi(t, \vec{x}) P = -\Phi(t, -\vec{x})$. This can derived if we notice
\begin{eqnarray}
PC \chi(x) CP = - CP \chi(x) PC,
\end{eqnarray}
or
\begin{eqnarray}
P \chi^C(x) P = -(P \chi(x) P)^C,
\end{eqnarray}
where $C$ and $P$ are the charge conjugate and parity operators. Therefore,
\begin{eqnarray}
C (\Phi \overline{\chi^C} \chi + \text{h.c.}) C &=& (\Phi \overline{\chi^C} \chi + \text{h.c.}), \nonumber \\
P (\Phi(x) \overline{\chi^C}(x) \chi(x) + \text{h.c.}) P &=& (\Phi(\tilde{x}) \ \overline{\chi^C}(\tilde{x}) \chi (\tilde{x}) + \text{h.c.}),
\end{eqnarray}
where $\tilde{x} = (t, -\vec{x})$ are the parity transformed coordinates. After we decompose $\chi$ into $\chi_1^4$ and $\chi_2^4$ through (\ref{Decomposition4Spinor}), it is easily known that the C-parities for $\chi_1$ and $\chi_2$ are ``$+$'' and ``$-$'', respectively.

The (C-)parity of the dark photon is similar to the visible photon, which is defined to be odd.

For a two-fermion system, the total C-parity is calculated to be $(-1)^{L+S}$ for the $\overline{\chi} \chi$/$\chi \overline{\chi}$ system, or equivalently, it is $(-1)^{i+j}$ for all the $| \chi_i \chi_j \rangle$ systems, while the total parity should be $(-1)^L$ for two same charged fermions, and $(-1)^{L+1}$ for opposite charged fermions. One can compare the angular momentums and the (C-)parity informations listed in the Tab.~\ref{TwoBodyDecay}.

For the two-boson systems $RR$, $II$, $IR$, $R\gamma^{\prime}$, $I\gamma^{\prime}$, or 
$\gamma^{\prime} \gamma^{\prime}$, the total C-parity can be directly acquired through multiplying 
all the C-parities of the particle components. They are $+$, $+$, $-$, $-$, $+$, $+$ respectively. 
While all of their parities are $(-1)^{L}$.

Then we are ready to present how we acquired the selection rules shown in Tab.~\ref{TwoBodyDecay}. 

For the $\overline{\chi} \chi$/$\chi \overline{\chi}$ initial states, the dark charge conservation could only permit the $\gamma^{\prime} \gamma^{\prime}$, and $\Phi^* \Phi$ final states. $\Phi^* \Phi$ can be decomposed into $RR$, $II$ and $IR$. From the C-parities of these boson pairs, we know $IR$ can only be decayed through $|\chi_1 \chi_2 \rangle - |\chi_2 \chi_1 \rangle$ initial states. For the 101$-$$+$ situation, it was prohibited by the parity conservation law. $RR$ and $II$ might appear in other combinations, however their orbital momentums are constrained to be even, so their parity are always +. Therefore, they are ruled out in the $^1S_0 (0^{+ -})$ case due to the parity conservation, and are forbidden in the 111$+$$+$ situation due to the angular momentum conservation. Finally, $\gamma^{\prime} \gamma^{\prime}$ channel was forbidden in all the $J=1$ initial states because of the Landau-Yang theorem. 

For the $\chi \chi$/$\overline{\chi} \overline{\chi}$ initial states, the dark charge conservation only permits the $\Phi^{(*)} \gamma^{\prime}$ final states. This can be composed into $R \gamma^{\prime}$ and $I \gamma^{\prime}$. $J=0$ situations are eliminated because a transverse photon cannot form a $J=0$ state together with a scalar. $R \gamma^{\prime}$ and $I \gamma^{\prime}$ were separately placed because of the C-parity conservation law. 

After the symmetry breaking, the parity-odd scalar $\Phi$ takes a vacuum expectation value. This means that the parity is spontaneously broken, and the bound state eigenstates can no longer be discriminated by the parity. Therefore, in Tab.~\ref{TwoBodyM2} and \ref{ThreeBodyM2}, we eliminate the parity. However, the selection rules discussed before $P$ breaking are very good approximations in calculating the dark matter annihilation S-matrices because of the small $v_{\Phi}$ we have adopted in this paper.

\acknowledgments
We thank  Peiwen Wu, Chen Zhang, Bin Zhu, Li-Gong Bian, Jason Evans, Feng Luo, 
Shigeki Matsumoto, Junmou Chen, Yu Jia, Yu Feng, Wenlong Sang and Gao-Liang Zhoufor helpful discussions. 
This work is supported in part by the Korea Research Fellowship Program through the National 
Research Foundation of Korea (NRF) funded by the Ministry of Science and ICT 
(2017H1D3A1A01014127) (YLT), and by NRF Research Grant NRF-2019R1A2C3005009 (PK), and by 	NRF Research Grant NRF-2018R1A2B6007000 (TM).




%



\newpage
\bibliographystyle{JHEP}
\bibliography{DMBS_Z2_strip}
\end{document}